\documentclass[aps,twocolumn,showpacs,superscriptaddress,groupedaddress]{revtex4-1}

\usepackage[all, warning]{onlyamsmath}
\usepackage{amssymb, amsmath}
\usepackage{graphicx}
\usepackage{dcolumn}
\usepackage{bm}
\usepackage[figuresright]{rotating}
\usepackage{fancyhdr}
\usepackage{enumitem}
\usepackage{siunitx}
\usepackage{indentfirst}
\usepackage{xspace}
\usepackage{epsfig}
\usepackage{grffile}
\usepackage[T1]{fontenc}
\usepackage[utf8]{inputenc}

\newcommand{\pp}{p+p}
\newcommand{\pPb}{p+\text{Pb}}
\newcommand{\gamp}{\gamma+p}
\newcommand{\pizero}{\pi^{0}}
\newcommand{\pt}{p_\text{T}}
\newcommand{\pz}{p_\text{z}}
\newcommand{\avept}{\langle p_\text{T} \rangle}
\newcommand{\sqs}{\sqrt{s}}
\newcommand{\snn}{\sqrt{s_\textrm{NN}}}
\newcommand{\elab}{E_\text{lab}}

\newcommand{\xbj}{x_{Bj}}
\newcommand{\xf}{x_{F}}
\newcommand{\xe}{x_{E}}
\newcommand{\ybeam}{y_\text{beam}}
\newcommand{\ylab}{y_\text{lab}}

\newcommand{\deltay}{\Delta y}

\newcommand{\nmf}{R_\textrm{pPb}}
\newcommand{\mev}[1]{\SI{#1}{\mega\electronvolt}}
\newcommand{\gev}[1]{\SI{#1}{\giga\electronvolt}}
\newcommand{\tev}[1]{\SI{#1}{\tera\electronvolt}}
\newcommand{\ev}[1]{\SI{#1}{\electronvolt}}
\newcommand{\hph}{\hphantom{0}}
\newcommand{\eone}{$\times10^{-1}$}
\newcommand{\etwo}{$\times10^{-2}$}
\newcommand{\ethr}{$\times10^{-3}$}
\newcommand{\efou}{$\times10^{-4}$}
\newcommand{\efiv}{$\times10^{-5}$}
\newcommand{\esix}{$\times10^{-6}$}

\newcommand{\etal}{\textsl{et al}.~}

\begin{document}

\title{Measurements of longitudinal and transverse momentum distributions for
neutral pions in the forward-rapidity region with the LHCf detector}
\newcommand{\firenzeinfn}{\affiliation{INFN Section of Florence, Italy}}
\newcommand{\firenzeuniv}{\affiliation{University of Florence, Italy}}
\newcommand{\firenzecnr}{\affiliation{IFAC-CNR, Italy}}
\newcommand{\nagoya}{\affiliation{Institute for Space-Earth Environmental Research, Nagoya University, Nagoya, Japan}}
\newcommand{\nagoyaphys}{\affiliation{Graduate school of Science, Nagoya University, Japan}}
\newcommand{\kmi}{\affiliation{Kobayashi-Maskawa Institute for the Origin of Particles and the Universe, Nagoya University, Nagoya, Japan}}
\newcommand{\polyteque}{\affiliation{Ecole-Polytechnique, Palaiseau, France}}
\newcommand{\waseda}{\affiliation{RISE, Waseda University, Japan}}
\newcommand{\cern}{\affiliation{CERN, Switzerland}}
\newcommand{\kanagawa}{\affiliation{Kanagawa University, Japan}}
\newcommand{\cataniainfn}{\affiliation{INFN Section of Catania, Italy}}
\newcommand{\cataniauniv}{\affiliation{University of Catania, Italy}}
\newcommand{\lbn}{\affiliation{LBNL, Berkeley, California, USA}}

\author{O.~Adriani}
\firenzeinfn
\firenzeuniv

\author{E.~Berti}
\firenzeinfn
\firenzeuniv

\author{L.~Bonechi}
\firenzeinfn

\author{M.~Bongi}
\firenzeinfn
\firenzeuniv

\author{R.~D'Alessandro}
\firenzeinfn
\firenzeuniv

\author{M.~Del~Prete}
\firenzeinfn
\firenzeuniv

\author{M.~Haguenauer}
\polyteque

\author{Y.~Itow}
\nagoya
\kmi

\author{T.~Iwata}
\waseda

\author{K.~Kasahara}
\waseda

\author{K.~Kawade}
\nagoya

\author{Y.~Makino}
\nagoya

\author{K.~Masuda}
\nagoya

\author{E.~Matsubayashi}
\nagoya

\author{H.~Menjo}
\nagoyaphys

\author{G.~Mitsuka}
\altaffiliation[Present address: ]{RIKEN BNL Research Center,\\
Brookhaven~National~Laboratory, USA}
\email[E-mail: ]{gaku.mitsuka@riken.jp}
\firenzeuniv

\author{Y.~Muraki}
\nagoya

\author{P.~Papini}
\firenzeinfn

\author{A.-L.~Perrot}
\cern

\author{S.~Ricciarini}
\firenzeinfn
\firenzecnr

\author{T.~Sako}
\nagoya
\kmi

\author{N.~Sakurai}
\altaffiliation[Present address: ]{Institute of Socio-Arts and Sciences,
Tokushima University, Tokushima, Japan}
\kmi

\author{T.~Suzuki}
\waseda

\author{T.~Tamura}
\kanagawa

\author{A.~Tiberio}
\firenzeinfn
\firenzeuniv

\author{S.~Torii}
\waseda

\author{A.~Tricomi}
\cataniainfn
\cataniauniv

\author{W.~C.~Turner}
\lbn

\author{M.~Ueno}
\nagoya

\author{Q.~D.~Zhou}
\nagoya

\collaboration{The LHCf Collaboration}
\noaffiliation

\date{\today}

\begin{abstract}

The differential cross sections for inclusive neutral pions as a function of
transverse and longitudinal momentum in the very forward rapidity region have
been measured at the Large Hadron Collider (LHC) with the Large Hadron Collider
forward detector (LHCf) in proton--proton collisions at $\sqs=2.76$ and
$\tev{7}$ and in proton--lead collisions at nucleon--nucleon center-of-mass
energies of $\snn = \tev{5.02}$. Such differential cross sections in
proton--proton collisions are compatible with the hypotheses of limiting
fragmentation and Feynman scaling. Comparing proton--proton with proton--lead
collisions, we find a sizable suppression of the production of neutral pions in
the differential cross sections after subtraction of ultra-peripheral
proton--lead collisions. This suppression corresponds to the nuclear
modification factor value of about 0.1--0.3.
The experimental measurements presented in this paper provide a benchmark for
the hadronic interaction Monte Carlo simulation codes that are used for the
simulation of cosmic ray air showers.

\end{abstract}


\maketitle

%
%
\section{Introduction}\label{sec:introduction}

Observations of high-energy cosmic rays with energy above $\ev{e14}$ provide key
information for a yet un-established origin(s) and acceleration mechanism(s) for
cosmic rays.
The compilation of current observations reveals kinks in the energy spectrum
that agree with the turning points in the mass composition~\cite{Kampert} at
$\sim 3\times\ev{e15}$ (the so called 'knee') and provide a consistent
description of the transition from Galactic to extragalactic cosmic rays at
$\sim 5\times\ev{e18}$ (the so called 'ankle').
In particular, a cut-off feature of ultrahigh-energy cosmic rays (UHECRs) at
$\sim 5\times\ev{e19}$ is supposed to existence of
Greisen-Zatsepin-Kuzmin~\cite{Greisen,Zatsepin} cut-off, while the source and
propagation of the UHECRs is still a mystery~\cite{Kotera}.
In order to grasp the experimental signature of the source of UHECRs, and to
understand consistent picture of transition from galactic component around
$\ev{e14}$, many extensive air-shower experiments, including on-going UHECR
observatories (i.e. Auger~\cite{Auger} and Telescope Array~\cite{TA})
have collected the data on energy spectrum, mass composition, and arrival
direction of UHECRs high energy cosmic rays over the past few
decades~\cite{CRene,CRmass,CRdir}.

It is important to note that critical parts of the analysis still depend on
Monte Carlo (MC) simulations of air shower development that are sensitive to the
choice of hadronic interaction models. Therefore different hadronic interaction
models, which simultaneously predict the soft and hard QCD interactions, provide
different viewpoints even using exactly the same data
compilation~\cite{Kampert,Ulrich}. Currently the lack of knowledge about forward
particle production in hadronic collisions at high energy hinders the
interpretation of observations of high-energy cosmic rays~\cite{Ulrich}.

Here it should be remarked that the LHC at CERN has so far reached $\tev{13}$
centre-of-mass energy in proton--proton ($\pp$) collisions.
This energy corresponds to the cosmic ray energy $9.0\times\ev{e16}$ in the
target rest frame which is well above the first turning point in the mass
composition of primary cosmic rays from proton dominated to light nuclei
dominated, namely the knee at approximately $3\times\ev{e15}$~\cite{Selvon}. The
data provided by the LHC in the forward region, defined as the fragmentation
region of a projectile particle, should thus provide a useful benchmark for the
MC simulation codes that are used for the simulation of air showers.

The energy in the laboratory frame converted from the collision energy in $\pp$
collisions at $\sqs = \tev{7}$ ($\elab = 2.6\times\ev{e16}$) is two orders of
magnitude lower than the ankle region where a transition from galactic to
extragalactic cosmic rays may occur.
However, extrapolation from the LHC energy range to a higher energy range can be
achieved by using a scaling law in the forward rapidity region. One possibility
for such a scaling law is the hypothesis of limiting
fragmentation~\cite{Limiting1,Limiting2,Limiting3}, which specifies that the
secondary particles will approach a limiting distribution of rapidity in the
rest frame of the target hadron. In this case the fragmentation of a colliding
hadron would occur independently of the center-of-mass energy and then the
differential cross sections as a function of rapidity (hereafter rapidity
distributions) in the fragmentation region, namely the forward rapidity region,
would form a limiting distribution.

Understanding particle production in nucleon--nucleus or nucleus--nucleus
interactions is also of importance for ultrahigh-energy cosmic ray interactions,
where parton density in nuclei is expected to be enhanced by $\propto A^{1/3}$.
The presence of a high gluon density in the nucleus is known to greatly modify
the absolute yield and the momentum distribution of the particles that are
produced~\cite{Albacete}.

The LHCf experiment~\cite{LHCfTDR} is designed to measure the hadronic
production cross sections of neutral particles at very forward angles in $\pp$
and proton--lead ($\pPb$) collisions. The LHCf experiment also provides a unique
opportunity to investigate all the effects mentioned in the previous paragraph,
namely, the limiting fragmentation, the Feynman scaling~\cite{Feynman}, and the
high parton density in nuclear target. In a previous
publication~\cite{LHCfpppi0} we presented the $\pizero$ production cross
sections as a function of the transverse momentum (hereafter $\pt$
distributions) in $\pp$ collisions at $\sqs=\tev{7}$. However tests of the
limiting fragmentation and the Feynman scaling predictions were not performed.
Conversely, in the analysis of this paper, the comparison of the LHCf data taken
in $\pp$ collisions at $\sqs=2.76$ and $\tev{7}$ respectively makes it possible
to perform these tests. In addition the analysis presented in this paper has
updates that lead to a deeper understanding of forward $\pizero$ production
compared to our previous publications~\cite{LHCfpppi0,LHCfpPbpi0}: the upper
range for $\pt$ analysis is extended to $\gev{1.0}$ and differential cross
sections as a function of longitudinal momentum (hereafter $\pz$ distributions)
as well as $\pt$ distributions are presented.

The paper is organized as follows. In Sec.~\ref{sec:detector}, the LHCf
detectors are described. Sections~\ref{sec:data} and \ref{sec:mc_simulation}
summarize the conditions for taking data and the MC simulation methodology,
respectively. In Sec.~\ref{sec:framework}, the analysis framework and the
factors that contribute to the systematic uncertainty of the results are
explained. In Sec.~\ref{sec:result} the analysis results are presented and
compared with the predictions of several hadronic interaction models. In
Sec.~\ref{sec:discussions} the analysis results for $\pp$ and $\pPb$ collisions
are described. Finally, concluding remarks are found in
Sec.~\ref{sec:conclusions}.

%
%
\section{The LHCf detector}\label{sec:detector}

Two independent detectors called LHCf Arm1 and LHCf Arm2 were assembled to study
$\pp$ and $\pPb$ collisions at the LHC~\cite{LHCfJINST}. In $\pp$ collisions at
$\sqs = \tev{7}$, both LHCf Arm1 and LHCf Arm2 detectors were operated to
measure the neutral secondary particles emitted into the positive and negative
large rapidity regions, respectively. In $\pp$ collisions at $\sqs = \tev{2.76}$
and $\pPb$ collisions at $\snn = \tev{5.02}$, only the LHCf Arm2 detector was
used to measure the neutral secondary particles emitted into the negative
rapidity region (the proton remnant side in $\pPb$ collisions). Here the
rapidity $y$ is defined as $y = \tanh^{-1}(\pz/E)$~\cite{PDG}.

The LHCf detectors each consist of two sampling and imaging calorimeters
composed of 44 radiation lengths of tungsten and 16 sampling layers of
\SI{3}{\milli\meter} thick plastic scintillators. The transverse sizes of the
calorimeters are 20$\times$\SI{20}{\square\milli\meter} and
40$\times$\SI{40}{\square\milli\meter} for Arm1, and
25$\times$\SI{25}{\square\milli\meter} and
32$\times$\SI{32}{\square\milli\meter} for Arm2.
The smaller and larger calorimeters are hereafter called the \textit{Small
Calorimeter} and the \textit{Large Calorimeter}, respectively. Four X-Y layers
of position-sensitive detectors are interleaved with the layers of tungsten and
scintillator in order to provide the transverse profiles of the showers.
Scintillating fiber (SciFi) belts~\cite{LHCfScifi} are used for Arm1 and silicon
microstrip sensors~\cite{LHCfSilicon} are used for Arm2. Readout pitches are
\SI{1}{\milli\meter} and \SI{0.16}{\milli\meter} for Arm1 and Arm2,
respectively. The Front Counters, additional components of the LHCf detectors,
are simple thin plastic scintillators (80$\times$\SI{80}{\square\milli\meter})
and are installed in front of the LHCf calorimeters.
They act as monitors for beam-beam collision rates with a higher detection
efficiency than the LHCf calorimeters.

The LHCf detectors were installed in the instrumentation slots of the target
neutral absorbers (TANs)~\cite{TAN} located $\pm$\SI{140}{\meter} from the ATLAS
interaction point (IP1) in the direction of the LHCb interaction point for Arm1
and in the direction of the ALICE interaction point for Arm2, and at a
zero-degree collision angle. The trajectories of charged particles produced at
IP1 and directed towards the TANs are deflected by the inner beam separation
dipole magnets D1 before reaching the TANs themselves. Consequently, only
neutral particles produced at IP1 enter the LHCf detectors. The vertical
positions of the LHCf detectors in the TANs are manipulated so that the LHCf
detectors cover the pseudorapidity range from 8.4 to infinity for a beam
crossing half angle of \SI{145}{\micro\radian}. The Small Calorimeter
effectively covers the zero-degree collision angle. Following $\pPb$ collision
operation, the LHCf detectors were removed from the TAN instrumentation slots in
April, 2013 in order to protect them from radiation damage when the LHC is
operated at high luminosity.

LHCf triggers are generated at three levels~\cite{LHCfIJMPA}. The first level
trigger is generated from beam pickup signals when a bunch passes IP1. A shower
trigger is then generated when signals from any successive three scintillation
layers in any calorimeter exceeded a predefined threshold. The shower trigger
threshold is chosen to detect photons greater than $\gev{100}$ with an
efficiency of $>\SI{99}{\percent}$. A second level trigger is generated when a
shower trigger has occurred and the data acquisition system is activated. The
highest level trigger, or third level trigger, is generated when a specified
combination of shower triggers, front counter triggers and data acquisition
trigger has occurred.
The live time efficiency of the data acquisition systems is defined as the ratio
of the number of second level triggers to the number of shower triggers. The
efficiency depends on the luminosity during the data taking period and is always
less than unity due to pileup. The final results shown are corrected for the
live time efficiency.

More details on the scientific goals of the experiment are given in
Ref.~\cite{LHCfTDR}. The performance of the LHCf detectors has been studied in
previous reports~\cite{LHCfSPS,LHCfIJMPA}.

%
%
\section{Experimental data taking conditions}\label{sec:data}

The experimental data used for the analysis in this paper were obtained at three
different collision energies and colliding particle configurations. Data taking
conditions are explained in the subsections below, ordered according to the
dates of the operation periods with the earliest first.

\subsection{$\pp$ collisions at $\sqs=\tev{7}$}\label{sec:pp7TeV}

The data in $\pp$ collisions at $\sqs=\tev{7}$ with a zero-degree beam crossing
angle were obtained from May 15 to 22, 2010 (LHC Fills 1104, 1107, 1112, and
1117). The events that were recorded during a luminosity optimization scan and a
calibration run were removed from the data sets for this analysis. The
integrated luminosities for the data analysis reported in this paper were
derived from the counting rate of the Front Counters~\cite{LHCfLumi} and were
$\SI{2.67}{\nano\barn^{-1}}$ (Arm1) and $\SI{2.10}{\nano\barn^{-1}}$ (Arm2) after
taking the live time efficiencies into account.

Pileup interactions in the same bunch crossing may increase the multi-hit events
that have more than one shower event in a single calorimeter, leading to a
potential bias in the momentum distributions of $\pizero$s. The contamination of
multi-hit events due to pileup interactions is estimated to be only
\SI{0.2}{\percent} and therefore produces a negligible effect~\cite{LHCfpppi0}.
Detailed discussions of background events from collisions between the beam and
residual gas molecules in the beam tube can be found in a previous
report~\cite{LHCfIJMPA}.

\subsection{$\pPb$ collisions at $\snn=\tev{5.02}$}\label{sec:pPb5.02TeV}

The data in $\pPb$ collisions were obtained at $\snn = \tev{5.02}$ with
\SI{145}{\micro\radian} beam crossing half angle and with only the Arm2 detector
recording data on the proton remnant side. The beam energies were \tev{4} for
protons and \tev{1.58} per nucleon for Pb nuclei.
Because of the asymmetric beam energies where the proton beam travels at
$\theta=\pi$ and the Pb beam at $\theta=0$, the nucleon--nucleon center-of-mass
in $\pPb$ collisions is shifted to rapidity $-0.465$ ($=1/2 \times \log((A_p
Z_\text{Pb})/(Z_p A_\text{Pb}))$ where $A$ and $Z$ are the mass and atomic
numbers, respectively~\cite{Salgado}).

Data used in this analysis were taken in two different fills; during LHC Fill
3478 on January 21, 2013 and during LHC Fill 3481 on January 21 and 22. The
integrated luminosity of the data was \SI{0.63}{\nano\barn^{-1}} after
correcting for the live time efficiencies of the data acquisition
systems~\cite{CERNLumi}. The trigger scheme was essentially identical to that
used in $\pp$ collisions at $\sqs = \tev{7}$. The bunch spacing in $\pPb$
collisions (\SI{200}{\nano\second}), which was smaller than the gate width for
analog to digital conversion in the LHCf data acquisition system
(\SI{500}{\nano\second}) created the possibility of integrating two or at most
three signal pulses from the pileup of successive $\pPb$ collisions. The
probability for this to occur was estimated from the timing distribution for
shower triggers and was less than $\SI{5}{\percent}$. Contamination by
successive collisions is not corrected for in this study, while it is considered
in the beam-related systematic uncertainty. The contamination of multi-hit
events due to pileup interactions is negligible (\SI{0.4}{\percent}).

It should be remarked that beam divergence causes a smeared beam spot at the
TAN, leading to a bias in the measured momentum distributions. The effect of a
non-zero beam spot size at the TAN was evaluated with MC simulations (see
Ref.~\cite{LHCfpPbpi0}). This effect is taken into account in the final results
reported for the $\pt$ and $\pz$ distributions.

\subsection{$\pp$ collisions at $\sqs=\tev{2.76}$}\label{sec:pp2.76TeV}

The data in $\pp$ collisions at $\sqs = \tev{2.76}$ were obtained with a
\SI{145}{\micro\radian} beam crossing half angle and beam energy \tev{1.38} for
each proton. Data used in this analysis were taken during LHC Fill 3563 on
February 13, 2013. The integrated luminosity for this data was
\SI{2.36}{\nano\barn^{-1}} after correcting for the live time efficiencies of
the data acquisition system~\cite{ATLASPbPb}. The trigger scheme, trigger
efficiency, and contamination of multi-hit events were mostly the same as the
$\pPb$ collision data at $\snn=\tev{5.02}$. The effects of beam divergence were
dealt with in the same way as was described for $\pPb$ collisions at
$\snn=\tev{5.02}$ (Sec.~\ref{sec:pPb5.02TeV}).

%
%
\section{Monte Carlo simulations methodology}\label{sec:mc_simulation}

MC simulations have been performed in two steps:\\
(I) Event generation in $\pp$ and $\pPb$ collisions at IP1
(Sec.~\ref{sec:mc_signal}) and (II) particle transport from IP1 to the LHCf
detectors and consequent simulation of the response of the LHCf detectors
(Sec.~\ref{sec:mc_detector}).

MC simulation events are generated following steps (I) and (II) and are used for
the validation of reconstruction algorithms, determination of cut criteria, and
determination of the response matrix for momentum distributions unfolding.
Conversely, MC simulations that are used only for comparison with the
measurement results in Sec.~\ref{sec:result} are limited to step (I) only, since
the final $\pt$ and $\pz$ distributions in Sec.~\ref{sec:result} are already
corrected for detector response and eventual reconstruction bias. The
statistical uncertainties of the MC simulations used in this paper are
negligibly small compared to the statistical uncertainties of the LHCf data.

\subsection{Collision event modeling}\label{sec:mc_signal}

Collision event modeling of $\pp$ hadronic interactions at $\sqs=2.76$ and
$\tev{7}$ are simulated and the resulting fluxes of secondary particles are
generated with several event generators: \textsc{dpmjet} 3.06~\cite{DPMJET},
\textsc{qgsjet} II-04~\cite{QGSJET}, \textsc{sibyll} 2.1~\cite{SIBYLL},
\textsc{epos lhc}~\cite{EPOS} and \textsc{pythia} 8.185~\cite{PYTHIA6,
PYTHIA8b}. Hereafter the version number for these event generators is omitted
for simplicity, unless otherwise noted.

In the analysis of this paper, we use the integrated interface \textsc{crmc}
1.5.3~\cite{CRMC} for executing the first four event generators, whereas the
fifth event generator, \textsc{pythia}, serves as its own front end for the
generation of proton--proton hadronic interaction events.

Events in $\pPb$ collisions are divided into two categories according to the
value of the impact parameter: (1) general hadronic interactions and (2)
ultraperipheral collisions (UPCs).
Category (1) occurs when the impact parameter between $p$ and Pb is smaller than
the sum of their radii. These inelastic $\pPb$ interactions at $\snn =
\tev{5.02}$ are simulated using the hadronic interaction models \textsc{dpmjet},
\textsc{qgsjet}, and \textsc{epos} with the \textsc{crmc} interface.
\textsc{sibyll} was not used because it only supports nuclei lighter than Fe.
\textsc{pythia} also does not support heavy ion collisions and thus was also not
used for $\pPb$ collisions.

Category (2) $\pPb$ UPCs occur when the impact parameter is larger than the sum
of $p$ and Pb radii. The UPC events are simulated by the combination of
\textsc{starlight}~\cite{STARLIGHT} for the virtual photon flux, \textsc{sophia}
2.1~\cite{SOPHIA} for low-energy photon--proton interactions, and either
\textsc{dpmjet} 3.05~\cite{DPMJET} or \textsc{pythia} 6.428~\cite{PYTHIA6} for
high-energy photon--proton interactions. The UPC simulation distributions used
in this analysis are taken from the average of two UPC simulations; one using
\textsc{dpmjet} 3.05 and the second using \textsc{pythia} 6.428 for the
high-energy photon--proton interaction. Differences between these two UPC
simulations are taken into account as a systematic uncertainty in the UPC
simulation. See Ref.~\cite{LHCfUPC} for more details.

In both $\pp$ and $\pPb$ collisions, the MC events used for the determination of
the response matrix for unfolding the momentum distributions
(Sec.~\ref{sec:correction}) are simulated by \textsc{pythia} at the requisite
beam energies. A single $\pizero$ with energy larger than \gev{100} and possible
associated background particles are selected from the secondary particles
produced. There is no significant dependence of the unfolding performance on the
choice of event generator for the MC simulation events that are used for the
response matrix. This was verified by repeating event simulations with three of
the event generators; \textsc{dpmjet}, \textsc{pythia} and \textsc{epos}.

In all of the MC simulations, the $\pizero$s from short-lived particles that
decay within \SI{1}{\meter} of IP1, e.g. $\eta$, $\rho$, $\omega$, etc.
($\lesssim\SI{10}{\percent}$ for each relative to all $\pizero$s), are accounted
for consistently in the treatment of LHCf data.
The \SI{145}{\micro\radian} beam crossing half angle is also taken into account
for $\pp$ collisions at $\sqs=\tev{2.76}$ and for $\pPb$ collisions at
$\snn=\tev{5.02}$.

\subsection{Simulation of particle transport from IP1 to the LHCf detector and
of the detector response}\label{sec:mc_detector}

Transport of secondary particles inside the beam pipe from IP1 to the TAN, the
electromagnetic and hadronic showers produced in the LHCf detector by the
particles arriving at the TAN and the detector response are simulated with the
\textsc{cosmos} and \textsc{epics} libraries~\cite{EPICS}.

Secondary particles produced by the interaction between IP1 collision products
and the beam pipe are also taken into account in this step. The secondary
particles from beam pipe interaction events generally have energy well below
$\gev{100}$ and thus provide no bias to the momentum distributions of collision
events that focus only on energies above $\gev{100}$. The survey data for
detector position and random fluctuations due to electrical noise are also taken
into account in this step. See Ref.~\cite{LHCfpppi0} for more details.

%
%
\section{Analysis framework}\label{sec:framework}

\subsection{$\pizero$ event reconstruction and selection}
\label{sec:reconstruction}

The standard reconstruction algorithms consist of four steps: hit position
reconstruction, energy reconstruction, particle identification, and $\pizero$
event selection.

\subsubsection{Position reconstruction}\label{sec:posreco}

Hit position reconstruction starts with a search for multi-hit and single hit
events. A multi-hit event is defined to have more than one photon registered in
a single calorimeter. A single-hit event is defined to have a single hit in each
of the two calorimeters in a given detector, Arm1 or Arm2.

Therefore multi-hit event candidates should have two or more distinct peaks in
the lateral-shower-impact-distribution of a given calorimeter and are then
identified using the TSpectrum algorithm~\cite{TSpectrum} implemented in
\textsc{root}~\cite{ROOT}.
TSpectrum provided the basic functionality for peak-finding in a spectrum with a
continuous background and statistical fluctuations.

The MC simulation estimated efficiencies for identifying multi-hit events are
larger than \SI{70}{\percent} and \SI{90}{\percent} for Arm1 and Arm2,
respectively~\cite{LHCfIJMPA}. Given the list of shower peak position candidates
that have been obtained above, the lateral distributions are fit to a Lorenzian
function~\cite{Lednev} to obtain more precise estimates of the shower peak
positions, heights, and widths. In the case of multi-hit events, two peaks are
fit using superimposed Lorenzian functions. Multi-hit events with three or more
peaks are rejected from the analysis.
Conversely, single-hit events, not having two or more identifiable peaks in a
single calorimeter but having a single hit in each calorimeter are correctly
selected with an efficiency better than \SI{98}{\percent} for true single-photon
events with energy greater than $\gev{100}$ for both Arm1 and Arm2.

\subsubsection{Energy reconstruction}\label{sec:enereco}

The photon energy is reconstructed using the measured energy deposited in the
LHCf calorimeters. The charge information in each scintillation layer is first
converted to a deposited energy by using the calibration coefficients obtained
from the SPS electron test beam data taken below \gev{200}~\cite{LHCfSPS}. The
sum of the energy deposited in the 2nd to 13th scintillation layers is then
converted to the primary photon energy using an empirical function. The
coefficients of the function are determined from the response of the
calorimeters to single photons using MC simulations. Corrections for shower
leakage effects and the light-yield collection efficiency of the scintillation
layers are carried out during the energy reconstruction
process~\cite{LHCfJINST}. In the case of multi-hit events, the reconstructed
energy based on the measured energy deposited is split into two energies,
primary and secondary. Fractions of the energy for the primary and secondary
hits are determined according to the peak height and width of the corresponding
distinct peaks in the lateral-shower-impact-distribution.

\subsubsection{Particle identification}\label{sec:pid}

Particle identification (PID) is applied in order to efficiently select pure
electromagnetic showers and to reduce hadron (predominantly neutron)
contamination. PID in the study of this paper depends only on the parameter
$L_{90\%}$. $L_{90\%}$ is defined as the longitudinal distance, in units of
radiation length ($X_0$), measured from the 1st tungsten layer of the
calorimeter to the position where the energy deposition integral reaches
\SI{90}{\percent} of the total shower energy deposition. Events with an
electromagnetic shower generally have a $L_{90\%}$ value smaller than 20\,$X_0$,
while events with a hadronic shower generally have $L_{90\%}$ larger than
20\,$X_0$. The threshold $L_{90\%}$ value as a function of the photon energy is
defined in order to keep the $\pizero$ selection efficiency at \SI{90}{\percent}
over the entire energy range of the individual photons. PID criteria are
determined by MC simulations for each calorimeter.

\subsubsection{$\pizero$ event selection}\label{sec:pi0selection}

\begin{figure}[htbp]
  \centering
  \includegraphics[width=4.1cm, keepaspectratio]{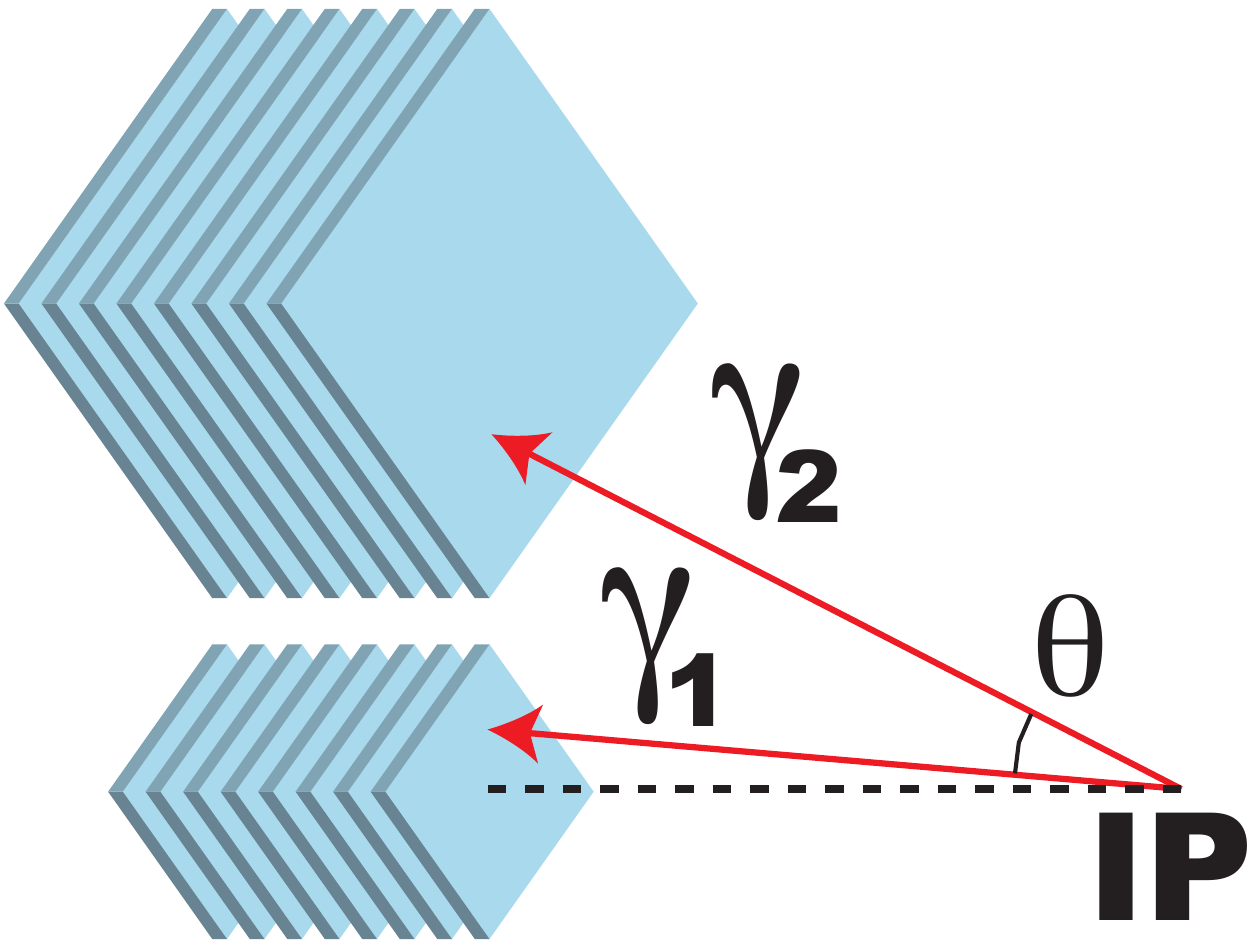}
  \includegraphics[width=4.1cm, keepaspectratio]{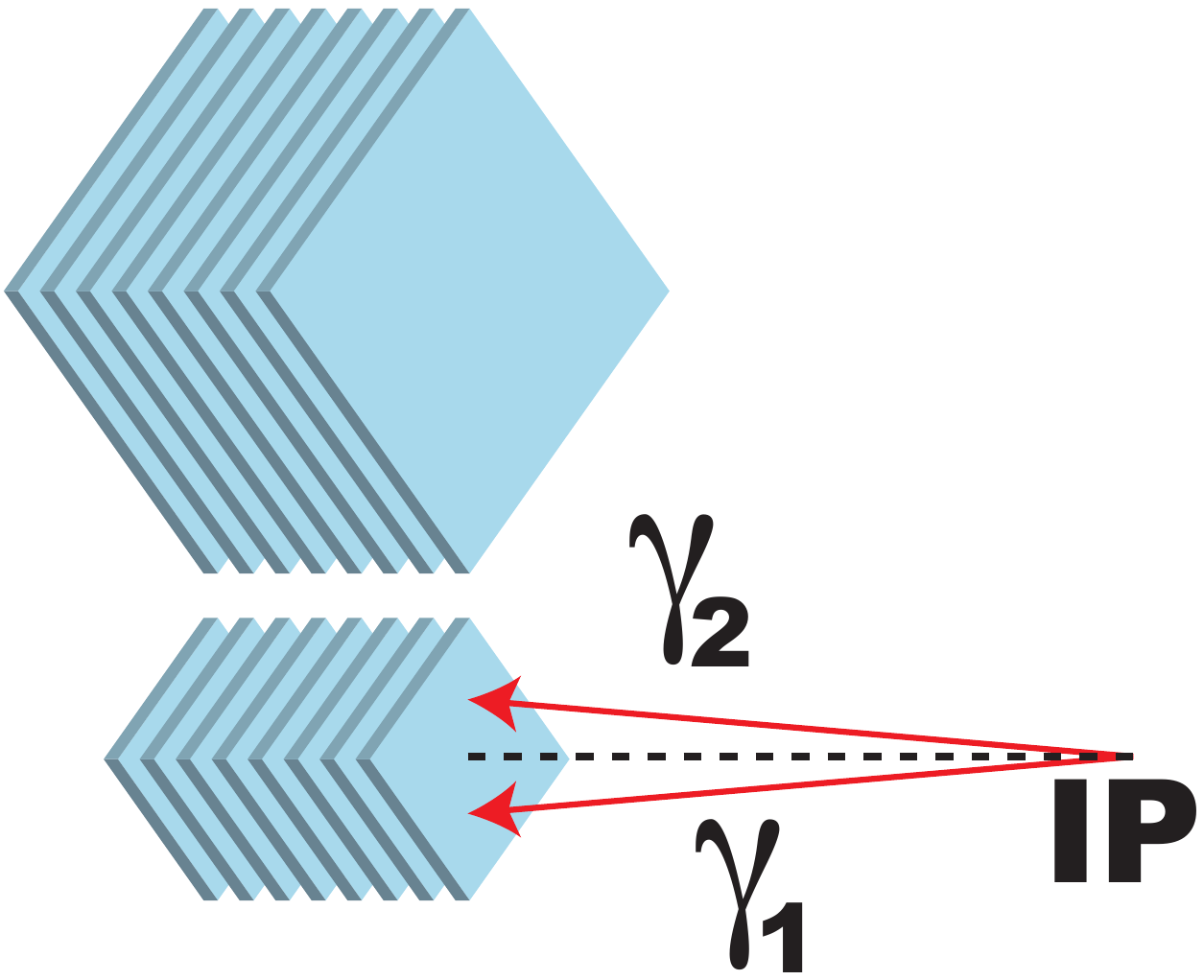}
  \caption{(color online). Observation of $\pizero$ decay by a LHCf detector.
  Left: Type-I $\pizero$ event having one photon entering each calorimeter.
  Right: Type-II $\pizero$ event having two photons entering one calorimeter,
  here entering the small calorimeter.}
  \label{fig:pi0type}
\end{figure}

The $\pizero$ are then identified by their decay into two photons, leading to
the distinct peak in the invariant mass distribution around the $\pizero$ rest
mass. The invariant mass of the two photons is calculated using the
reconstructed photon energies and incident positions. The $\pizero$ events used
in the analysis of this paper are classified into two categories:
\textit{Type-I} $\pizero$ and \textit{Type-II} $\pizero$ events. A Type-I event
is defined as having a single photon in each of the two calorimeters of Arm1 or
Arm2 (the left panel of Fig.~\ref{fig:pi0type}). A Type-II event is defined as
having two photons in the same calorimeter (the right panel of
Fig.~\ref{fig:pi0type}). Note that Type-II events were not used in the previous
analyses~\cite{LHCfpppi0,LHCfpPbpi0}, and thus are taken into account for the
first time in this paper. As detailed in Sec.~\ref{sec:correction}, the phase
spaces covered by Type-I and Type-II events are complementary. In particular,
the inclusion of Type-II events extends the $\pt$ upper limit for analysis from
$\gev{0.6}$ in the previous analyses to $\gev{1.0}$.

\begin{figure}[htbp]
  \centering
  \includegraphics[width=8.4cm, keepaspectratio]{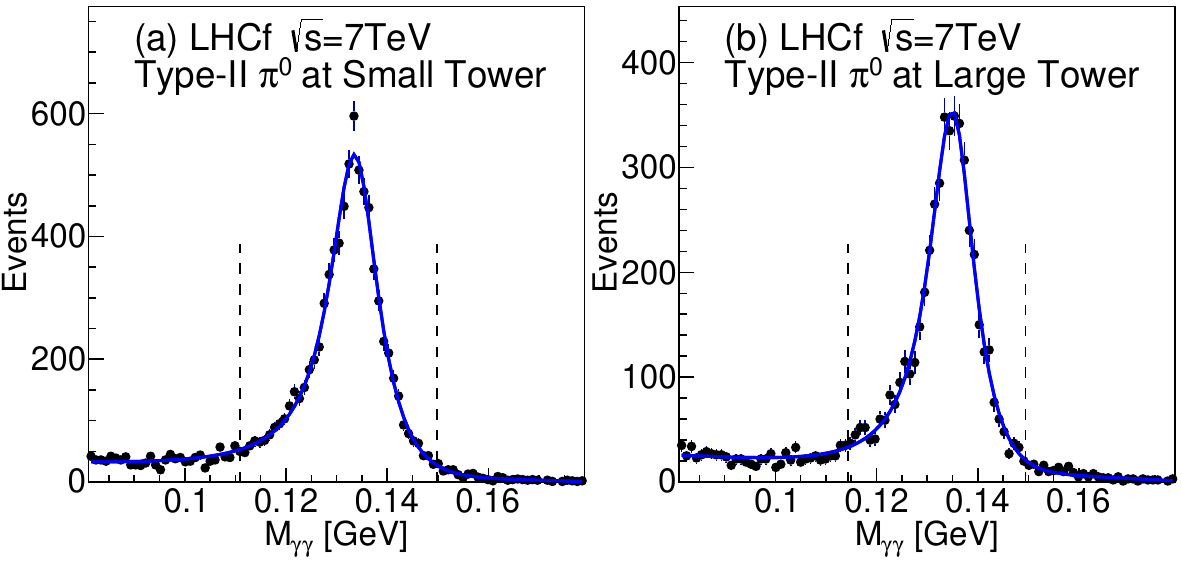}
  \caption{(color online). Reconstructed invariant mass distributions in $\pp$
  collisions at $\sqs=\tev{7}$. Left: Type-II $\pizero$ events in the Arm2 small
  calorimeter. Right: Type-II $\pizero$ events in the Arm2 large calorimeter. The
  solid curves show the best-fit composite physics model to the invariant mass
  distributions.}
  \label{fig:mass-fit}
\end{figure}

Figure~\ref{fig:mass-fit} shows the reconstructed two-photon invariant mass
($M_{\gamma\gamma}$) distributions of LHCf data in the rapidity range $8.8 < y <
10.8$.
The left and right panels of Fig.~\ref{fig:mass-fit} show the distributions for
Type-II events in the Arm2 small calorimeter and Arm2 large calorimeter
respectively. The sharp peaks around \mev{135} are due to $\pizero$ events.
The distributions in Fig.~\ref{fig:mass-fit} are based only on data from $\pp$
collisions at $\sqs=\tev{7}$ during LHC Fill 1104.
Similar invariant mass distributions are obtained from other fills and from
Arm1. Kinematic quantities of the $\pizero$s (four-momenta, $\pt$, $\pz$ and
rapidity) are reconstructed by using the photon energies and incident positions
measured by the LHCf calorimeters, and are used for producing the $\pt$ and
$\pz$ distributions. The projected position of the proton beam axis on the LHCf
detector (beam center) is used in order to derive the correct $\pt$ and $\pz$
values of each event. The beam center position is obtained from the LHCf
position-sensitive detectors of Arm1 and Arm2 for each fill.

\begin{figure}[htbp]
  \begin{center}
  \includegraphics[width=8.5cm, keepaspectratio]{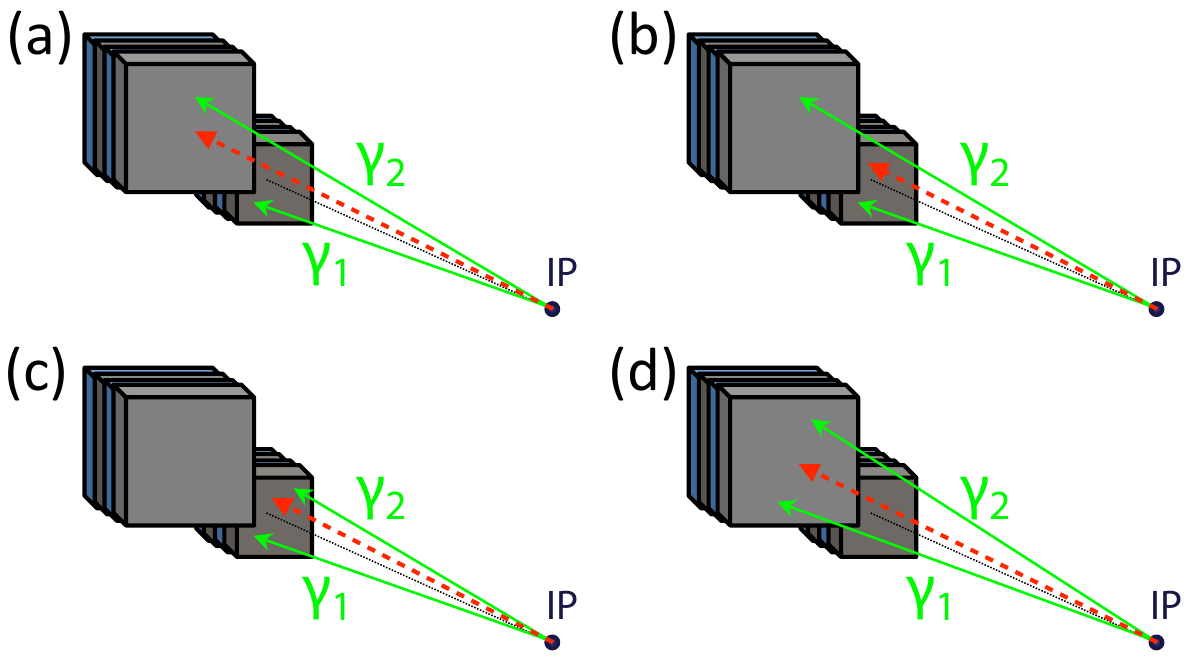}
  \caption{(color online). Diagrams of all multi-hit events that are rejected.
  Panels (a) and (b) show the multi-hit Type-I $\pizero$ events and panels (c)
  and (d) show the multi-hit Type-II $\pizero$ events. Red and green arrows
  indicate a background particle not originating in a $\pizero$ decay and two
  photons originating in a $\pizero$ decay, respectively.}
  \label{fig:multipi0}
  \end{center}
\end{figure}

The $\pizero$ event selection criteria that are applied prior to the
reconstruction of the $\pizero$ kinematics are summarized in
Table.~\ref{tbl:eventselection}. Type-I events accompanied by at least one
additional background particle in one of the two calorimeters (usually a photon
or a neutron) and not originating in a $\pizero$ decay are denoted as multi-hit
$\pizero$ events and are rejected as background events. Similarly, Type-II
events accompanied by at least one additional background particle in the
calorimeter used for $\pizero$ identification are rejected.
Figure~\ref{fig:multipi0} shows diagrams of all types of multi-hit events that
are rejected. Panels (a) and (b) show the multi-hit Type-I $\pizero$ events and
panels (c) and (d) show the multi-hit Type-II $\pizero$ events. Red and green
arrows indicate a background particle not originating in a $\pizero$ decay and
two photons originating in a $\pizero$ decay, respectively.
The final inclusive production rates reported in this paper are corrected for
these cut efficiencies and will be discussed in Sec.~\ref{sec:correction}.

\begin{table}
    \caption{Summary of criteria for selection of the $\pizero$ sample.}
    \begin{ruledtabular}
    \begin{tabular}{l l}
      \multicolumn{2}{c}{Type-I $\pizero$ events} \\
      \cline{1-2}
      Incident position & Within \SI{2}{\milli\meter} from the edge of
      calorimeter \\
      Energy threshold  & $E_\text{photon} > \gev{100}$    \\
      Number of hits    & Single-hit in each calorimeter  \\
      PID               & Photon-like in each calorimeter \\ \\
      \multicolumn{2}{c}{Type-II $\pizero$ events} \\
      \cline{1-2}
      Incident position & Within \SI{2}{\milli\meter} from the edge of
      calorimeter \\
      Energy threshold  & $E_\text{photon} > \gev{100}$    \\
      Number of hits    & Two hits \\
      PID               & Photon-like\\
    \end{tabular}
    \end{ruledtabular}
    \label{tbl:eventselection}
\end{table}

\subsection{Corrections for experimental effects}\label{sec:correction}

The raw $\pt$ and $\pz$ distributions of $\pizero$s are corrected for: (1)
contamination by background events, (2) reconstruction inefficiency and the
smearing caused by finite position and energy resolutions, (3) geometrical
acceptance and the branching ratio of $\pizero$ decay, and (4) the efficiency of
the multi-hit $\pizero$ cut.
We now discuss each of these corrections in some detail.

\subsubsection{Background contamination}\label{sec:bgsubtract}

First, the background contamination of the $\pizero$ events from hadronic
events, and from the coincidence of two photons not originating from the decay
of a single $\pizero$ are estimated using a sideband method~\cite{LHCfpppi0}. As
shown in Fig.~\ref{fig:mass-fit} for instance, the reconstructed two-photon
invariant mass distributions of LHCf data are fit to a composite physics model
(solid blue curve). The model consists of an asymmetric Gaussian distribution
for the $\pizero$ signal component and a third order Chebyshev polynomial
function for the background component. The fit is performed over the two photon
invariant mass range $0.08 < M_{\gamma\gamma} < \gev{0.18}$. The $\pizero$
signal window is defined by the two dashed vertical lines in
Fig.~\ref{fig:mass-fit} that are placed $\pm 3\sigma$ from the mean value.
Here the mean value and the standard deviation are obtained from the best-fit
asymmetric Gaussian distribution. The background window is defined as the region
within $\pm 6\sigma$ distance from the peak value and excluding the $\pizero$
signal window. The fraction of the background component included in the
$\pizero$ signal window can be estimated using the ratio of the integral of the
best-fit third order Chebyshev function over the $\pizero$ signal window divided
by the integral over the $\pizero$ signal and background windows. The width of
the asymmetric Gaussian function comes from the detector response, predominantly
from shower leakage near the edges of the calorimeters. The reconstructed energy
is corrected for shower leakage.

\subsubsection{Reconstruction inefficiency and smearing in position and energy
resolution}\label{sec:unfolding}

Second, a spectrum unfolding is performed to simultaneously correct for both the
reconstruction inefficiency and the smearing caused by finite position and
energy resolution. The contamination by background events that has been
estimated by the sideband method is taken into account in the unfolding process.
We follow basically same unfolding procedure as in the previous
analyses~\cite{LHCfpppi0,LHCfpPbpi0}, although the unfolding algorithm is based
on a fully Bayesian unfolding method~\cite{Choudalakis} instead of an iterative
Bayesian unfolding method~\cite{Dagostini}. The calculation of the ``a
posteriori'' probability in multi-dimensional space (the measured spectrum
multiplied by the true spectrum) is achieved using a Markov Chain Monte Carlo
simulation~\cite{Caldwell}. The convergence of the Markov Chain Monte Carlo
simulation is ensured by the Gelman-Rubin test~\cite{Gelman}. Production of the
MC events used for the calculation of the response matrix for the unfolding is
explained in Sec.~\ref{sec:mc_signal}.

\subsubsection{Geometric acceptance and branching ratio corrections}
\label{sec:geometry}

\begin{figure}[htbp]
  \begin{center}
  \includegraphics[width=8.5cm, keepaspectratio]{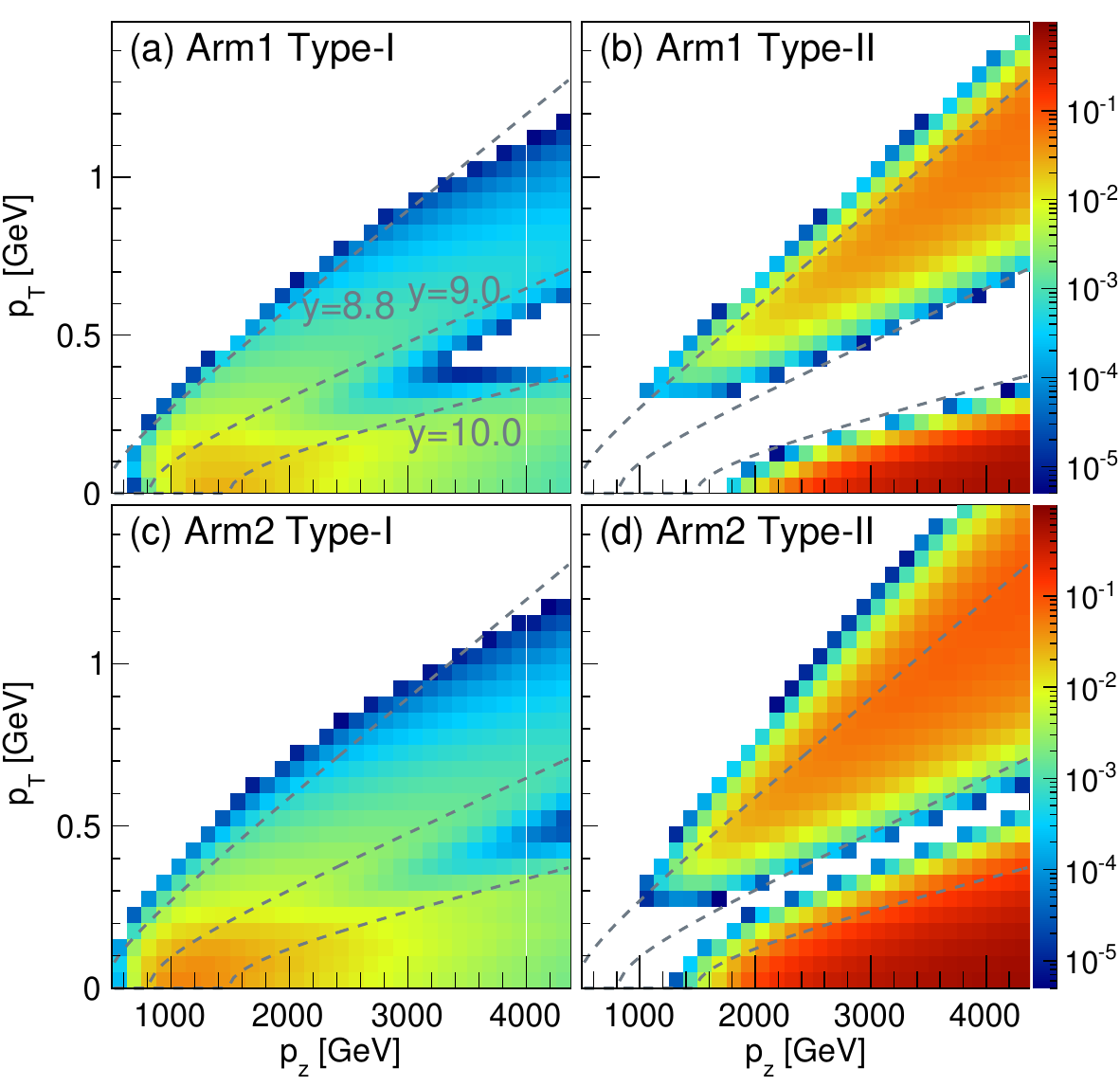}
  \caption{(color online). The acceptance map of $\pizero$ detection by the LHCf
  detectors in $\pz$--$\pt$ phase space: Arm1 Type-I (left top), Arm1 Type-II
  (right top), Arm2 Type-I (left bottom), and Arm2 Type-II (right bottom).
  The fiducial area cuts and energy threshold ($E_{\text{photon}}>\gev{100}$) are
  taken into account. Dashed curves indicate lines of constant rapidity
  $\pizero$s, $y$ = 8.8, 9.0, and 10.0 reading from top to bottom.}
  \label{fig:acceptance}
  \end{center}
\end{figure}

Thirdly, the limiting aperture of the LHCf calorimeters is estimated by using MC
simulations. The procedure for performing MC simulations is given in
Ref.~\cite{LHCfpppi0}. Figure~\ref{fig:acceptance} shows the acceptance
efficiency as a function of the $\pizero$ $\pz$ and $\pt$. The acceptance
efficiency has been obtained by taking the ratio of the $\pz$--$\pt$
distribution of $\pizero$s that are within the aperture of the LHCf calorimeters
divided by the distribution of all simulated $\pizero$s. The fiducial
cuts~\cite{LHCfpppi0} and reconstructed energy cut (both of the $\pizero$ decay
photons must have $E>\gev{100}$) are also applied to the accepted $\pizero$
events. Dashed curves in Fig.~\ref{fig:acceptance} indicate lines of constant
$\pizero$ rapidity. The acceptance efficiencies in Fig.~\ref{fig:acceptance} are
purely kinematic and do not depend upon a particular hadronic interaction model.
The aperture correction is achieved by dividing, point by point, the
distributions before the acceptance correction by the acceptance efficiency.
The branching ratio inefficiency is due to $\pizero$ decay into channels other
than two photons. The branching ratio for $\pizero$ decay into two photons is
\SI{98.8}{\percent} and is taken into account by increasing the $\pizero$
acceptance efficiency by \SI{1.2}{\percent}.

\subsubsection{Loss of events due to the multi-hit $\pizero$ cut}
\label{sec:multipi0}

Fourth, in order that the reported $\pizero$ distributions represent inclusive
cross sections it is necessary to correct the data for the loss of events due to
the multi-hit cut (Sec.~\ref{sec:pi0selection}).
The correction factor is defined as $f^\textrm{multihit}_i =
(N^\textrm{multi}_i+N^\textrm{single}_i)/N^\textrm{single}_i$, where
$N^\textrm{multi}_i$ and $N^\textrm{single}_i$ are the number of expected
multihit and single-hit $\pizero$ events in the $i$-th bin respectively. The
factors $f^\textrm{multihit}_i$ are estimated using hadronic interaction models
introduced in Sec.~\ref{sec:mc_signal} and are in the range
$1.0<f^\textrm{multihit}_i<1.1$ over all the $\pt$ and $\pz$ bins. LHCf $\pt$
and $\pz$ distributions are then multiplied by the average of these factors for
the various interaction models and their contribution to the systematic
uncertainty is derived from the observed variations amongst the interaction
models.
Consequently, the single-hit $\pizero$ distributions are corrected to represent
inclusive $\pizero$ production distributions. All the procedures just described
have been verified using the MC simulations introduced in
Sec.~\ref{sec:mc_signal}.

\subsection{Systematic uncertainties}\label{sec:systerror}

Systematic uncertainties are determined by three factors: (1) possible biases in
event reconstruction, (2) uncertainty of the LHC machine conditions, and (3) an
interaction model dependence.

\subsubsection{Systematic uncertainties in event reconstructions and unfolding of distributions}
\label{sec:systreco}

Uncertainties related to biases in event reconstruction are mainly due to five
causes: (1) single-hit/multi-hit separation, (2) PID, (3) energy scale
uncertainty, (4) position-dependent corrections for both shower leakage and the
light yield of the calorimeters, and (5) unfolding of distributions. For the
first four terms, we follow the same approaches to estimate the systematic
uncertainties as we used in the previous study~\cite{LHCfpppi0}.

Concerning the unfolding process, the uncertainty is estimated by adding the
following three components in quadrature. First, the uncertainty due to a
possible dependence of the unfolding procedure on the shape of the $\pt$ or
$\pz$ distributions to be unfolded is estimated from MC simulations; we estimate
the variation of the ratios of the unfolded distributions to the true
distributions among the three true distributions predictions by \textsc{dpmjet},
\textsc{qgsjet}, and \textsc{epos}. The second component is a dependence of the
unfolding procedure on the event generator used in the generation of the
response matrix for unfolding, which is negligible as we mentioned in
Sec.~\ref{sec:mc_signal}.
Finally, the third component is the systematic uncertainty in the unfolding
algorithm itself. This is evaluated by comparing two unfolded distributions, one
obtained by a fully Bayesian unfolding method and the second obtained by the
iterative Bayesian unfolding method. The uncertainty in the first component is
$\SI{10}{\percent}$ over the all $\pt$ and $\pz$ bins, and the uncertainties in
the other two components make no significant contribution. Thus we assign
$\SI{10}{\percent}$ for the systematic uncertainty in the unfolding of $\pt$ and
$\pz$ distributions.

\subsubsection{Systematic uncertainties in the LHC machine conditions}
\label{sec:systmachine}

The LHC machine conditions introduce systematic uncertainties in beam position
and luminosity. The beam position at the LHCf detectors varies from fill to fill
owing to variations of the beam transverse position and the crossing angles at
IP1. The beam center positions at the LHCf detectors obtained for LHC Fills 1089
to 1134 by the LHCf position-sensitive detectors and by the beam position
monitors (BPMSW) installed $\pm \SI{21}{\meter}$ from IP1~\cite{BPM} are
consistent with each other within $\pm$\SI{1}{\milli\meter}.
The systematic shifts to the $\pt$ and $\pz$ distributions are then evaluated by
taking the ratios of distributions with the beam center displaced by
$\pm\SI{1}{\milli\meter}$ to distributions with no displacement present.
The evaluated systematic shifts to the $\pt$ and $\pz$ distributions are
5--$\SI{20}{\percent}$ depending on the $\pt$ and $\pz$ values.

The uncertainty in the luminosity depends on the collision configuration. For
the data in $\pp$ collisions at $\sqs=\tev{7}$, the luminosity value used for
the analysis is derived from the counting rate of the Front Counters.
Considering the uncertainties in both the calibration of the Front Counters $\pm
\SI{3.4}{\percent}$ and in the beam intensity measurement $\pm
\SI{5.0}{\percent}$ during the Van der Meer scans, we estimate an uncertainty of
$\pm \SI{6.1}{\percent}$ in the luminosity for $\pp$ collisions at
$\sqs=\tev{7}$~\cite{LHCfLumi}.
For the $\pp$ collision data at $\sqs=\tev{2.76}$ and $\pPb$ collision data at
$\snn=\tev{5.02}$, LHCf data were taken simultaneously with data taken by the
ATLAS experiment. The luminosity values used for this data analysis were then
provided by the LHCf Front Counters and also by the ATLAS collaboration.

The luminosity uncertainties in $\pp$ collisions at $\sqs=\tev{2.76}$ and in
$\pPb$ collisions at $\snn=\tev{5.02}$ are estimated to be $\pm
\SI{3.1}{\percent}$~\cite{ATLASPbPb} and $\pm
\SI{20}{\percent}$~\cite{CERNLumi}, respectively.

Pileup of successive $\pPb$ collisions due to the small bunch spacing
(\SI{200}{\nano\second}) relative to the data acquisition time
(\SI{500}{\nano\second}) amounts to $<\SI{5}{\percent}$ systematic uncertainty of
$\pt$ and $\pz$ distributions (see Sec.~\ref{sec:pPb5.02TeV}), and may provide a
slight shift of the absolute normalization for the $\pt$ and $\pz$
distributions. This effect is not corrected for in this study, but is taken into
account as uncertainty related to the LHC machine condition.

\subsubsection{Systematic uncertainties depending on the interaction models used in the MC simulations}
\label{sec:systmc}

The analysis in this paper unavoidably relies on the predictions given by MC
simulations. First, we correct LHCf data for the loss of multi-hit $\pizero$
events (Sec.~\ref{sec:multipi0}). The correction factors $f^\textrm{multihit}$
show a systematic uncertainty of less than $\SI{10}{\percent}$ among the
hadronic interaction models. Second, for $\pPb$ collisions only, the
contamination from UPC induced $\pizero$ events in LHCf data is derived from MC
simulations (Sec.~\ref{sec:mc_signal}). The comparison of the predicted $\pt$
and $\pz$ distributions of $\pizero$s between two UPC MC simulations, one using
\textsc{dpmjet} 3.05 and the other one using \textsc{pythia} 6.428 for
high-energy photon--proton interaction, show a systematic uncertainty of roughly
3--$\SI{20}{\percent}$.

In summary, there are 10 systematic uncertainties. The first four (1)
Single/multihit selection, (2) PID, (3) energy scale and (4) position-dependent
correction are explained in Ref.~\cite{LHCfpppi0} and we follow the same
approaches as we used in Ref.~\cite{LHCfpppi0}. The remaining six systematic
uncertainties and the text containing their explanations are:
(5) Unfolding uncertainty is explained and evaluated in Sec.~\ref{sec:systreco},
(6) Offset of beam axis is explained in the 1st paragraph of
Sec.~\ref{sec:systmachine}, 5--$\SI{20}{\percent}$ shifts in $\pt$ or $\pz$
distributions are obtained, (7) Luminosity uncertainty is explained in the 2nd
paragraph of Sec.~\ref{sec:systmachine}.
(8) Contamination of successive $\pPb$ collisions is explained in the 3rd
paragraph of Sec.~\ref{sec:systmachine}, (This uncertainty is due to
contamination and thus only a positive error is quoted.) (9) The uncertainty in
multihit $\pizero$ events $\pm \SI{10}{\percent}$, and (10) the uncertainty in
UPC $\pm$(3--\SI{20}{\percent}) are found in Sec.~\ref{sec:systmc}.
Table~\ref{tbl:systematicerror} summarizes the systematic uncertainties of the
$\pizero$ $\pt$ and $\pz$ distributions.

\begin{table}
  \centering
    \caption{Summary of the systematic uncertainties. Numerical values indicate the
    maximum variation of bin contents in the $\pt$ and $\pz$ distributions due to
    systematic uncertainties. Note that the uncertainty in contamination of
    successive $\pPb$ collisions and in UPC $\pizero$ simulation pertain only to
    $\pPb$ collisions.}
    \begin{ruledtabular}
    \begin{tabular}{c c}
      Single/multihit selection         & $\pm$\SI{3}{\percent}      \\
      Particle identification           & $\pm$(0--\SI{20}{\percent})\\
      Energy scale                      & $\pm$(5--\SI{20}{\percent})\\
      Position-dependent correction     & $\pm$(5--\SI{30}{\percent})\\
      Unfolding                         & $\pm$(5--\SI{10}{\percent})\\
      Offset of beam axis               & $\pm$(5--\SI{20}{\percent})\\
      Luminosity ($\pp$ at \tev{7})     & $\pm$\SI{6.1}{\percent}    \\
      Luminosity ($\pp$ at \tev{2.76})  & $\pm$\SI{3.1}{\percent}    \\
      Luminosity ($\pPb$ at \tev{5.02}) & $\pm$\SI{20}{\percent}     \\
      Contamination of successive $\pPb$ collisions &  <\SI{5}{\percent}\\
	  Multihit $\pizero$ correction     & <\SI{10}{\percent}         \\
	  UPC $\pizero$ simulation          & $\pm$(3--\SI{20}{\percent})\\
    \end{tabular}
    \end{ruledtabular}
    \label{tbl:systematicerror}
\end{table}

%
%
\section{Analysis results}\label{sec:result}

\subsection{Results in $\pp$ collisions at $\sqs=\tev{7}$}
\label{sec:result7TeV}

The inclusive production rate of neutral pions as a function of $\pt$ and $\pz$
is given by the expression~\cite{PDG}
\begin{equation}
    \frac{1}{\sigma_\text{inel}} E \frac{d^{3}\sigma}{dp^3} \Rightarrow
    \frac{1}{N_\text{inel}}\frac{d^2N(\pt, y)}{2\pi\pt d\pt dy} =
    \frac{1}{N_\text{inel}}E\frac{d^2N(\pt, \pz)}{2\pi\pt d\pt d\pz}.
	\label{eq:cross_pp}
\end{equation}
\noindent $\sigma_\text{inel}$ is the inelastic cross section for $\pp$
collisions at $\sqs=\tev{7}$. $E d^{3}\sigma / dp^3$ is the inclusive cross
section for $\pizero$ production. The number of inelastic collisions,
$N_\text{inel}$, used for the production rate normalization is calculated from
$N_\text{inel}$ = $\sigma_\text{inel} \int {\cal L} dt$, taking the inelastic
cross section $\sigma_\text{inel}=\SI{73.6}{\milli\barn}$~\cite{LHCfpppi0}.
The uncertainty in $\sigma_\text{inel}$ is estimated to be $\pm
\SI{3.0}{\milli\barn}$ by comparing the values of $\sigma_\text{inel}$ reported
in Refs.~\cite{TOTEMinel1,TOTEMinel2,TOTEMinel3,ATLASinel}.

Using the integrated luminosities $\int {\cal L} dt$, reported in
Sec.~\ref{sec:pp7TeV}, $N_\text{inel}$ is \num{2.67+-0.11e8} for Arm1 and
\num{2.10+-0.09e8} for Arm2. $d^2N(\pt, y)$ is the number of $\pizero$s produced
within the transverse momentum interval $d\pt$ and the rapidity interval $dy$.
Similarly $d^2N(\pt, \pz)$ is the number of $\pizero$s produced within $d\pt$
and the longitudinal momentum interval $d\pz$.

Experimental $\pt$ and $\pz$ distributions measured independently with the Arm1
and Arm2 detectors are combined following a pull method~\cite{Pull} and the
final $\pt$ and $\pz$ distributions are then obtained by minimizing the value of
the chi-square function defined by
  \begin{eqnarray}
    \chi^2 =
    \sum_{i=1}^{n}
    \sum_{a=1}^{5}
    \left(
    \frac{R_{a, i}^\text{obs}
    (1 + \mathit{S}_{a, i}) - R_i^\text{comb}
    }
    {\sigma_{a, i}}
    \right)^2
    + \chi^2_\text{penalty},
    \label{eq:combine_chi2}
  \end{eqnarray}
where the index $i$ represents the $\pt$ or $\pz$ bin number running from 1 to
$n$ (the total number of $\pt$ or $\pz$ bins), and the index $a$ indicates the
type of distributions: $a=1$ Arm1 Type-I events, $a=2$ Arm1 Type-II events with
the Large Calorimeter, $a=3$ Arm2 Type-I events, $a=4$ Arm2 Type-II events with
the Small Calorimeter, and $a=5$ Arm2 Type-II events with the Large Calorimeter.
Note that Arm1 Type-II events with the Small Calorimeter are not used for this
analysis since the energy reconstruction accuracy for these events is still
being investigated.
$R_{a, i}^\text{obs}$ is the inclusive production rate in the $i$th bin of the
$a$th distribution, which corresponds to the second and third terms
Eq.~(\ref{eq:cross_pp}).
$R_i^\text{comb}$ is the inclusive production rate in the $i$th bin obtained by
combining all $R_{a, i}^\text{obs}$'s for $a=$1--5.
$\sigma_{a, i}$ is the uncertainty of $R_{a, i}^\text{obs}$. The $\sigma_{a, i}$
are calculated by quadratically adding the statistical uncertainty and the
systematic uncertainty in the energy scale. The energy scale uncertainty has
been estimated with test beam data taken at the SPS and is uncorrelated
bin-by-bin unlike the other systematic uncertainties~\cite{LHCfpppi0}.
The systematic correction $\mathit{S}_{a, i}$ modifies the number of events in
the $i$th bin of the $a$th distribution:
\begin{equation}
    \mathit{S}_{a, i}
    = \sum_{j=1}^{7}f_{a,i}^{j}\varepsilon_{a}^{j}.
    \label{eq:combine_syst}
\end{equation}
The coefficient $f_{a, i}^{j}$ is the systematic shift of the $i$th bin content
of the $a$th distribution due to the $j$th systematic uncertainty term. The
systematic uncertainty consists of seven uncertainties related to the
single-hit/multihit separation, the PID, the energy scale (owing to the
invariant mass shift of the measured $\pizero$ events), the position-dependent
correction, the unfolding procedure, the beam center position, and the loss of
multihit $\pizero$ events.
These uncertainties are assumed to be fully uncorrelated between the Arm1 and
Arm2 detectors, while correlations between Type-I and Type-II events and bin-bin
correlations have been accounted for.
The coefficients $\varepsilon_{a}^{j}$, which should follow a Gaussian
distribution, can be varied, within the constraints of the penalty term given by
\begin{equation}
    \chi^2_\text{penalty} =
    \sum_{j=1}^{7}
    \sum_{a=1}^{5}
    |\varepsilon_{a}^{j}|^2,
    \label{eq:combine_penalty}
\end{equation}
to achieve the minimum $\chi^2$ value for each chi-square test.
Note that the uncertainty in the luminosity determination,
$\pm\SI{3.1}{\percent}$--$\pm\SI{20}{\percent}$, not included in
Eq.~(\ref{eq:combine_syst}) and Eq.~(\ref{eq:combine_penalty}), can cause
independent shifts of all the $\pt$ and $\pz$ distributions.

\begin{figure*}[htbp]
  \begin{center}
  \includegraphics[width=18cm, keepaspectratio]{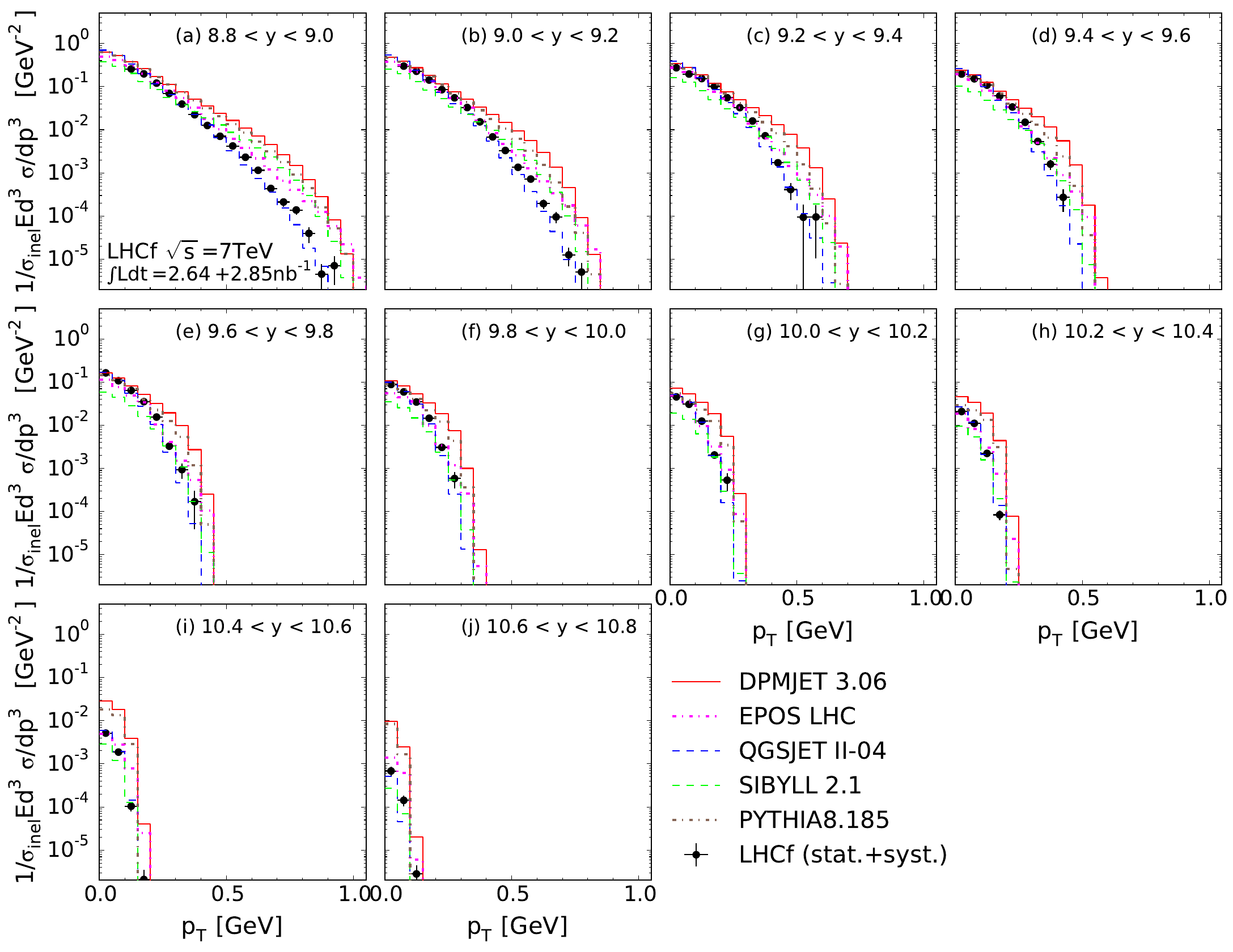}
  \caption{(color online). LHCf $\pt$ distributions (filled circles) in $\pp$
  collisions at $\sqs=\tev{7}$. Error bars indicate the total statistical and
  systematic uncertainties. The predictions of hadronic interaction models are
  shown for comparison: \textsc{dpmjet} (solid red line), \textsc{qgsjet}
  (dashed blue line), \textsc{sibyll} (dotted green line), \textsc{epos}
  (dashed-dotted magenta line), and \textsc{pythia} (dashed-double-dotted brown
  line).}
  \label{fig:pt_7TeV}
  \end{center}
\end{figure*}

The LHCf $\pt$ distributions (filled circles) are obtained from the best-fit
$R^\text{comb}$ and are shown in Fig.~\ref{fig:pt_7TeV}.
The \SI{68}{\percent} confidence intervals incorporating the statistical and
systematic uncertainties, except for the luminosity uncertainty, are indicated
by the error bars. LHCf $\pt$ distributions are corrected for the influences of
the detector response, event selection efficiencies and geometrical acceptance
efficiencies, and thus LHCf $\pt$ distributions can be compared directly to the
predicted $\pt$ distributions from hadronic interaction models.
For comparison, the predictions from various hadronic interaction models are
also shown in Fig.~\ref{fig:pt_7TeV}: \textsc{dpmjet} (solid red line),
\textsc{qgsjet} (dashed blue line), \textsc{sibyll} (dotted green line),
\textsc{epos} (dashed-dotted magenta line), and \textsc{pythia} (default
parameter set, dashed-double-dotted brown line).
For these hadronic interaction models, the inelastic cross section used for the
production rate normalization is taken from the predefined value in each model.

\begin{figure*}[htbp]
  \begin{center}
  \includegraphics[width=18cm, keepaspectratio]{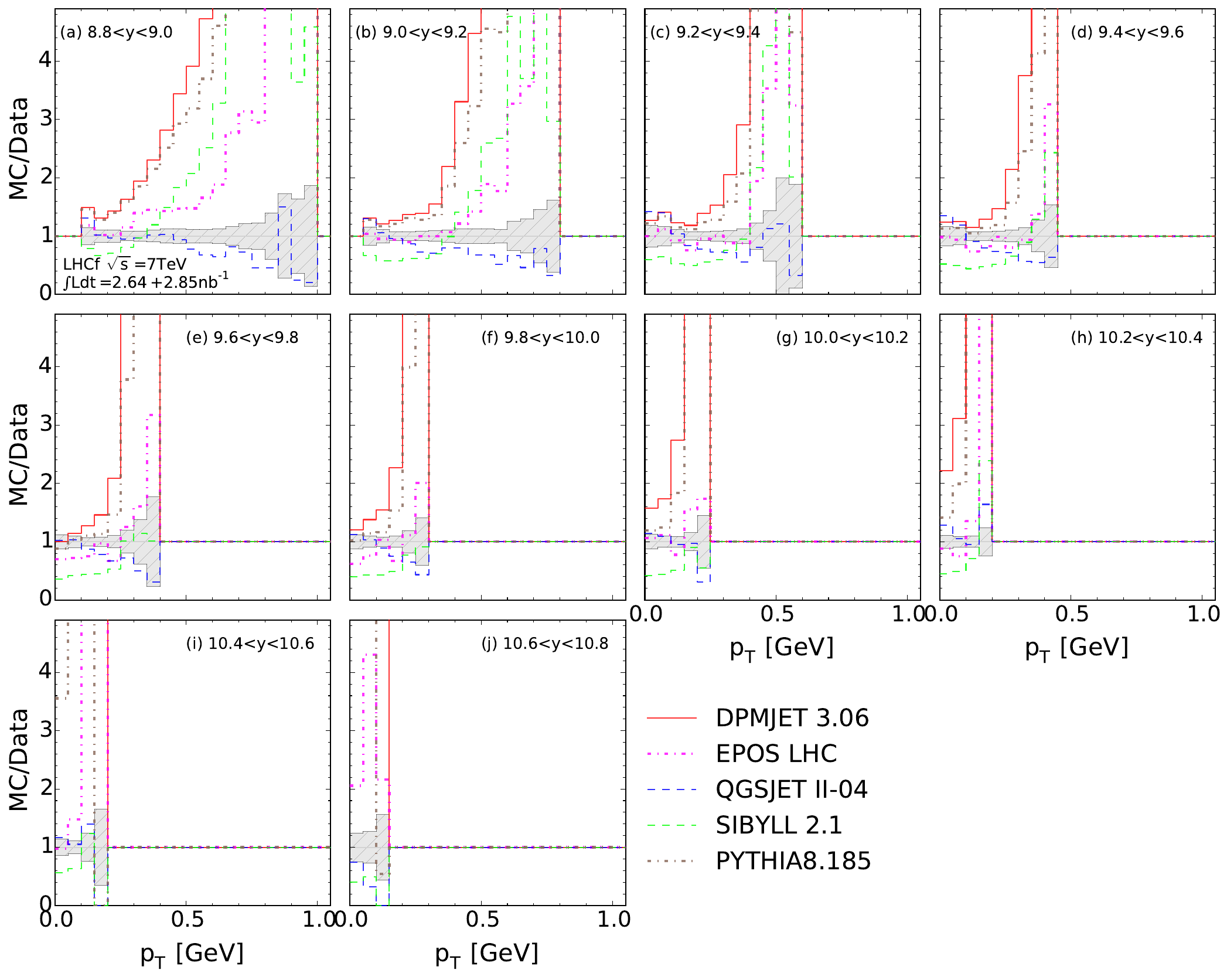}
  \caption{(color online). Ratios of LHCf $\pt$ distributions to the $\pt$
  distributions predicted by hadronic interaction models in $\pp$ collisions at
  $\sqs=\tev{7}$ are shown by solid red line (\textsc{dpmjet}), dashed blue line
  (\textsc{qgsjet}), dotted green line (\textsc{sibyll}), dashed-dotted magenta
  line (\textsc{epos}), and dashed-double-dotted brown line (\textsc{pythia}).
  Shaded areas indicate the range of total uncertainties of the LHCf $\pt$
  distributions.}
  \label{fig:pt_7TeV_r}
  \end{center}
\end{figure*}

Figure~\ref{fig:pt_7TeV_r} presents the ratios of the inclusive production rates
predicted by the hadronic interaction models listed above to those obtained by
LHCf data.
Shaded areas have been taken from the statistical and systematic uncertainties.
In Fig.~\ref{fig:pt_7TeV_r}, the denominator and the numerators, namely the
inclusive production rate for LHCf data and for the hadronic interaction models,
respectively, are properly normalized by the inelastic cross section for each,
thus we do not apply any other normalization to the ratios.
The inclusive production rates of $\pizero$s measured by LHCf and the ratios of
$\pizero$ production rate of MC simulation to data are summarized in Appendix.

In the comparisons in Fig.~\ref{fig:pt_7TeV} and \ref{fig:pt_7TeV_r},
\textsc{qgsjet} has good overall agreement with LHCf data, while \textsc{epos}
produces a slightly harder distribution than the LHCf data for $\pt>\gev{0.5}$.
These two models are based on the parton-based Gribov-Regge
approach~\cite{Gribov1,Regge} and are tuned by using the present LHC data
(ALICE, ATLAS, CMS, and TOTEM)~\cite{QGSJET,EPOS}.
The prediction of \textsc{sibyll} agrees well with the LHCf data for $8.8<y<9.2$
and $\pt<\gev{0.4}$, while the absolute yield of \textsc{sibyll} is about half
that of the LHCf data for $y>9.2$.
The predictions of \textsc{dpmjet} and \textsc{pythia} are compatible with LHCf
data for $9.0<y<9.8$ and $\pt<\gev{0.2}$, while for $\pt>\gev{0.2}$ they become
significantly harder than both LHCf data and the other model predictions.
Generally the harder distributions appearing in \textsc{sibyll},
\textsc{dpmjet}, and \textsc{pythia} can be attributed to the baryon/meson
production mechanism that is used by these models. For example the popcorn
approach~\cite{Andersson,Eden} implemented in the Lund model is known to produce
hard distributions of forward mesons~\cite{Drescher2}. Indeed, by only changing
the tuning parameters of the popcorn approach in \textsc{dpmjet} one obtains
softer meson distributions and consequently $\pt$ distributions that are
compatible with LHCf data. However such a crude tune may bring disagreements
between the model predictions and other experimental results, e.g. forward
neutron $\pz$ and $\pt$ distributions.

\begin{figure*}[htbp]
  \begin{center}
  \includegraphics[width=16cm, keepaspectratio]{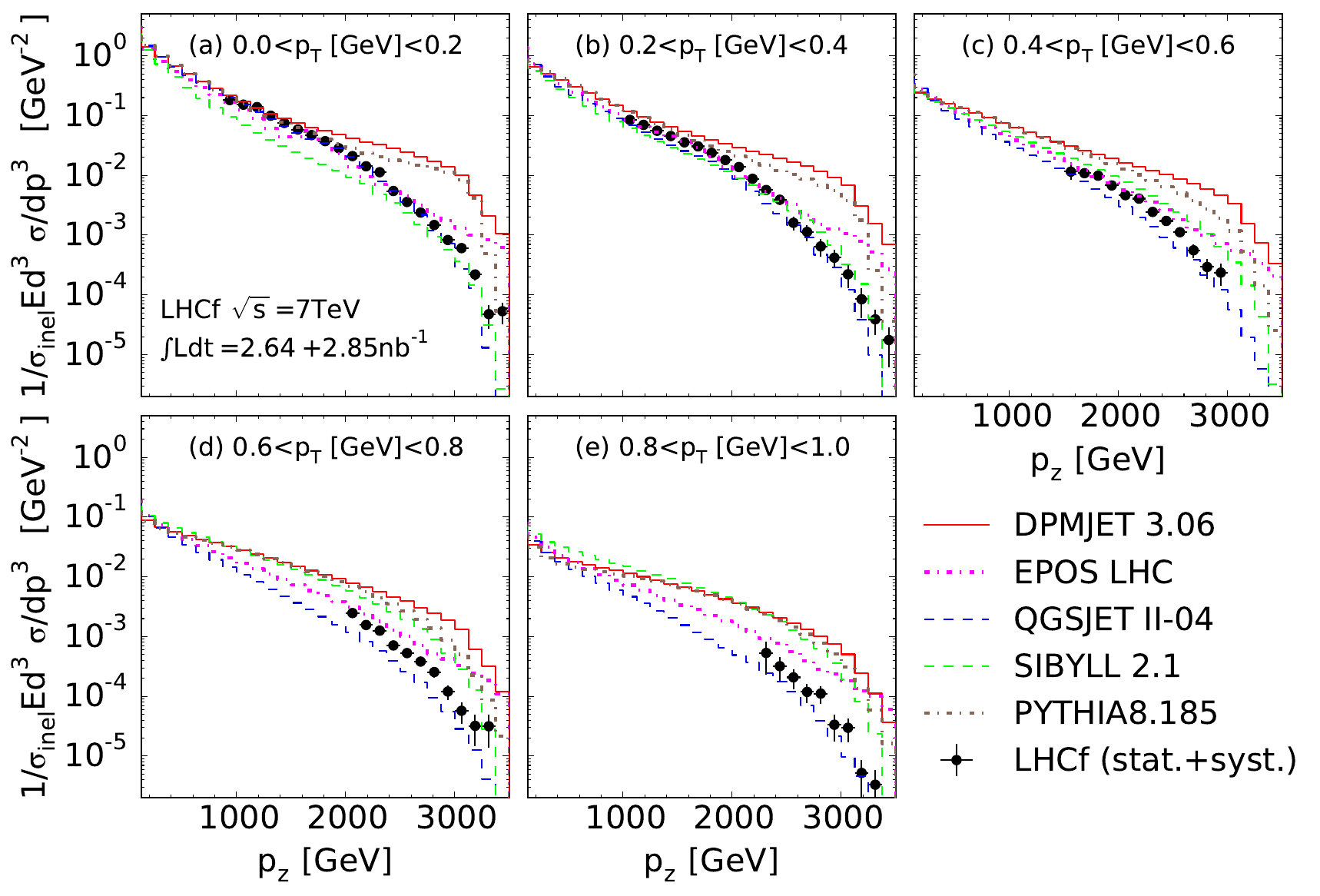}
  \caption{(color online). LHCf $\pz$ distributions (filled circles) in $\pp$
  collisions at $\sqs=\tev{7}$. Error bars indicate the total statistical and
  systematic uncertainties. The predictions of hadronic interaction models are
  shown for comparison: \textsc{dpmjet} (solid red line), \textsc{qgsjet}
  (dashed blue line), \textsc{sibyll} (dotted green line), \textsc{epos}
  (dashed-dotted magenta line), and \textsc{pythia} (dashed-double-dotted brown
  line).}
  \label{fig:pz_7TeV}
  \end{center}
\end{figure*}

\begin{figure*}[htbp]
  \begin{center}
  \includegraphics[width=18cm, keepaspectratio]{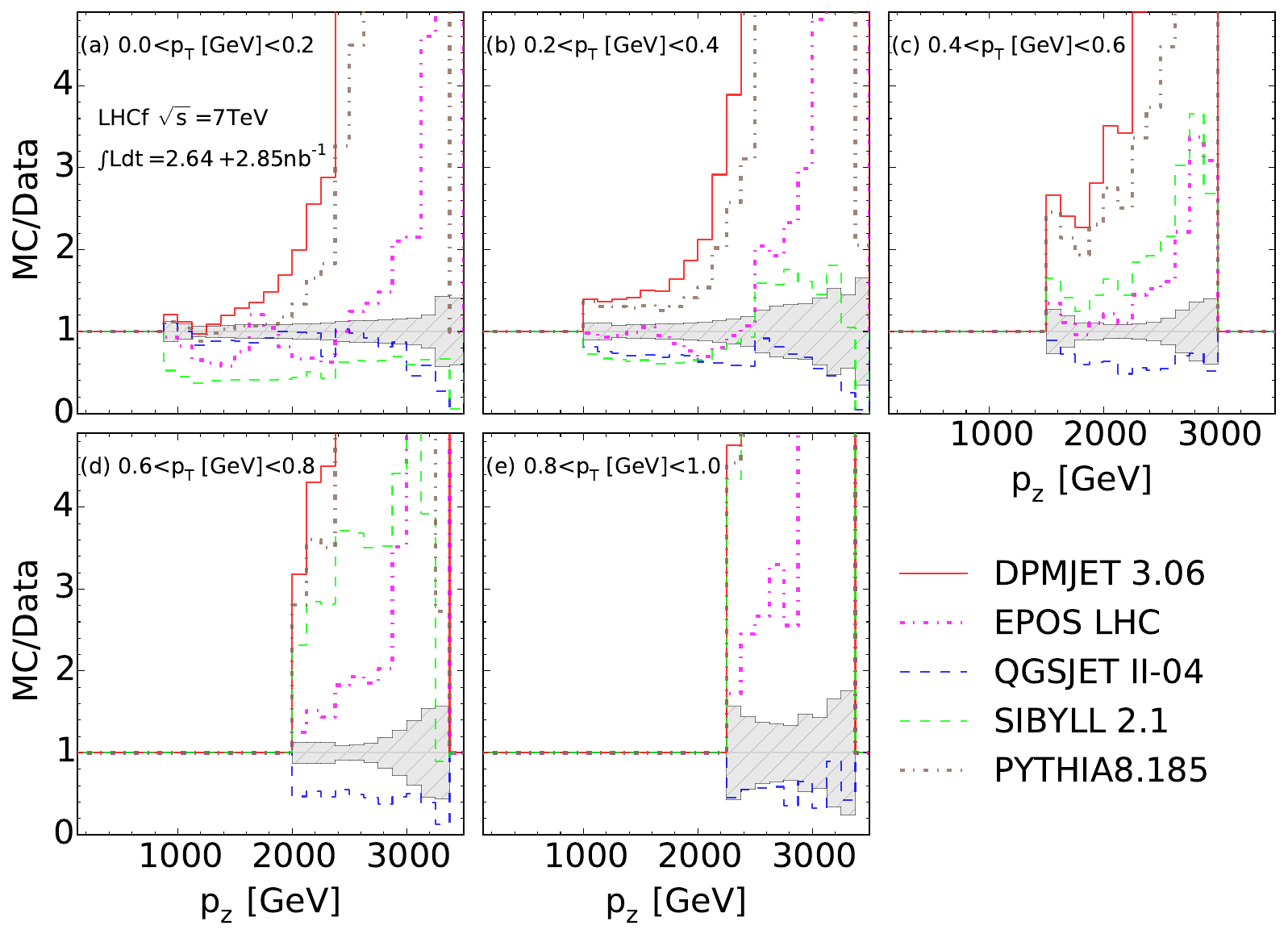}
  \caption{(color online). Ratios of LHCf $\pz$ distributions to the $\pz$
  distributions predicted by hadronic interaction models in $\pp$ collisions at
  $\sqs=\tev{7}$ are shown by solid red line (\textsc{dpmjet}), dashed blue line
  (\textsc{qgsjet}), dotted green line (\textsc{sibyll}), dashed-dotted magenta
  line (\textsc{epos}), and dashed-double-dotted brown line (\textsc{pythia}).
  Shaded areas indicate the range of total uncertainties of the LHCf $\pz$
  distributions.}
  \label{fig:pz_7TeV_r}
  \end{center}
\end{figure*}

The LHCf $\pz$ distributions are shown in Fig.~\ref{fig:pz_7TeV}. The $\pz$
distributions predicted by various hadronic interaction models are also shown in
Fig.~\ref{fig:pz_7TeV}.
Figure~\ref{fig:pz_7TeV_r} presents the ratios of $\pz$ distributions predicted
by the hadronic interaction models to the LHCf $\pz$ distributions. Shaded areas
have been taken from the statistical and systematic uncertainties.
The same conclusions for the comparisons are obtained as those found for
Fig.~\ref{fig:pt_7TeV} and \ref{fig:pt_7TeV_r}.
There is again an overall agreement between LHCf data and the \textsc{qgsjet}
prediction, especially for $0.0<\pt<\gev{0.2}$. The \textsc{epos} prediction is
compatible with LHCf data for $\pt<\tev{2}$, while showing a hard slope for
$\pt>\tev{2}$ in all $\pt$ regions.
The predictions by \textsc{dpmjet} and \textsc{pythia} agree with LHCf data for
$\pt<\gev{0.2}$ and $\pz<\tev{1.6}$, while showing a harder distribution for the
higher $\pz$ regions. \textsc{sibyll} predicts a smaller production of
$\pizero$s for $\pt<\gev{0.2}$ and becomes similar with \textsc{dpmjet} and
\textsc{pythia} with increasing $\pt$.

\subsection{Results in $\pp$ collisions at $\sqs=\tev{2.76}$}
\label{sec:result2TeV}

The inclusive production rates of $\pizero$s as a function of $\pt$ and $\pz$
are given by Eq.~(\ref{eq:cross_pp}).
Using the inelastic cross section
$\sigma_\text{inel}=\SI{62.5+-5.0}{\milli\barn}$~\cite{PDG} and the integrated
luminosities reported in Sec.~\ref{sec:pp2.76TeV}, $N_\text{inel}$ is calculated
as \num{1.60+-0.13e8}.
The uncertainty on $\sigma_\text{inel}$ is estimated by comparing the
$\sigma_\text{inel}$ value with the present experimental
result~\cite{ALICEinel}.
Note that only the LHCf Arm2 detector was operated in $\pp$ collisions at $\sqs
= \tev{2.76}$ and that only Type-I events are used for the analysis since
Type-II event kinematics are outside the calorimeter acceptance for $\sqs =
\tev{2.76}$.

\begin{figure*}[htbp]
  \begin{center}
  \includegraphics[width=18cm, keepaspectratio]{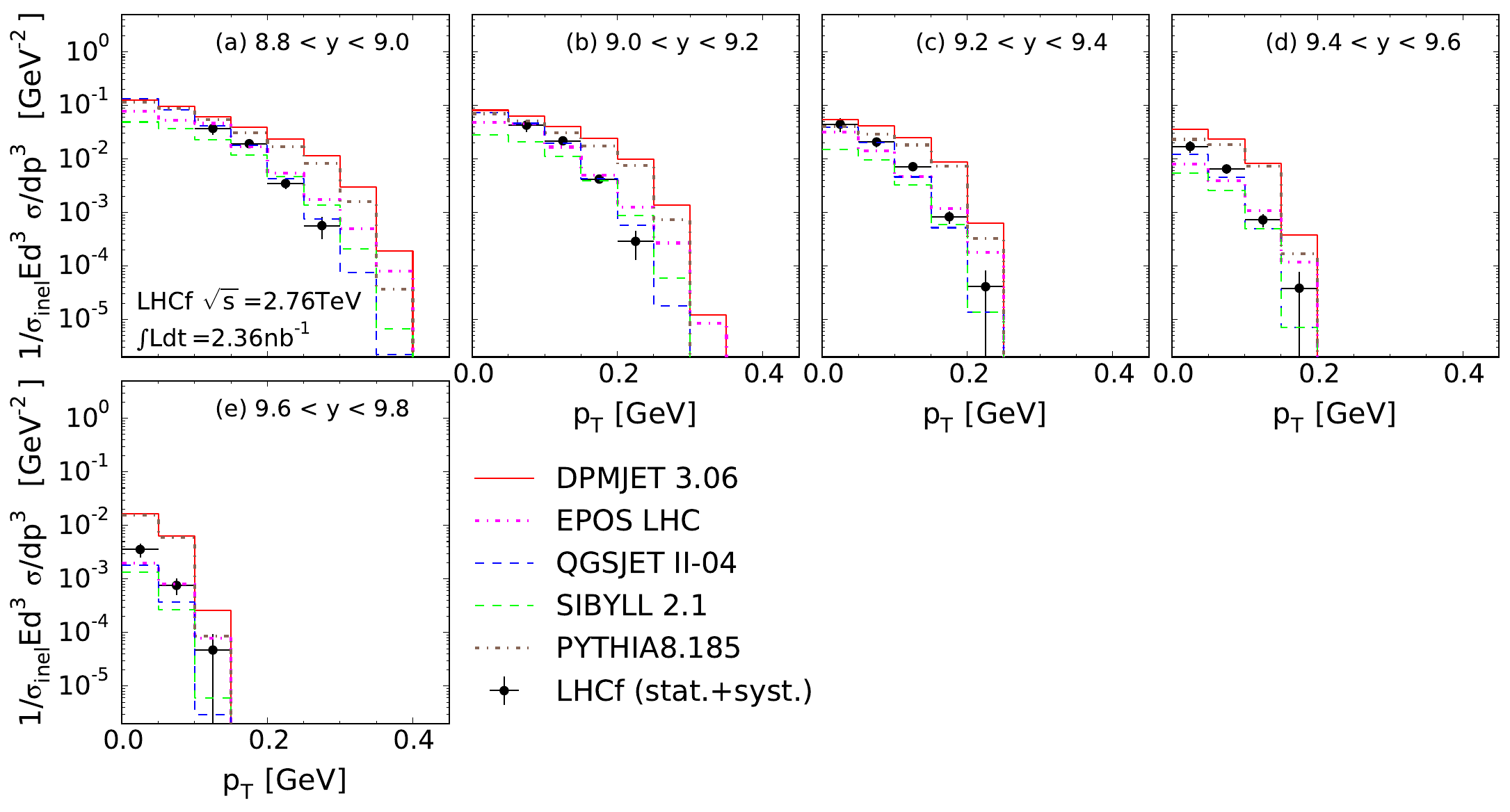}
  \caption{(color online). LHCf $\pt$ distributions (filled circles) in $\pp$
  collisions at $\sqs=\tev{2.76}$. Error bars indicate the total statistical and
  systematic uncertainties. The predictions of hadronic interaction models are
  shown for comparison: \textsc{dpmjet} (solid red line), \textsc{qgsjet}
  (dashed blue line), \textsc{sibyll} (dotted green line), \textsc{epos}
  (dashed-dotted magenta line), and \textsc{pythia} (dashed-double-dotted brown
  line).}
  \label{fig:pt_2TeV}
  \end{center}
\end{figure*}

\begin{figure*}[htbp]
  \begin{center}
  \includegraphics[width=18cm, keepaspectratio]{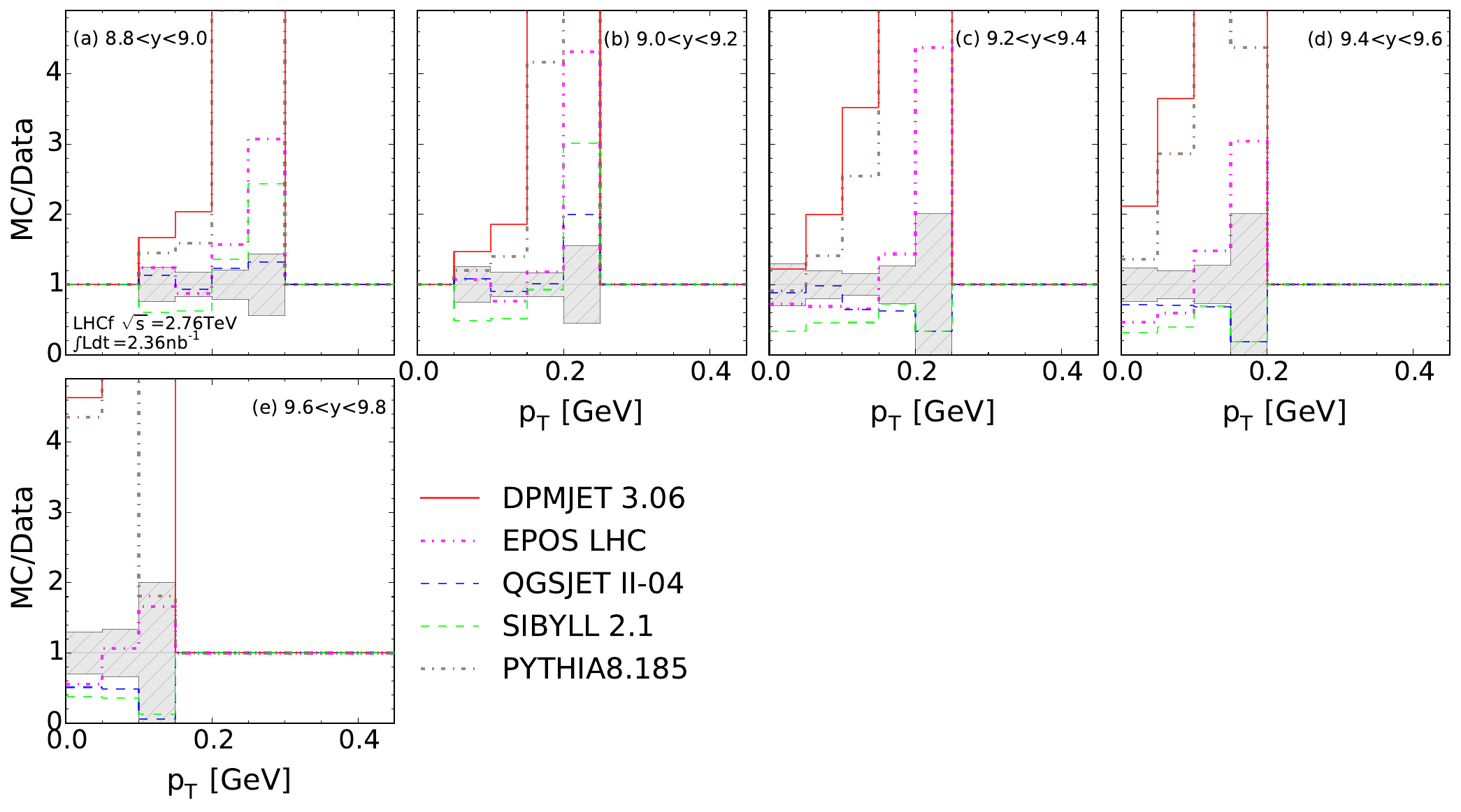}
  \caption{(color online). Ratios of LHCf $\pt$ distributions to the $\pt$
  distributions predicted by hadronic interaction models in $\pp$ collisions at
  $\sqs=\tev{2.76}$ are shown by solid red line (\textsc{dpmjet}), dashed blue
  line (\textsc{qgsjet}), dotted green line (\textsc{sibyll}), dashed-dotted
  magenta line (\textsc{epos}), and dashed-double-dotted brown line
  (\textsc{pythia}). Shaded areas indicate the range of total uncertainties of
  the $\pt$ spectra.}
  \label{fig:pt_2TeV_r}
  \end{center}
\end{figure*}

LHCf $\pt$ distributions are shown in Fig.~\ref{fig:pt_2TeV}. The $\pt$
distributions predictions for the hadronic interaction models are also shown in
Fig.~\ref{fig:pt_2TeV} for comparison. Figure~\ref{fig:pt_2TeV_r} presents the
ratios of $\pt$ distributions predicted by the hadronic interaction models to
the LHCf $\pt$ distributions.
\textsc{qgsjet} provides the best agreement with LHCf data, although it is
slightly softer than the LHCf data for $y>9.2$. The prediction of \textsc{epos}
shows a harder behavior than both \textsc{qgsjet} and LHCf data.
\textsc{sibyll} tends to have generally a smaller $\pizero$ yield and a harder
distribution compared to \textsc{qgsjet} and \textsc{epos}, leading to the
smaller and larger yields with respect to LHCf data in the $\pt$ regions below
and above $\gev{0.1}$.
\textsc{dpmjet} and \textsc{pythia} predict larger $\pizero$ yields than both
LHCf data and other models over the entire rapidity range. The same discussion
on the popcorn model in the previous Section~\ref{sec:result7TeV} can be applied
to the predictions of \textsc{sibyll}, \textsc{dpmjet}, and \textsc{pythia}.

\begin{figure*}[htbp]
  \begin{center}
  \includegraphics[width=18cm, keepaspectratio]{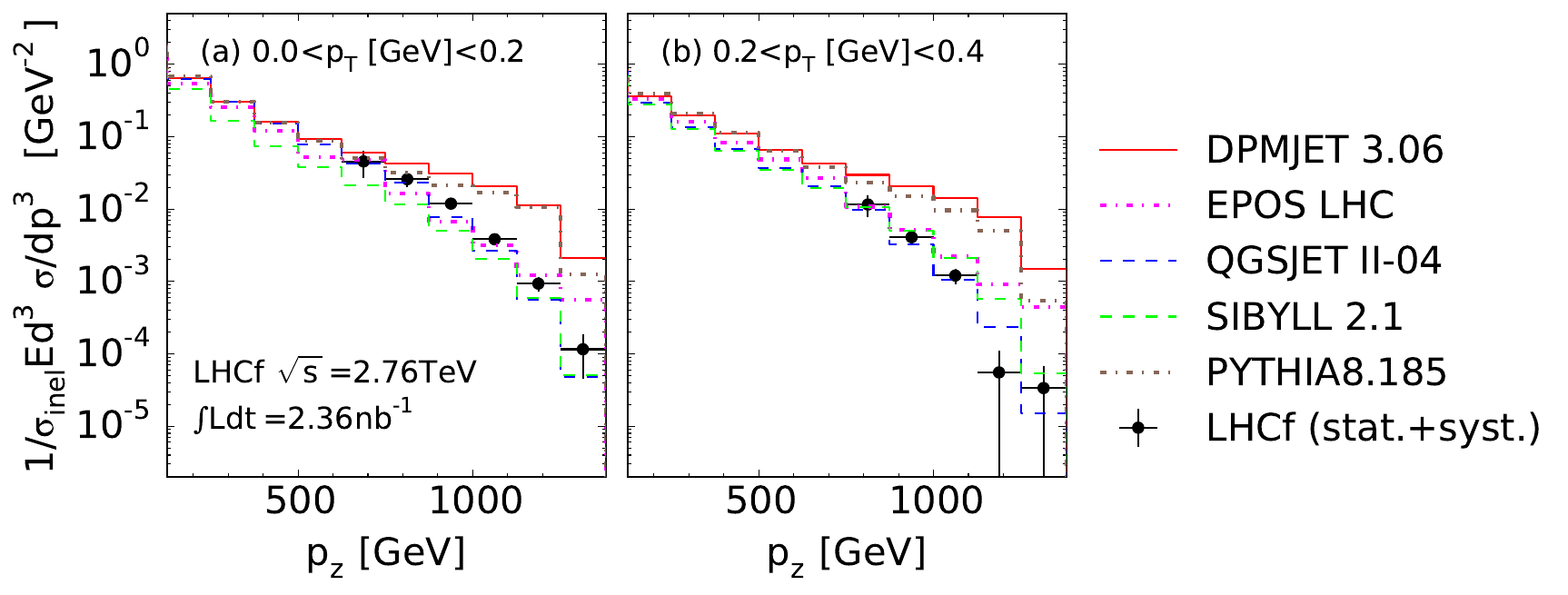}
  \caption{(color online). LHCf $\pz$ distributions (filled circles) in $\pp$
  collisions at $\sqs=\tev{2.76}$. Error bars indicate the total statistical and
  systematic uncertainties. The predictions of hadronic interaction models are
  shown for comparison: \textsc{dpmjet} (solid red line), \textsc{qgsjet}
  (dashed blue line), \textsc{sibyll} (dotted green line), \textsc{epos}
  (dashed-dotted magenta line), and \textsc{pythia} (dashed-double-dotted brown
  line).}
  \label{fig:pz_2TeV}
  \end{center}
\end{figure*}

\begin{figure*}[htbp]
  \begin{center}
  \includegraphics[width=18cm, keepaspectratio]{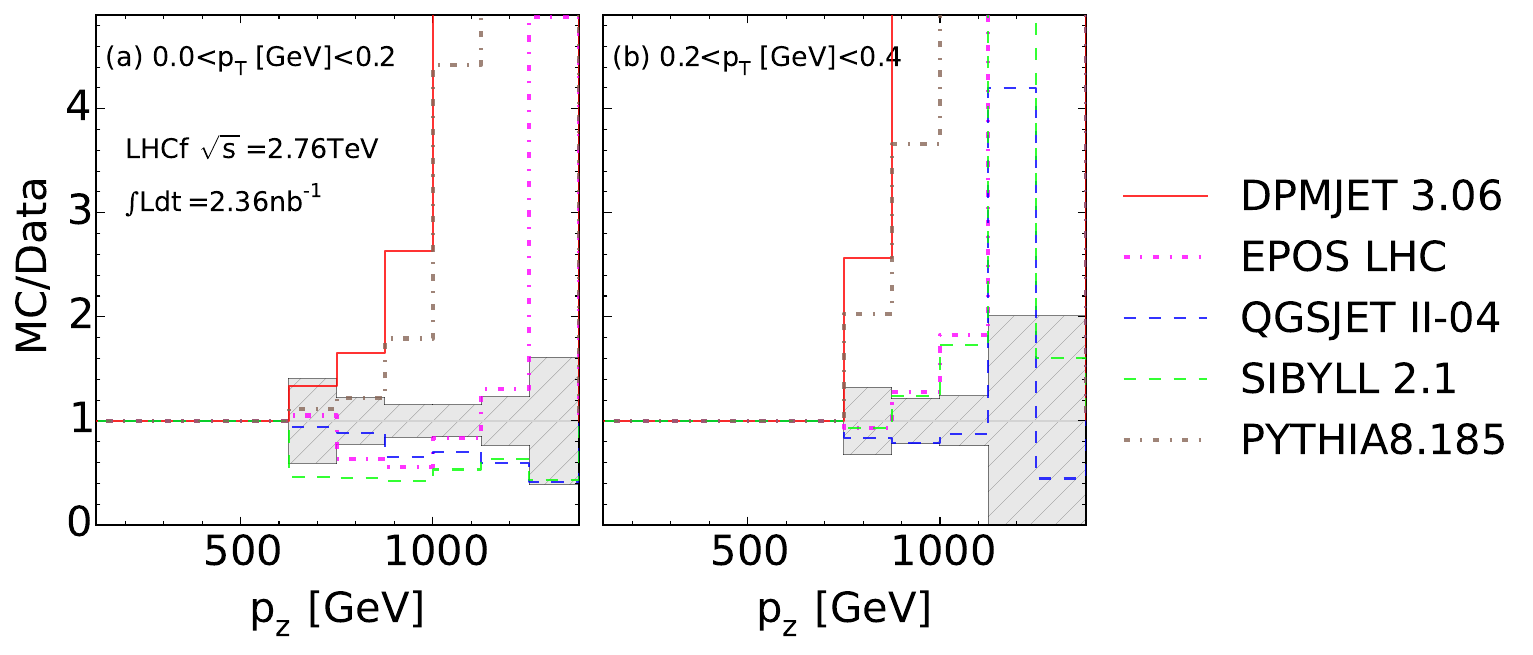}
  \caption{(color online). Ratios of LHCf $\pz$ distributions to the $\pz$
  distributions predicted by hadronic interaction models in $\pp$ collisions at
  $\sqs=\tev{2.76}$ are shown by solid red line (\textsc{dpmjet}), dashed blue
  line (\textsc{qgsjet}), dotted green line (\textsc{sibyll}), dashed-dotted
  magenta line (\textsc{epos}), and dashed-double-dotted brown line
  (\textsc{pythia}). Shaded areas indicate the range of total uncertainties of
  the $\pz$ distributions.}
  \label{fig:pz_2TeV_r}
  \end{center}
\end{figure*}

LHCf $\pz$ distributions are shown in Fig.~\ref{fig:pz_2TeV}.
Figure~\ref{fig:pz_2TeV_r} presents the ratios of $\pz$ distributions predicted
by the hadronic interaction models to LHCf $\pz$ distributions.
The same tendencies found in Fig.~\ref{fig:pz_7TeV} are present here also
\textsc{qgsjet} gives the best agreement for $0.0<\pt<\gev{0.4}$ and
\textsc{epos} has a harder behavior especially for $0.2<\pt<\gev{0.4}$.
The predictions of \textsc{dpmjet} and \textsc{pythia} are significantly harder
than LHCf data for $\pt<\gev{0.4}$ and show poor overall agreement with LHCf
data. This can be explained by the popcorn model in a way similar to the harder
$\pt$ distributions of the \textsc{sibyll}, \textsc{dpmjet} and \textsc{pythia}
models found in Fig.~\ref{fig:pz_7TeV} and the preceding Section.

\subsection{Results in $\pPb$ collisions at $\snn=\tev{5.02}$}
\label{sec:resultpPb}

The inclusive $\pizero$ production rate in $\pPb$ collisions is given as
\begin{eqnarray}
    \frac{1}{\sigma^\text{pPb}_\text{inel}} E\frac{d^{3}\sigma^\text{pPb}}{dp^3}
    &=&
    \frac{1}{N^\text{pPb}_\text{inel}}
    \frac{d^2 N^\text{pPb}(\pt, \ylab)}{2\pi \pt d\pt d\ylab} \nonumber \\
    &=&
    \frac{1}{N^\text{pPb}_\text{inel}}
    E\frac{d^2 N^\text{pPb}(\pt, \pz)}{2\pi\pt d\pt d\pz},
    \label{eq:cross_pPb}
\end{eqnarray}
\noindent where $\sigma^\text{pPb}_\text{inel}$ is the inelastic cross section,
$E d^{3}\sigma^\text{pPb} / dp^3$ is the inclusive cross section of $\pizero$
production in $\pPb$ collisions at $\snn = \tev{5.02}$, and $\ylab$ is the
rapidity in the detector reference frame.
The number of inelastic $\pPb$ collisions, $N^\text{pPb}_\text{inel}$, used for
normalizing the production rates is calculated from $N^\text{pPb}_\text{inel}$ =
$\sigma^\text{pPb}_\text{inel} \int {\cal L} dt$, assuming the inelastic $\pPb$
cross section $\sigma^\text{pPb}_\text{inel} =
\SI{2.11+-0.11}{\barn}$~\cite{dEnterria}.
The value for $\sigma^\text{pPb}_\text{inel}$ is derived from the inelastic
$\pp$ cross section $\sigma^\text{pp}_\text{inel}$ and the Glauber multiple
collision model~\cite{dEnterria,Glauber}.
The uncertainty on $\sigma^\text{pPb}_\text{inel}$ is estimated by comparing the
$\sigma^\text{pPb}_\text{inel}$ value with other calculations and experimental
results presented in Refs.~\cite{ALICEpPb,CMSpPb}.
Using the integrated luminosities described in Sec.~\ref{sec:data},
$N^\text{pPb}_\text{inel}$ is \num{9.33+-0.47e7}.
Note that only the LHCf Arm2 detector (proton remnant side) was operated in
$\pPb$ collisions at $\snn = \tev{5.02}$.

\begin{figure*}[htbp]
  \begin{center}
  \includegraphics[width=18cm, keepaspectratio]{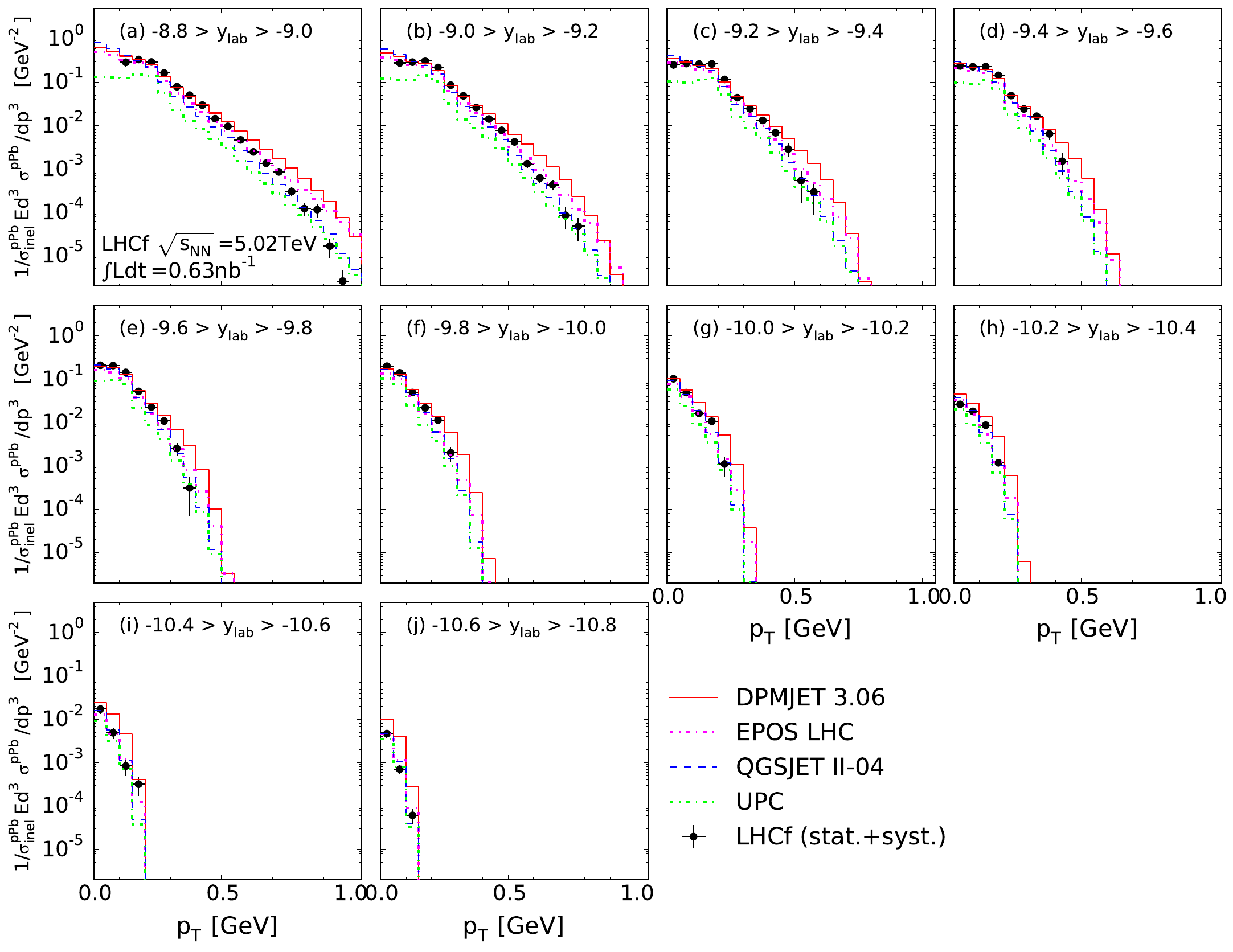}
  \caption{(color online). LHCf $\pt$ distributions (filled circles) in $\pPb$
  collisions at $\snn=\tev{5.02}$. Error bars indicate the total statistical and
  systematic uncertainties. The predictions of hadronic interaction models are
  shown for comparison: \textsc{dpmjet} (solid red line), \textsc{qgsjet}
  (dashed blue line), and \textsc{epos} (dashed-dotted magenta line).}
  \label{fig:pt_pPb}
  \end{center}
\end{figure*}

Figure~\ref{fig:pt_pPb} shows LHCf $\pt$ distributions with both statistical and
systematic errors (filled circles and error bars). The $\pt$ distributions in
$\pPb$ collisions at $\snn=\tev{5.02}$ predicted by the hadronic interaction
models, \textsc{dpmjet} (solid red line), \textsc{qgsjet} (dashed blue line),
and \textsc{epos} (dotted magenta line), are also shown in the same figure for
comparison. The expected UPC contribution discussed in Sec.~\ref{sec:mc_signal}
is added to the hadronic interaction model predictions for consistency with the
treatment of LHCf data, and the UPC $\pt$ distribution is shown for reference
(dashed-double-dotted green line).

\begin{figure*}[htbp]
  \begin{center}
  \includegraphics[width=18cm, keepaspectratio]{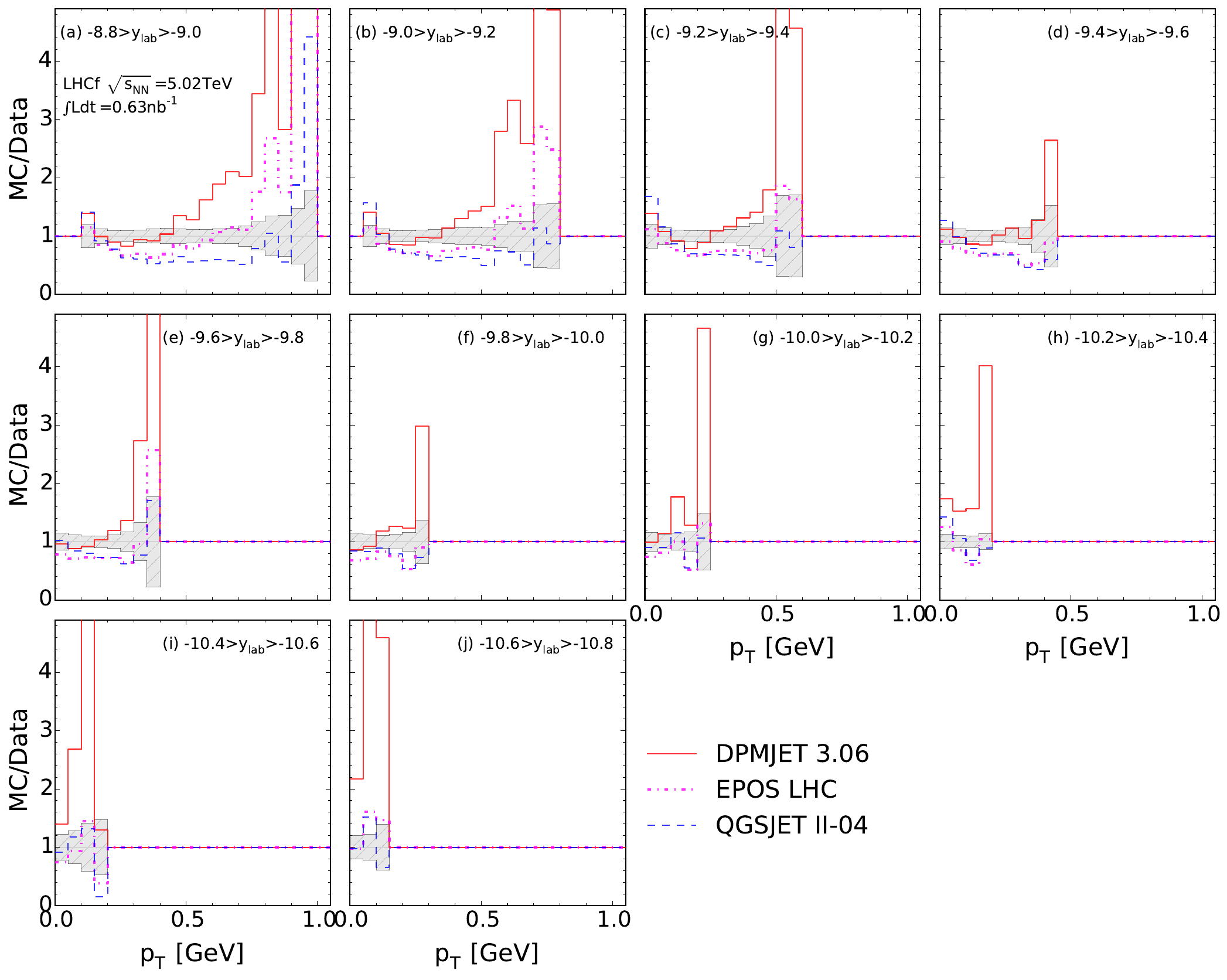}
  \caption{(color online). Ratios of LHCf $\pt$ distributions to the $\pt$
  distributions predicted by hadronic interaction models in $\pPb$ collisions at
  $\snn=\tev{5.02}$ are shown by solid red line (\textsc{dpmjet}), dashed blue
  line (\textsc{qgsjet}), and dashed-dotted magenta line (\textsc{epos}). Shaded
  areas indicate the range of total uncertainties of the $\pt$ distributions.}
  \label{fig:pt_pPb_r}
  \end{center}
\end{figure*}

In Fig.~\ref{fig:pt_pPb}, \textsc{dpmjet} shows good agreement with LHCf data at
$-8.8>\ylab>-10.0$ and $\pt<\gev{0.3}$, while showing a harder behavior at
$-8.8>\ylab>-9.2$ and $\pt>\gev{0.5}$. \textsc{qgsjet} and \textsc{epos} predict
relatively similar distributions to each other and show better agreement with
LHCf data for $\pt>\gev{0.4}$ than \textsc{dpmjet}. The characteristic bump at
$\ylab>-9.8$ and $0.1\lesssim\pt\lesssim\gev{0.2}$, which is absent in $\pp$
collisions, originates from the channel $\gamp \to \pizero + p$ via baryon
resonances in UPCs. In fact the UPC simulation reproduces such a bump.
Figure~\ref{fig:pt_pPb_r} presents the ratios of LHCf $\pt$ distributions to the
$\pt$ distributions predicted by hadronic interaction models taking the UPC
contribution into account in the $\pt$ distributions.

\begin{figure*}[htbp]
  \begin{center}
  \includegraphics[width=18cm, keepaspectratio]{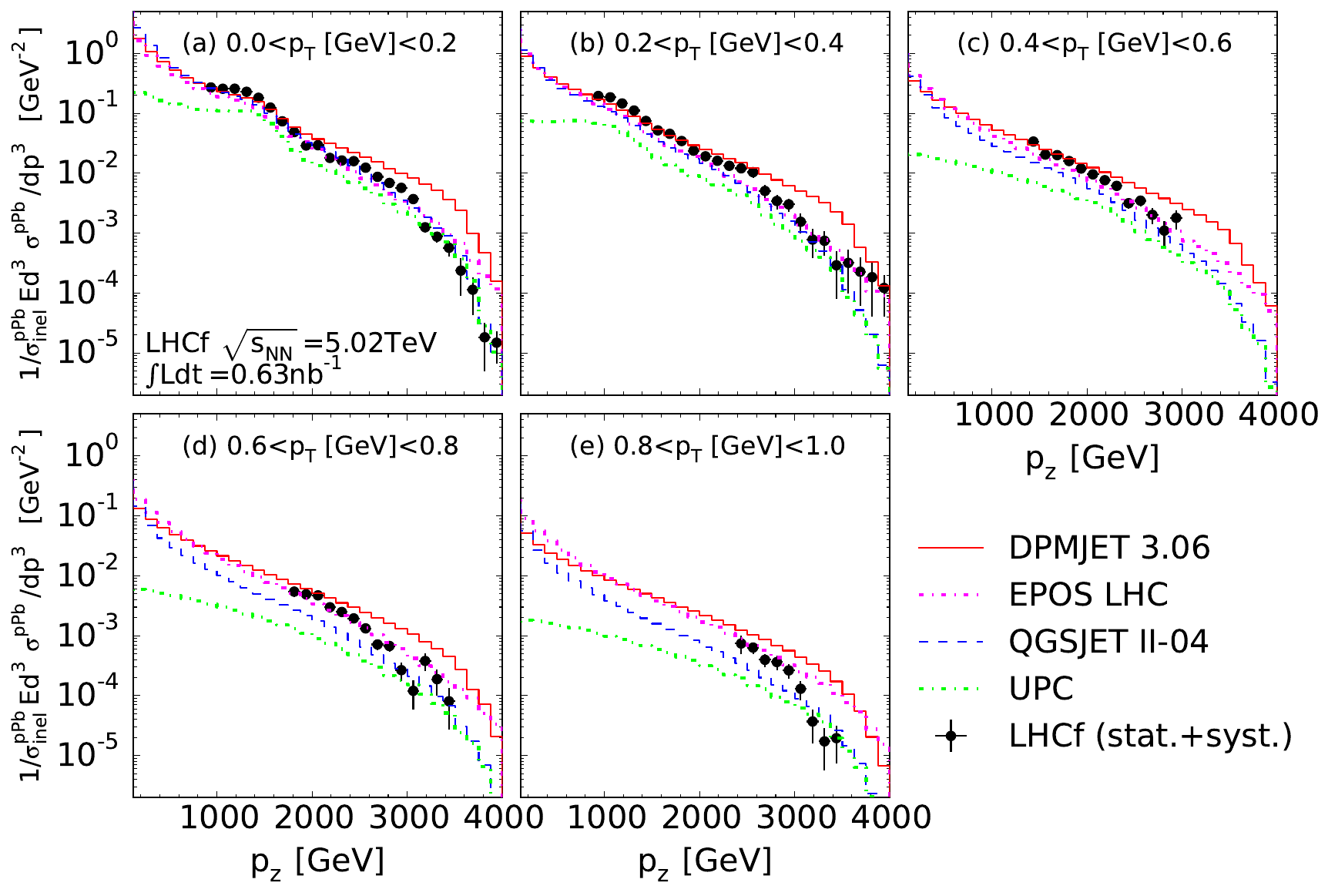}
  \caption{(color online). LHCf $\pz$ distributions (filled circles) in $\pPb$
  collisions at $\snn=\tev{5.02}$. Error bars indicate the total statistical and
  systematic uncertainties. The predictions of hadronic interaction models are
  shown for comparison: \textsc{dpmjet} (solid red line), \textsc{qgsjet}
  (dashed blue line), and \textsc{epos} (dashed-dotted magenta line).}
  \label{fig:pz_pPb}
  \end{center}
\end{figure*}

\begin{figure*}[htbp]
  \begin{center}
  \includegraphics[width=18cm, keepaspectratio]{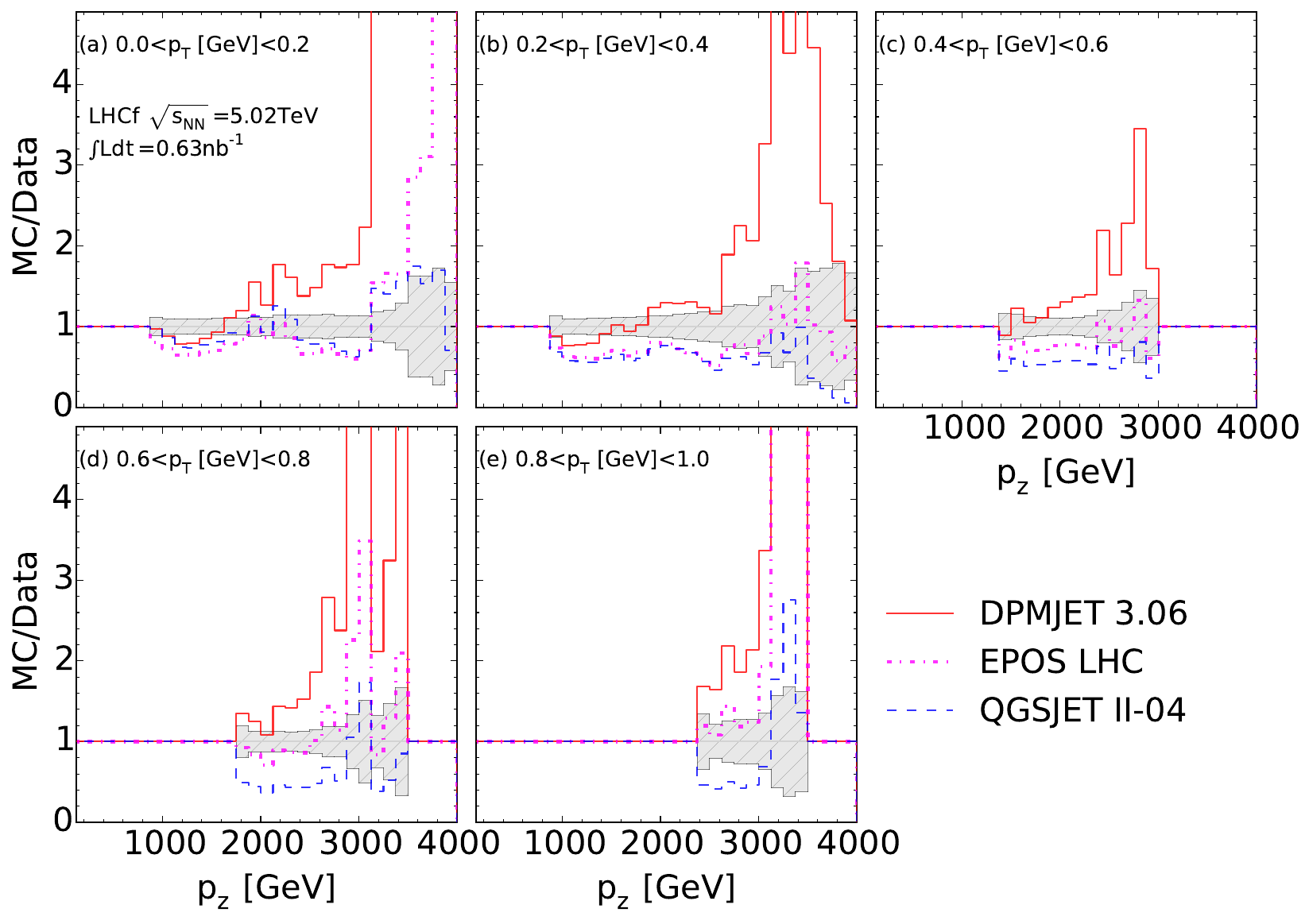}
  \caption{(color online). Ratios of LHCf $\pz$ distributions to the $\pz$
  distributions predicted by hadronic interaction models in $\pPb$ collisions at
  $\snn=\tev{5.02}$ are shown by solid red line (\textsc{dpmjet}), dashed blue
  line (\textsc{qgsjet}), and dashed-dotted magenta line (\textsc{epos}). Shaded
  areas indicate the range of total uncertainties of the $\pt$ distributions.}
  \label{fig:pz_pPb_r}
  \end{center}
\end{figure*}

The $\pz$ distributions are shown in Fig.~\ref{fig:pz_pPb}.
Figure~\ref{fig:pz_pPb_r} presents the ratios of LHCf $\pz$ distributions to the
$\pz$ distributions predicted by the hadronic interaction models.
A similar tendency to that found in $\pp$ collisions at $\sqs = \tev{7}$ is
found for LHCf data relative to model predictions.
Concerning the comparison of hadronic interaction models with LHCf data,
\textsc{qgsjet} shows a very good agreement at $\pt<\gev{0.2}$. However at
$\pt>\gev{0.2}$, there are no models giving a consistent description of LHCf
data within uncertainty over all $\pz$ bins, although \textsc{epos} shows a
certain compatibility with LHCf data for $\pt>\gev{0.4}$ and for $\pz<\tev{3}$.
The \textsc{dpmjet} predictions agree with LHCf data at $\pt<\gev{0.6}$ and
$\pz<\tev{2}$, while showing a harder distribution at higher $\pz$ similar to
$\pp$ collisions.
Again note the characteristic bump found in the LHCf data at $\pz\sim\tev{1.2}$
and $\pt<\gev{0.4}$, originating from the channel $\gamp \to \pizero + p$ via
baryon resonances in UPCs.

%
%
\section{Comparisons of the LHCf measurements among
different colliding hadrons and energies}\label{sec:discussions}

\subsection{Average transverse momentum}\label{sec:average_pt}

According to the scaling law proposed in Ref.~\cite{Amati}, the average
transverse momentum, denoted $\avept$, as a function of rapidity should be
independent of the center-of-mass energy in the projectile fragmentation region.
Here we obtain and compare the $\avept$ distributions as functions of rapidity
for $\pp$ and $\pPb$ collisions. In the study of this paper, $\avept$ is
obtained by three methods discussed below.
The first two methods use analytic distributions that are fit to the LHCf data
and the third uses numerical integration of the LHCf data.

\begin{figure}[htbp]
  \begin{center}
  \includegraphics[width=8.8cm, keepaspectratio]{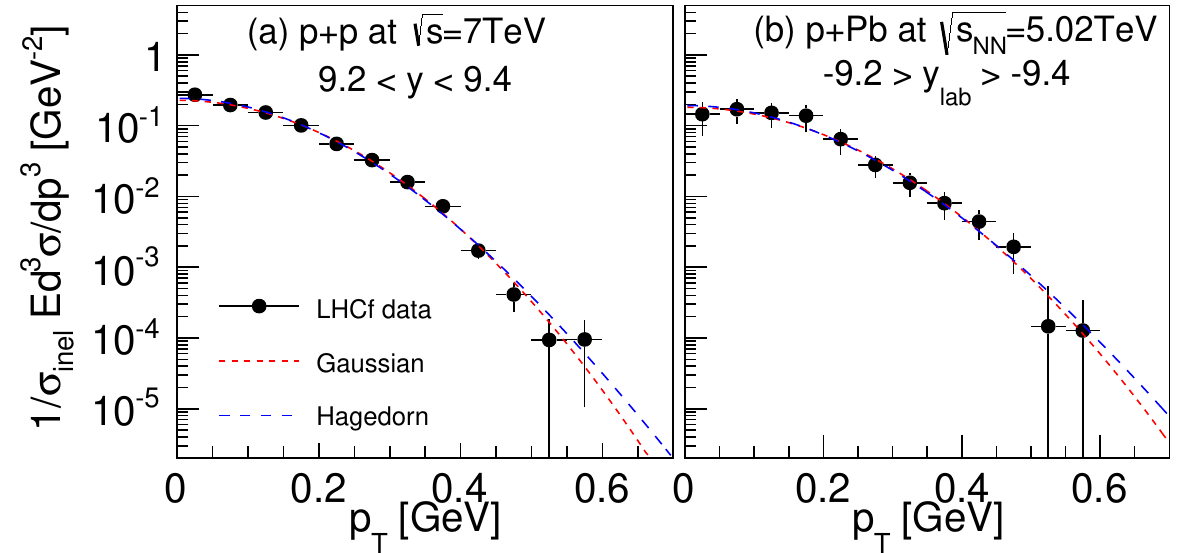}
  \caption{(color online). LHCf $\pt$ distributions (filled black
  circles), the best-fit Gaussian distributions (dotted red curve), and the
  best-fit Hagedorn functions (dashed blue curve). Left: the data for $\pp$
  collisions at $\sqs=\tev{7}$. Right: the data for $\pPb$ collisions at
  $\snn=\tev{5.02}$.}
  \label{fig:pt_fit}
  \end{center}
\end{figure}

The first method uses the fit of an empirical Gaussian distribution to the LHCf
$\pt$ distributions for each rapidity range in Fig.~\ref{fig:pt_7TeV},
\ref{fig:pt_2TeV}, and \ref{fig:pt_pPb}. The second method uses a Hagedorn
function.
Here we pay attention to the fact that soft scattering dominates the measured
$\pizero$ events for $\pt<\sim\gev{1}$ thus excluding from the analysis a
power-law distribution that is used predominantly for hard scattering at
$\pt>\sim\gev{1}$.
These methods do not necessarily require that the measured $\pt$ distribution be
available down to $\gev{0.0}$, although the best-fit distribution may then
include a systematic uncertainty in its fit~\cite{Hagedorn}.
Detailed descriptions of the parametrization and derivation of $\avept$ by using
the best-fit Gaussian distribution can be found in Ref.~\cite{LHCfpppi0}.
In a Hagedorn function~\cite{Hagedorn}, the invariant cross section of
identified hadrons, namely $\pizero$s in this paper, with a given mass
$\widehat{m}$\,[GeV] and temperature $\widehat{T}$\,[GeV] can be written as
\begin{equation}
    \frac{1}{\sigma_\text{inel}} E \frac{d^{3}\sigma}{dp^3} =
	A \cdot \sqrt{\pt^2+\widehat{m}^2} \sum_{n=1}^{\infty} K_1
	\left(n\frac{\sqrt{\pt^2+\widehat{m}^2}}{\widehat{T}} \right),
	\label{eq:hagedorn}
\end{equation}
where $A$\,[$\si{\per\cubic\giga\electronvolt}$] is a normalization factor and
$K_1$ is the modified Bessel function. Approximately half of the $\pizero$
measured with the LHCf detector are daughters from the decay of parent baryons
and mesons and are not directly produced. Thus the measured $\pt$ distribution
is no longer a thermal distribution of prompt $\pizero$s and so we set
$\widehat{m}$ as a free parameter as well as $A$ and $\widehat{T}$ in the fit of
a Hagedorn function to the $\pt$ distribution.
Equation~(\ref{eq:hagedorn}) converges by $n \approx 5$ and the computation is
in fact stopped at $n=10$.
The $\avept$ value is calculated by using the modified Bessel functions
$K_{5/2}$ and $K_2$ as functions of the ratio of the best-fit $\widehat{m}$ and
$\widehat{T}$ values~\cite{Hagedorn},
\begin{equation}
	\avept = \sqrt{\frac{\pi \widehat{m} \widehat{T}}{2}}
	\frac{\sum_{n=1}^{\infty}K_{5/2}(n(\widehat{m}/\widehat{T}))}
	{\sum_{n=1}^{\infty}K_{2}(n(\widehat{m}/\widehat{T}))}.
	\label{eq:hagedornavept}
\end{equation}
For reference, Fig.~\ref{fig:pt_fit} shows LHCf $\pt$ distributions (filled
black circles) and the best fits of the Gaussian distributions and the Hagedorn
functions.
The left panel of Fig.~\ref{fig:pt_fit} shows the results for $9.2<y<9.4$ in
$\pp$ collisions at $\sqs=\tev{7}$. The best-fit Gaussian distribution (dotted
red curve) and Hagedorn function (dashed blue curve) to the LHCf data mostly
overlap each other and give compatible $\avept$ values.
The right panel of Fig.~\ref{fig:pt_fit} shows the results for $-9.2>\ylab>-9.4$
in $\pPb$ collisions at $\snn=\tev{5.02}$. Note that the $\pt$ distribution for
LHCf data is plotted after subtraction of the UPC component where the systematic
uncertainty in the simulation of UPC events has been taken into account.
The best-fit Gaussian distribution and the Hagedorn function reproduce the LHCf
$\pt$ distributions within the total uncertainties and are also compatible with
each other.

Finally, for the third method, $\avept$ is obtained by numerically integrating
the $\pt$ distributions in Fig.~\ref{fig:pt_7TeV}, \ref{fig:pt_2TeV}, and
\ref{fig:pt_pPb}. The LHCf $\pt$ distributions in $\pPb$ collisions have already
had the UPC component subtracted. In this approach, $\avept$ is calculated only
in the rapidity range where the $\pt$ distributions are available down to
$\gev{0.0}$.
The high-$\pt$ tail that extends beyond the data ($\pt\gg\avept$) has a
negligible contribution to $\avept$. The final $\avept$ values obtained in this
analysis, denoted $\avept_\textrm{LHCf}$, have been determined by averaging the
$\avept$ values calculated with the three above described independent
approaches: Gaussian, Hagedorn and numerical integration. The uncertainty of
$\avept_\textrm{LHCf}$ for each rapidity bin is assigned to fully cover the
minimum and maximum $\avept$ values obtained by the three approaches. The
$\avept_\textrm{LHCf}$ values are summarized in Table.~\ref{tbl:avept}.

\begin{figure}[htbp]
  \begin{center}
  \includegraphics[width=7cm, keepaspectratio]{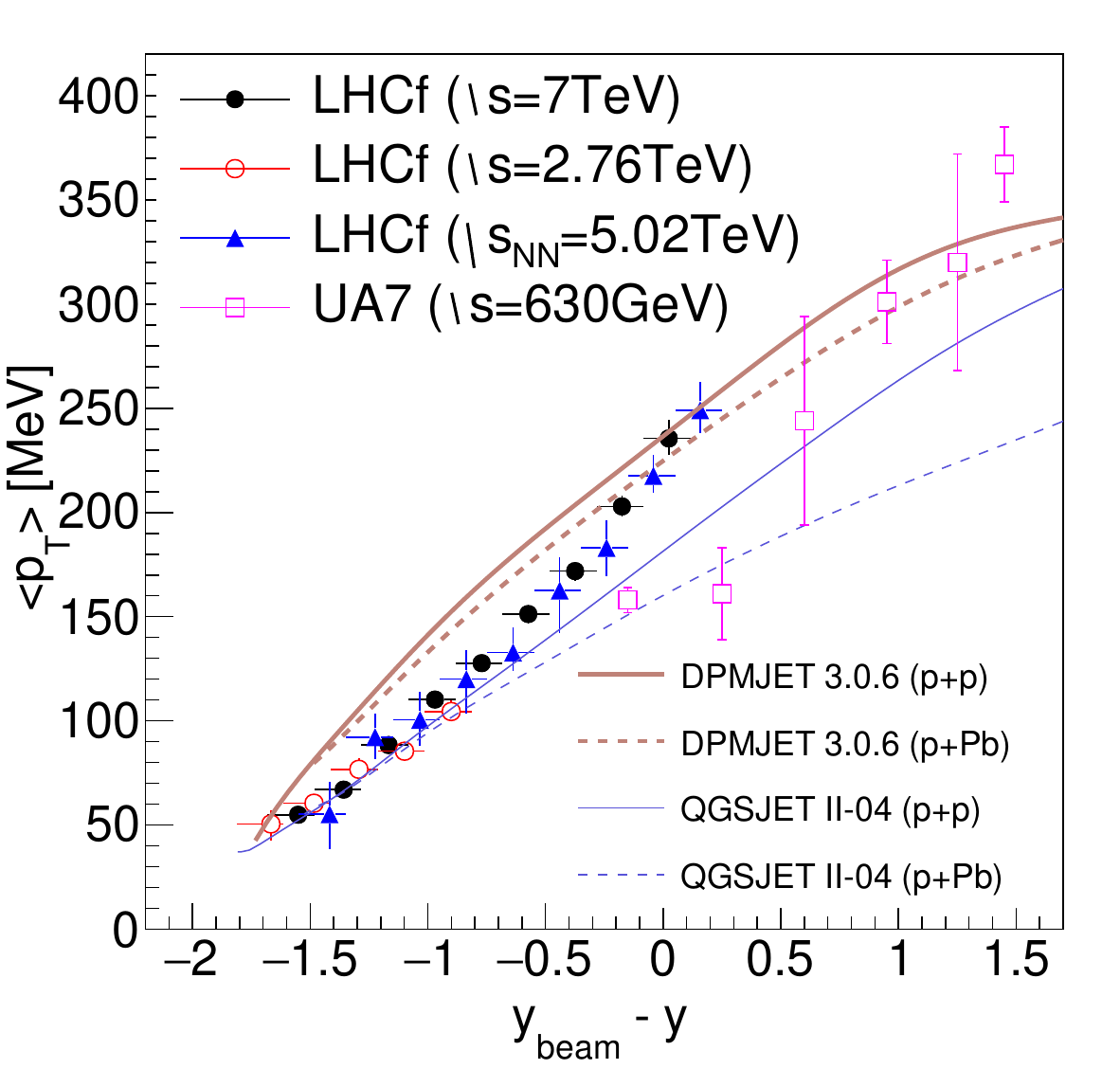}
  \caption{(color online). Average $\pt$ as a function of rapidity loss $\deltay
  = \ybeam-y$. Open red circles and filled black circles indicate LHCf data in
  $\pp$ collisions at $\sqs=2.76$ and $\tev{7}$, respectively. The results of
  the UA7 experiment (open magenta box) at Sp$\bar{\text{p}}$S ($p+\bar{p}$
  collisions at $\sqs=\gev{630}$) and the predictions from \textsc{dpmjet}
  (thick lines) and \textsc{qgsjet} (thin lines) are added for reference.}
  \label{fig:pt_scale}
  \end{center}
\end{figure}

In Fig.~\ref{fig:pt_scale}, $\avept$ in $\pp$ collisions at $\sqs=2.76$ and
$\tev{7}$, and in $\pPb$ collisions at $\snn=\tev{5.02}$ are presented as a
function of rapidity loss $\deltay \equiv \ybeam - y$, where $\ybeam$ is the
beam rapidity for each collision energy. The shift in rapidity by $\ybeam$
allows a direct comparison to be made between the $\avept$ results at different
collision energies.
We see that for $\deltay > -1.3$, $\avept$ at $\sqs=\tev{2.76}$ (open red
circles) has slightly smaller values than at $\sqs=\tev{7}$ (filled black
circles), although the two sets of data are mostly compatible at the
$\pm\SI{10}{\percent}$ level.
For reference, the Sp$\bar{\text{p}}$S UA7 results for $p+\bar{p}$ collisions at
$\sqs=\gev{630}$~\cite{UA7}  (open magenta squares) show a rapid roll off of
$\avept$ at low $\deltay$ compared to LHCf data.
The LHCf and UA7 results are particularly incompatible for $-0.3<\deltay<0.3$.
The comparison of the LHCf data with the UA7 results indicates that $\avept$ may
depend on the center-of-mass energy. However, in order to firmly confirm a
center-of-mass energy dependence of $\avept$, we need to have experimental data
at a lower collision energy, e.g., $\sqs<\tev{1}$ and with a wider range of
rapidity.
Approved plans are underway to obtain this data by installing the LHCf detector
at the RHIC ZDC position~\cite{RHICf}.
The $\avept$ values obtained from $\pPb$ collisions at $\snn=\tev{5.02}$ (filled
blue triangles) are consistent with those from $\pp$ collisions at
$\sqs=\tev{7}$ within the systematic uncertainties present.
The predictions from \textsc{dpmjet} (thick solid red line) and \textsc{qgsjet}
(thin solid blue line) in $\pp$ collisions at $\sqs=\tev{7}$ and $\pPb$
collisions at $\snn=\tev{5.02}$ have been added to Fig.~\ref{fig:pt_scale} for
reference. The predictions at $\sqs=\tev{2.76}$ are excluded in
Fig.~\ref{fig:pt_scale}, since these curves mostly overlap with those at
$\tev{7}$.
LHCf data in $\pp$ collisions at $\sqs=\tev{7}$ are close to the predictions
made by \textsc{dpmjet} at large $\deltay$ (small $y$) and become close to those
made by \textsc{qgsjet} at small $\deltay$ (large $y$). These relations between
LHCf data and the model predictions are consistent with the $\pt$ distributions
shown in Fig.~\ref{fig:pt_7TeV} and \ref{fig:pt_2TeV}.
The prediction from \textsc{dpmjet} (thick dashed red line) for $\pPb$
collisions at $\snn=\tev{5.02}$ is compatible with the LHCf result for
$-0.3<\deltay<0.2$, which is derived from the good agreement of this model with
LHCf data at $-8.8>\ylab>-10.0$ and $\pt<\gev{0.3}$.
Conversely, the prediction obtained from \textsc{qgsjet} (thin dashed blue line)
is smaller than LHCf data for $\deltay>-0.5$ and approaches the LHCf results on
decreasing $\deltay$. This tendency was already found in Fig.~\ref{fig:pt_pPb};
the prediction from \textsc{qgsjet} shows an overall agreement with LHCf $\pt$
distributions at $\ylab<-9.8$.

\begin{table}[htbp]
    \begin{center}
    \caption{The average $\pizero$ transverse momenta for the rapidity range
    $8.8<y<10.6$ in $\pp$ collisions at $\sqs=2.76$ and $\tev{7}$ and for the
    rapidity range $-8.8>\ylab>-10.6$ in $\pPb$ collisions at
    $\snn=\tev{5.02}$.}
    \begin{ruledtabular}
    \begin{tabular}{l c c c}
    \multicolumn{1}{c}{Rapidity~\footnote{The rapidity values for $\pPb$
    collisions are in the detector reference frame and must be multiplied by -1.}} &
    \multicolumn{3}{c}{$\avept_\textrm{LHCf}$ [MeV]} \\
    \cline{2-4}
    & $\pp~\tev{2.76}$ & $\pp~\tev{7}$ & $\pPb~\tev{5.02}$ \\
    \hline
$[$8.8, 9.0$]$  &    $103.5\pm7.5$\hph & $242.8\pm8.6$    & $244.5\pm43.2$    \\
$[$9.0, 9.2$]$  & \hph$78.5\pm7.8$\hph & $208.5\pm6.1$    & $223.1\pm12.7$    \\
$[$9.2, 9.4$]$  & \hph$76.4\pm5.7$\hph & $182.6\pm4.3$    & $189.9\pm7.6$\hph \\
$[$9.4, 9.6$]$  & \hph$60.3\pm5.2$\hph & $160.2\pm3.8$    & $173.8\pm17.2$    \\
$[$9.6, 9.8$]$  & \hph$50.4\pm10.4$    & $132.3\pm3.4$    & $138.1\pm18.7$    \\
$[$9.8, 10.0$]$ &                 & $113.9\pm3.4$    & $113.0\pm6.3$\hph \\
$[$10.0,10.2$]$ &              & \hph$87.3\pm3.9$    & $112.2\pm15.4$    \\
$[$10.2,10.4$]$ &              & \hph$67.5\pm3.0$ & \hph$90.7\pm6.7$\hph \\
$[$10.4,10.6$]$ &              & \hph$55.6\pm3.1$ & \hph$61.0\pm6.6$\hph \\
    \end{tabular}
    \end{ruledtabular}
    \label{tbl:avept}
    \end{center}
\end{table}

\subsection{Limiting fragmentation}\label{sec:limitfrag}

The hypothesis of limiting fragmentation~\cite{Limiting1,Limiting2,Limiting3}
asserts that the number of fragments of a colliding hadron will follow a
limiting rapidity distribution in the rest frame of the target hadron. In this
case the rapidity distribution of the secondary particles in the forward
rapidity region would be independent of the center-of-mass energy.
In this paper, a test of the limiting fragmentation hypothesis is performed by
using LHCf data in $\pp$ collisions at $\sqs=2.76$ and $\tev{7}$.

The normalized rapidity distribution of $\pizero$s,
$(1/\sigma_\text{inel})(d\sigma/dy)$, in this analysis can be obtained by using
very similar methods that were used for the derivation of the average $\pt$ in
Sec.~\ref{sec:average_pt}.

The first method uses the fit of an empirical distribution to the LHCf $\pt$
distributions in Fig.~\ref{fig:pt_7TeV} and \ref{fig:pt_2TeV} in each rapidity
range.
As we discussed in Sec.~\ref{sec:average_pt}, two distributions are chosen to
parametrize the $\pt$ distributions: a Gaussian distribution and a Hagedorn
function.
The rapidity distribution is derived by integrating the best-fit Gaussian
distribution and Hagedorn function along the $\pt$ axis from $\gev{0.0}$ to
infinity.

The rapidity distribution can also be obtained by numerically integrating the
$\pt$ distributions in Fig.~\ref{fig:pt_7TeV} and \ref{fig:pt_2TeV}.
In this approach, the derivation of the $(1/\sigma_\text{inel})(d\sigma/dy)$
value is possible only in the rapidity range where the $\pt$ distributions are
available down to \gev{0.0}.
Again, the final rapidity distribution is derived by averaging the rapidity
distributions obtained by the above three methods.
The estimated uncertainty is obtained from the minimum and maximum values for
each rapidity bin.

\begin{figure}[htbp]
  \begin{center}
  \includegraphics[width=7cm, keepaspectratio]{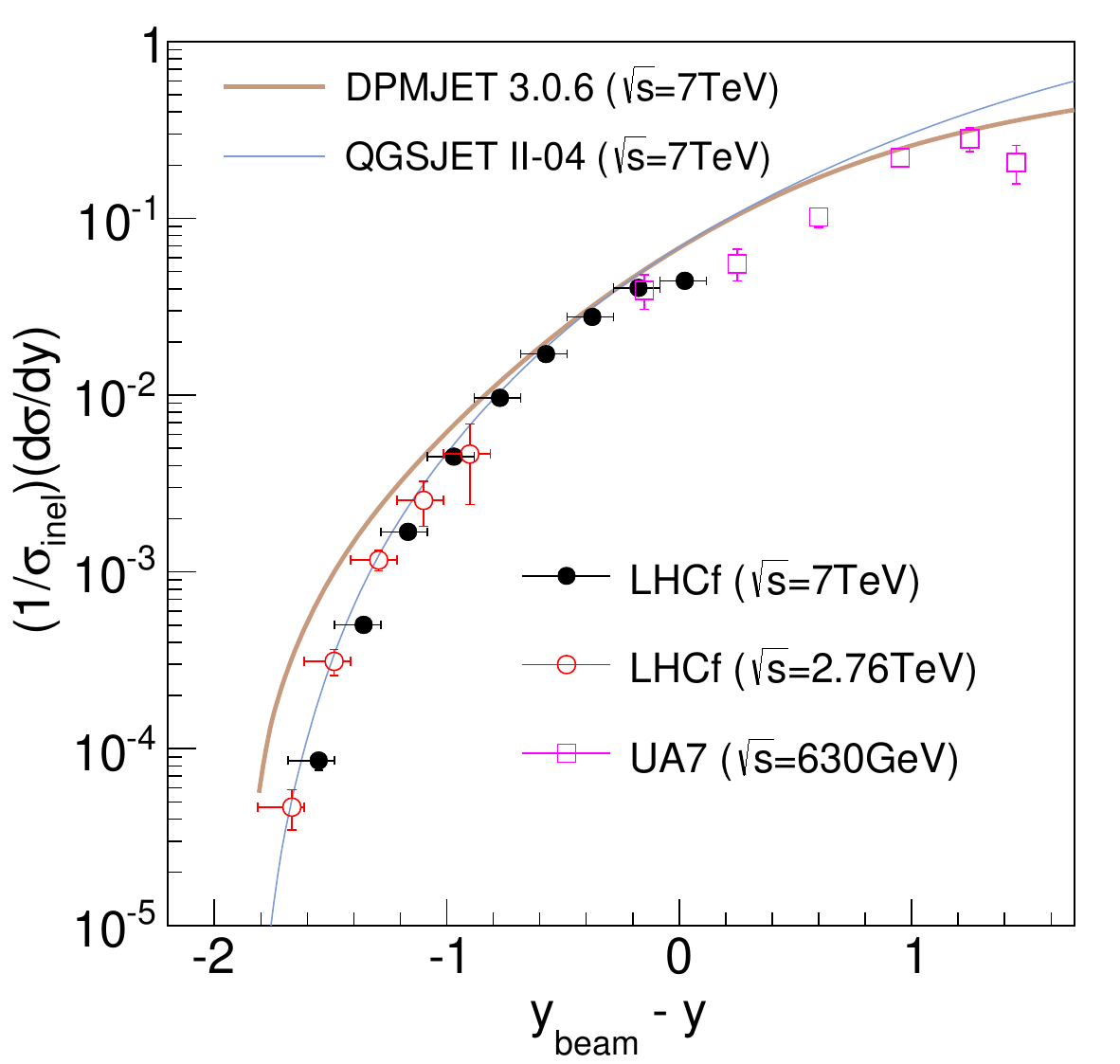}
  \caption{(color online). The $\pizero$ yield in each rapidity interval as a
  function of rapidity loss $\deltay = \ybeam-y$. Open red circles and filled
  black circles indicate LHCf data in $\pp$ collisions at $\sqs=2.76$ and
  $\tev{7}$, respectively. The results of the UA7 experiment (open magenta
  squares) at Sp$\bar{\text{p}}$S ($p+\bar{p}$ collisions at $\sqs=\gev{630}$)
  and the predictions by \textsc{dpmjet} (thick red line) and
  \textsc{qgsjet} (thin blue line) are added for reference.}
  \label{fig:dndy}
  \end{center}
\end{figure}

Figure~\ref{fig:dndy} shows the rapidity distributions as functions of the
rapidity loss $\deltay$ (i.e., $\ybeam - y$) in $\pp$ collisions at
$\sqs=\tev{2.76}$ (open red circles) and $\tev{7}$ (filled black circles).
The rapidity distributions for both collision energies mostly appear to lie
along a common curve in the rapidity range $-1.8<\deltay<-0.8$. LHCf data are
consistent at the $\pm\SI{15}{\percent}$ level with the hypothesis of limiting
fragmentation in the very forward region.

For comparison the experimental results from the UA7 experiment~\cite{UA7} are
also shown in Fig.~\ref{fig:dndy}. The extrapolated LHCf curve at $\tev{7}$ to
higher $\deltay$ (i.e., lower $y$) could be compatible with the UA7 results, at
least for $\deltay \lesssim 0.5$.

The predictions of \textsc{dpmjet} (thick red curve) and \textsc{qgsjet} (thin
blue curve) at $\sqs=\tev{7}$ have been added to Fig.~\ref{fig:dndy} for
reference. The predictions at $\sqs=\tev{2.76}$ have been omitted, since these
curves mostly overlap with those at $\tev{7}$ since limiting fragmentation holds
in \textsc{dpmjet} and \textsc{qgsjet}. The best agreement with LHCf data at
$\sqs=2.76$ and $\tev{7}$ is obtained by the \textsc{qgsjet} model.
The \textsc{dpmjet} predictions generally give a larger $\pizero$ yield and a
harder $\pt$ distribution especially for $y>9.8$ at $\sqs=\tev{7}$ and for
$y>9.4$ at \tev{2.76}.

\subsection{Feynman scaling}\label{sec:pzscale}

In Ref.~\cite{Feynman} Feynman proposed that the production cross sections of
secondary particles as functions of the Feynman-x variable (defined by $\xf
\equiv 2\pz/\sqs$) were independent of the incident energy in the forward
region.
If the so-called Feynman scaling holds, the differential cross section as a
function of $\xf$ (hereafter $\xf$ distribution)
$(\xf/\sigma_\text{inel})(d\sigma/d\xf)$ should be independent of the
center-of-mass energy for $\xf \gtrsim 0.2$. Here the rapidity distribution
introduced in Sec.~\ref{sec:limitfrag} can be rewritten as
\begin{equation}
	\frac{1}{\sigma_\text{inel}}\frac{d\sigma}{dy} =
	\frac{E}{\sigma_\text{inel}}\frac{d\sigma}{d\pz} =
	\frac{\xe}{\sigma_\text{inel}}\frac{d\sigma}{d\xf},
	\label{eq:scaling}
\end{equation}
where $\xe \equiv 2E/\sqs$ and $dy=d\pz / E$ are used for the second form.
Considering $\pz \approx E$ in the forward region, $\xe$ can be considered as
$\xf$ and thus the right hand side of Eq.~(\ref{eq:scaling}) becomes
approximately $(\xf/\sigma_\text{inel})(d\sigma/d\xf)$.
Consequently, the limiting fragmentation hypothesis that states
$(1/\sigma_\text{inel})(d\sigma/dy)$ is independent of the center-of-mass energy
in each rapidity bin can be rewritten as Feynman scaling which states
$(\xf/\sigma_\text{inel})(d\sigma/d\xf)$ is independent of the center-of-mass
energy in each $\xf$ bin.
In this paper, we test the Feynman scaling hypothesis by comparing LHCf data in
$\pp$ collisions at $\sqs=2.76$ and \tev{7}.

\begin{figure}[htbp]
  \begin{center}
  \includegraphics[width=7cm, keepaspectratio]{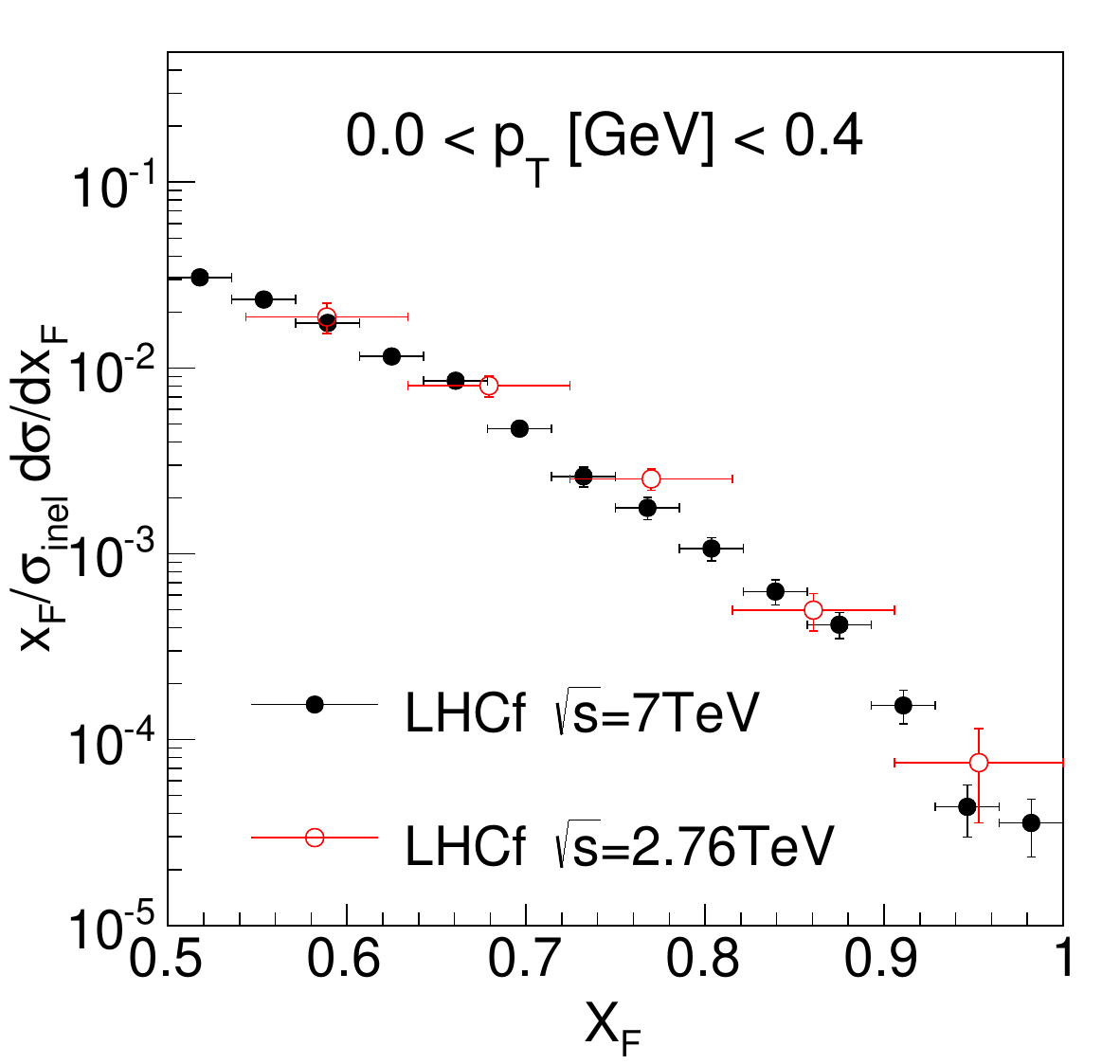}
  \caption{(color online). The $\pizero$ yield at $0.0<\pt<\gev{0.4}$ as a
  function of $\xf$. Open red circles and filled black circles indicate LHCf
  data in $\pp$ collisions at $\sqs=2.76$ and $\tev{7}$, respectively.}
  \label{fig:xf2}
  \end{center}
\end{figure}

\begin{figure}[htbp]
  \begin{center}
  \includegraphics[width=8.4cm, keepaspectratio]{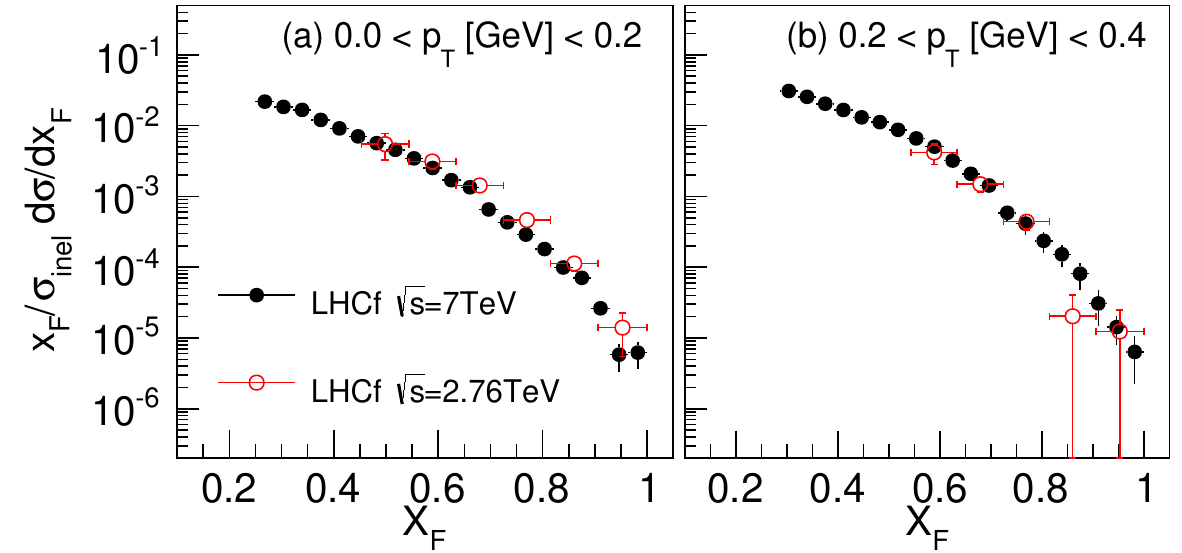}
  \caption{(color online). The $\pizero$ yield in each $\pt$ range as a function
  of $\xf$. Left: the distributions for $0.0<\pt<\gev{0.2}$. Right:
  the distributions for $0.2<\pt<\gev{0.4}$. Open red circles and filled black
  circles indicate LHCf data in $\pp$ collisions at $\sqs=2.76$ and $\tev{7}$,
  respectively.}
  \label{fig:xf}
  \end{center}
\end{figure}

In Fig.~\ref{fig:xf2}, we compare the $\xf$ distributions in the $\pt$ range
$0.0<\pt<\gev{0.4}$. Other $\pt$ ranges are excluded from the comparison, since
LHCf data at $\sqs=\tev{2.76}$ are unavailable outside this range.
The $\xf$ distributions at $\sqs=2.76$ and \tev{7} are compatible with each
other at the $\pm\SI{20}{\percent}$ level.
In Fig.~\ref{fig:xf}, we further compare the $\xf$ distributions for the reduced
$\pt$ ranges: $0.0<\pt<\gev{0.2}$ and $0.2<\pt<\gev{0.4}$.
At $0.0<\pt<\gev{0.2}$, only the bin $0.73<\xf<0.82$ at $\sqs=\tev{2.76}$
deviates from the one at \tev{7} by $\SI{30}{\percent}$, while the other bins
are consistent within their uncertainties.
At $0.2<\pt<\gev{0.4}$, all bins at $\sqs=\tev{2.76}$ are consistent with the
ones at \tev{7}, except for the bin $0.82<\xf<0.91$ that has a smaller
($\SI{40}{\percent}$) cross section than at \tev{7}, although there is a large
uncertainty at \tev{2.76}.
Overall the $\xf$ distributions at $\sqs=2.76$ and \tev{7} indicate that Feynman
scaling holds at the $\pm\SI{20}{\percent}$ level at these center-of-mass
energies in the very forward region.

Besides a test of the Feynman scaling, we find in Fig.~\ref{fig:xf} that the
yield of $\pizero$s at $\sqs=\tev{2.76}$ relative to $\tev{7}$ is slightly
larger for $0.0<\pt<\gev{0.2}$ and slightly smaller for $0.2<\pt<\gev{0.4}$.
This tendency means that the $\pt$ distributions at $\sqs=\tev{2.76}$ are softer
than those at \tev{7}, leading to the small $\avept$ values at \tev{2.76}
relative to those at \tev{7} as already found in Fig.~\ref{fig:pt_scale}.

\subsection{$\pt$ dependence of the $\xf$ distributions}\label{sec:xffit}

In hadronic interactions at large rapidities, partons from the projectile and
target hadrons generally have large and small momentum fractions respectively,
since the momentum fraction that the parton itself carries relative to the
parent projectile and target hadrons, i.e., the Bjorken-$x$ variable or $\xbj$,
is proportional to $e^{\pm y}$ ($+y$ for projectile and $-y$ for target).
Here we note that a parton (dominantly gluon) density, rapidly increases with
decreasing $\xbj$ when $\xbj<0.01$ with the target approaching the blackbody
limit where the gluon density is saturated. In the blackbody regime, the partons
cannot go through the target nuclear medium without interaction and suffer
transverse momenta transfers proportional to the saturation momentum scale
$Q_s$. The $Q_s$ values in the very forward region are $\sim\gev{1}$ in $\pp$
collisions and $\sim\gev{10}$ in $\pPb$ collisions, although the calculation of
$Q_s$ itself suffers from both theoretical and experimental uncertainties and is
also dependent on the impact parameter of the colliding hadrons~\cite{Albacete}.

In the $\pt$ region below $Q_s$, the $\xf$ distribution in the forward region
can be asymptotically written~\cite{Berera} as
\begin{equation}
	\frac{\xf}{\sigma_\text{inel}}\frac{d\sigma}{d\xf} \propto (1-\xf)^\alpha.
	\label{eq:xf}
\end{equation}
\noindent where $\alpha$ is the leading exponent.
In the blackbody regime, the $\xf$ distribution of the leading hadron is
strongly suppressed and thus $\alpha$ increases relative to the value found for
a dilute target.
Conversely, $\alpha$ decreases with increasing $\pt$ when $\pt$ approaches or
exceeds the saturation momentum scale $Q_s$.

\begin{figure}[htbp]
  \begin{center}
  \includegraphics[width=7cm, keepaspectratio]{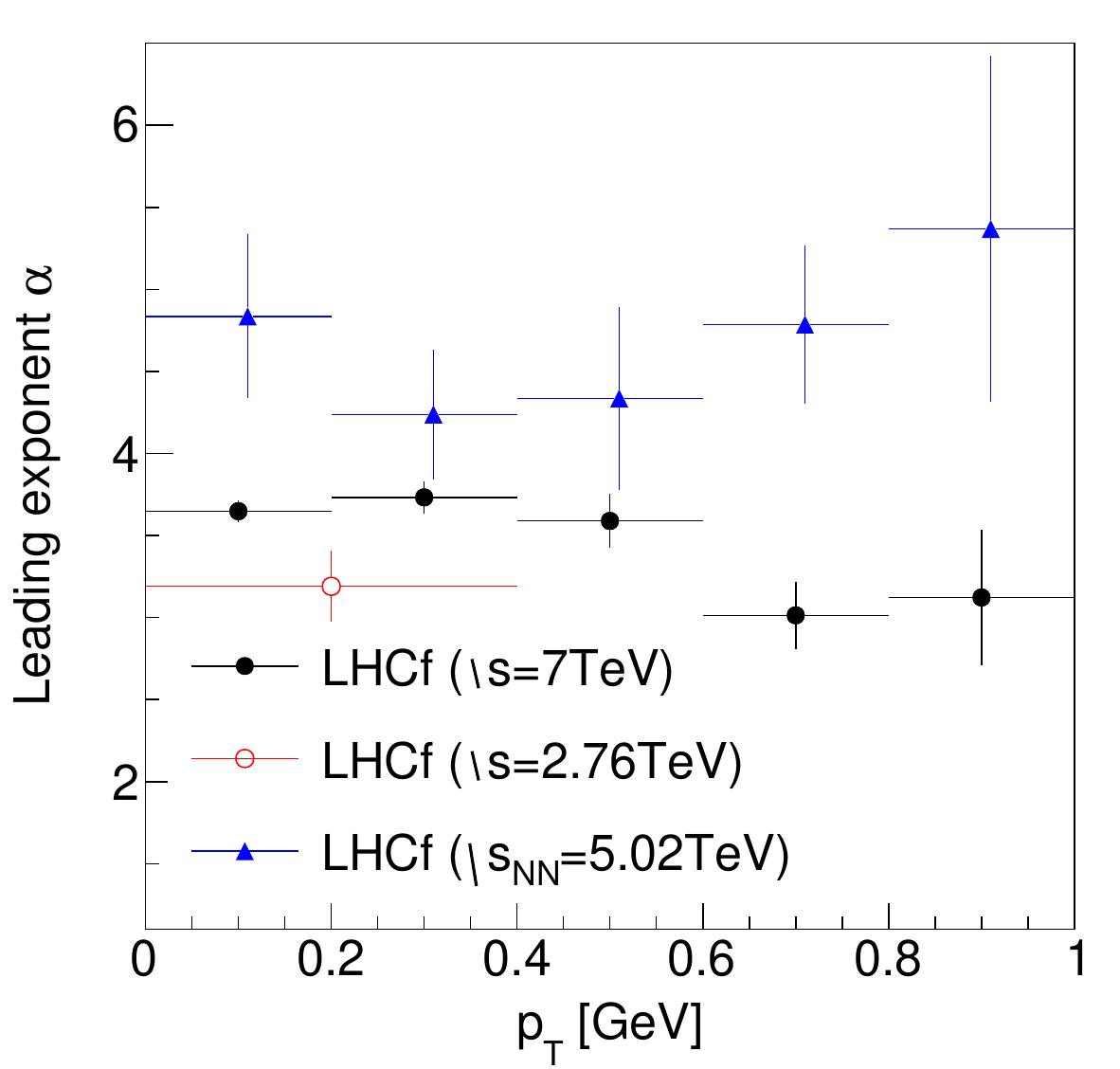}
  \caption{(color online). The best-fit leading exponent of (1-$\xf$) as a
  function of $\pt$. Open red circles and filled black circles indicate LHCf
  data in $\pp$ collisions at $\sqs=2.76$ and $\tev{7}$, respectively.
  Filled blue triangles indicate LHCf data in $\pPb$ collisions at
  $\snn=\tev{5.02}$.}
  \label{fig:xffit}
  \end{center}
\end{figure}

Figure~\ref{fig:xffit} shows the best-fit leading exponent $\alpha$ in each
$\pt$ range in $\pp$ and $\pPb$ collisions. The leading exponent in $\pp$
collisions at $\sqs=\tev{7}$ (filled black circles) is $\alpha \approx 3.7$ at
$\pt<\gev{0.6}$ and decreases to $\alpha \approx 3.0$ at $0.6<\pt<\gev{1.0}$.
The reduction of $\alpha$ with increasing $\pt$ can be understood as much of the
target staying in the blackbody regime for $\pt<\gev{0.6}$ and then gradually
escaping from the blackbody regime for $\pt>\gev{0.6}$. The leading exponent in
$\pp$ collisions at $\sqs=\tev{2.76}$ (open red circles) is slightly smaller
than that at \tev{7} though with large uncertainty. The comparison between
$\sqs=\tev{2.76}$ and \tev{7} may indicate that the upper $\pt$ limit of the
measurement at \tev{2.76} is near the saturation momentum at \tev{2.76} and that
the suppression due to the gluon density is weaker than at \tev{7}, although the
calculated $Q_s$ at \tev{2.76} is only slightly different from the $Q_s$ at
\tev{7}.
The leading exponents in $\pPb$ collisions at $\snn=\tev{5.02}$ (filled blue
triangles) are rather flat along the $\pt$ axis, within uncertainties that are
generally larger than those in $\pp$ collisions.
This may indicate that the saturation momentum in $\pPb$ collisions is well
above the measured $\pt$ range and also that the $\xf$ distributions in $\pPb$
collisions are suppressed relative to those for $\pp$ collisions.

\subsection{Nuclear modification factor}\label{sec:nmf}

The effects of high gluon density in the target are inferred from the comparison
of the leading exponent $\alpha$ between in $\pp$ and $\pPb$ collisions (see the
preceding Sec.~\ref{sec:xffit}). Here we introduce the nuclear modification
factor that quantifies the $\pt$ spectra modification caused by nuclear effects
in $\pPb$ collisions with respect to $\pp$ collisions.
The nuclear modification factor $\nmf$ is defined as
\begin{equation}
    \nmf \equiv
    \frac{\sigma^\textrm{pp}_\textrm{inel}}
    {\langle N_\textrm{coll} \rangle \sigma^\textrm{pPb}_\textrm{inel}}
    \frac{Ed^3\sigma^\textrm{pPb}/dp^3}{Ed^3\sigma^\textrm{pp}/dp^3},
    \label{eq:nmf}
\end{equation}
\noindent where $E d^{3}\sigma^\text{pPb} / dp^3$ and $E d^{3}\sigma^\text{pp} /
dp^3$ are the inclusive cross sections of $\pizero$ production in $\pPb$
collisions at $\snn=\tev{5.02}$ and in $\pp$ collisions at $\sqs=\tev{5.02}$,
respectively. These cross sections are obtained from Eq.~(\ref{eq:cross_pPb})
and Eq.~(\ref{eq:cross_pp}), with the subtraction of the expected UPC
contribution applied to the cross section for $\pPb$ collisions.
The uncertainty in the inelastic cross section $\sigma^\text{pPb}_\text{inel}$
is estimated to be $\pm\SI{5}{\percent}$~\cite{LHCfpPbpi0}. The average number
of binary nucleon--nucleon collisions in a $\pPb$ collision, $\langle
N_\textrm{coll} \rangle = 6.9$, is obtained from MC simulations using the
Glauber model~\cite{dEnterria}. The uncertainty of
$\sigma^\textrm{pp}_\textrm{inel}/\langle N_\textrm{coll} \rangle$ is estimated
by varying the parameters in the calculation with the Glauber model and is of
the order of $\pm\SI{3.5}{\percent}$~\cite{LHCfpPbpi0}. Finally the quadratic
sum of the uncertainties in $\sigma^\text{pPb}_\text{inel}$ and
$\sigma^\textrm{pp}_\textrm{inel}/\langle N_\textrm{coll} \rangle$ is added to
$\nmf$.

Since there is no LHCf data for $\pp$ collisions at exactly $\sqs=\tev{5.02}$,
$E d^{3}\sigma^\text{pp} / dp^3$ is derived by scaling the $\pt$ distributions
taken in $\pp$ collisions to other collision energies.
The derivation follows three steps.
First, the $\avept$ at $\sqs=\tev{5.02}$ is estimated by interpolating the
measured $\avept$ values at \tev{7}.
The uncertainty of the interpolated $\avept$ values is estimated to be
$\pm\SI{10}{\percent}$ according to the differences between the measured
$\avept$ values at $\sqs=2.76$ and \tev{7} for $-1.7<\deltay<-0.8$ (see
Fig.~\ref{fig:pt_scale}).
Second, the absolute normalization of the $\pt$ distribution value in each
rapidity range for $\sqs=\tev{5.02}$, i.e.,
$(1/\sigma_\text{inel})(d\sigma/dy)$, is determined by interpolating the
rapidity distribution at \tev{7} (see Fig.~\ref{fig:dndy}).
The uncertainty of the absolute normalization is estimated to be
$\pm\SI{15}{\percent}$ according to the discussion in Sec.~\ref{sec:limitfrag}
and is taken into account in the interpolated normalization.
Finally, the $\pt$ distributions at $\sqs=\tev{5.02}$ are produced by assuming
that the $\pt$ distribution follows either a Gaussian distribution or a Hagedorn
function and by using the $\avept$ values obtained in the first step and the
normalization obtained in the second step. The difference of the $\pt$
distribution produced using a Gaussian distribution or a Hagedorn function gives
the systematic uncertainty.
Note that the rapidity shift $-0.465$ explained in Sec.~\ref{sec:pPb5.02TeV} is
also taken into account for the $\pt$ distribution in $\pp$ collisions at
$\sqs=\tev{5.02}$.

\begin{figure*}[htbp]
  \begin{center}
  \includegraphics[width=16cm,keepaspectratio]{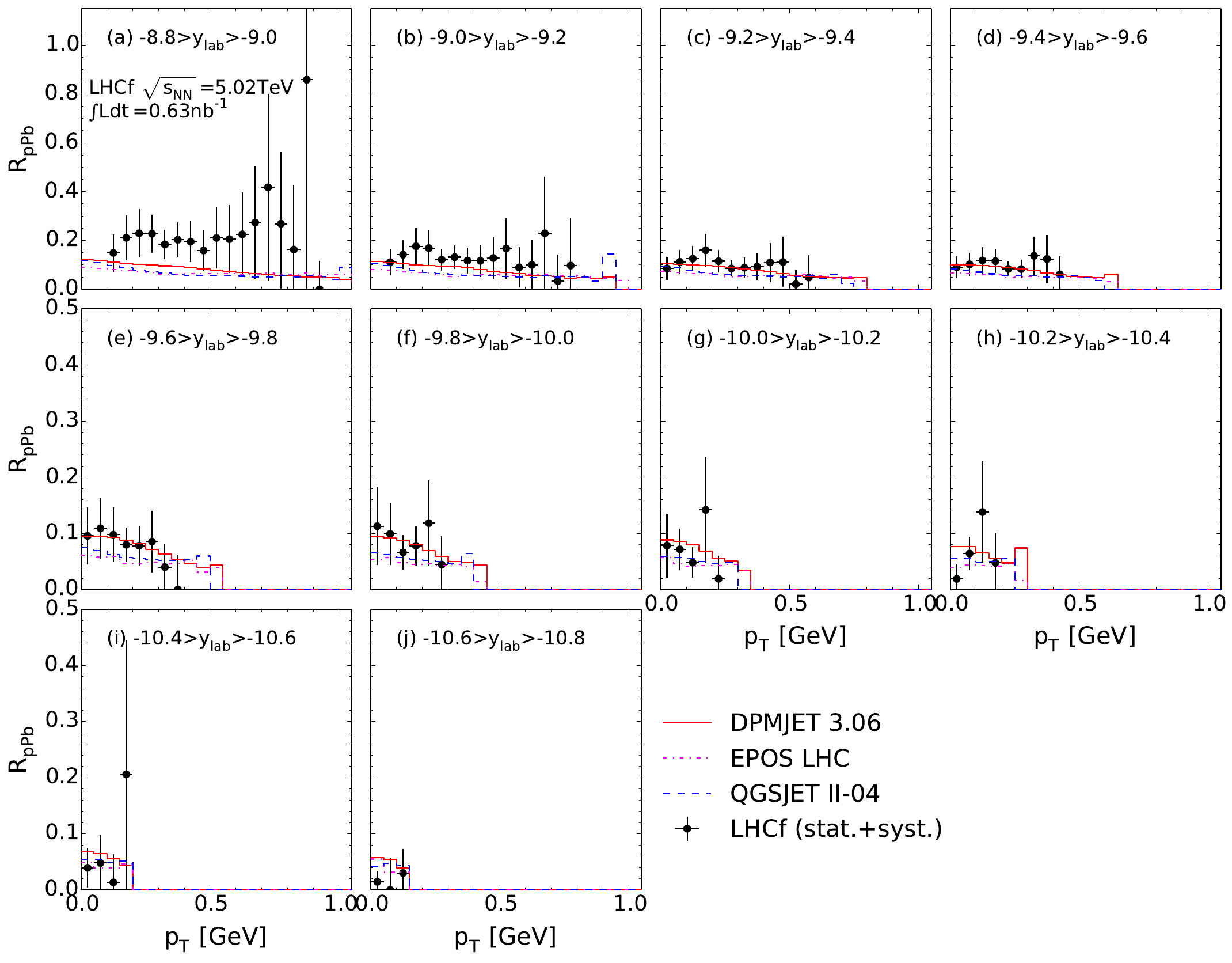}
  \caption{(color online). The nuclear modification factor for $\pizero$s.
  Filled circles indicate the factors obtained from LHCf data. Error bars
  indicate the total uncertainties incorporating both statistical and systematic
  uncertainties. Other lines are the predictions from hadronic interaction
  models: \textsc{dpmjet} (solid red line), \textsc{qgsjet} (dashed blue line),
  and \textsc{epos} (dashed-dotted magenta line).}
  \label{fig:nmf}
  \end{center}
\end{figure*}

Figure~\ref{fig:nmf} shows the nuclear modification factors $\nmf$ obtained from
LHCf data and the predictions from the hadronic interaction models,
\textsc{dpmjet} (solid red curve), \textsc{qgsjet} (dashed blue curve), and
\textsc{epos} (dotted magenta curve).
LHCf data show a strong suppression with $\nmf$ equal to 0.3 at $\ylab\sim-8.8$
down to $<0.1$ at $\ylab\sim-10.8$, although a large uncertainty is present due
to systematic uncertainties in the estimation of the $\avept$ values in $\pp$
collisions at $\sqs=\tev{5.02}$.
All hadronic interaction models, which employ different approaches for the
nuclear effects, predict small values of $\nmf \lesssim 0.15$. Within the
uncertainties the hadronic interaction models show an overall good agreement
with $\nmf$ estimated from LHCf data.

%
%
\section{Conclusions}\label{sec:conclusions}

The inclusive production of $\pizero$s was measured with the LHCf detector in
$\pp$ collisions at $\sqs=2.76$ and \tev{7} and in $\pPb$ collisions at
$\snn=\tev{5.02}$.
In $\pp$ collisions at $\sqs=\tev{7}$, differential cross sections as a function
of $\pt$ and $\pz$ for the $\pizero$s were measured by two independent LHCf
detectors, Arm1 and Arm2, with consistent results. Conversely, only the LHCf
Arm2 detector was used in $\pp$ collisions at $\sqs=\tev{2.76}$ and in $\pPb$
collisions.

In $\pp$ collisions, \textsc{qgsjet} II-04 shows an overall agreement with LHCf
data, while \textsc{epos lhc} distributions have a slightly harder behavior than
LHCf data for $\pt>\gev{0.5}$.
\textsc{dpmjet} 3.06 and \textsc{pythia} 8.185 have in general shown a harder
momentum distributions and a poor agreement with LHCf data.
In $\pPb$ collisions, \textsc{dpmjet} 3.06 showed good agreement with LHCf data
for $-8.8>\ylab>-10.0$ and $\pt<\gev{0.3}$, while \textsc{qgsjet} II-04 and
\textsc{epos lhc} did better reproducing the LHCf data for $\pt>\gev{0.4}$ than
\textsc{dpmjet} 3.06.

The average values of $\pt$, denoted $\avept$, at $y>8.8$ in $\pp$ collisions
and at $\ylab>8.8$ in $\pPb$ collisions were calculated using the LHCf $\pt$
distributions.
The $\avept$ values obtained have been shown to be independent of the
center-of-mass energy at the $\SI{10}{\percent}$ level.
Tests of limiting fragmentation and Feynman scaling hypotheses using LHCf data
in $\pp$ collisions show that both hypotheses hold in the forward region at the
$\SI{15}{\percent}$--$\SI{20}{\percent}$ level.
The leading exponent $\alpha$ and the nuclear modification factor $\nmf$ derived
from LHCf data indicate a strong suppression of $\pizero$ production from the
nuclear target relative to that from the nucleon target.
Within the uncertainties all of the hadronic interaction models presented gave
an overall good agreement with $\nmf$ estimated by LHCf data.
According to the analysis in this paper, we expect that the number of particles
leading to an electromagnetic component in air showers would follow the limiting
rapidity distribution and Feynman scaling hypotheses.
Combining the results for forward $\pizero$s in this paper with the recent
results for forward neutrons in Ref.~\cite{LHCfneutron} strongly constrain
models for air shower production at the TeV scale.

As a future prospect, additional analyses using correlations between forward
$\pizero$s and other particles (e.g., two-particle angular correlations) are
needed to reach a better understanding of the forward meson production mechanism
and the strong suppression of $\pizero$ production in $\pPb$ collisions compared
to $\pp$ collisions.
The ATLAS and LHCf Collaborations have taken $\pp$ data at $\sqs=\tev{13}$ and
$\pPb$ data at $\snn=\tev{5.02}$ with common triggers. This data could provide
the possibility for performing analyses of two particle correlations.

%
%
\begin{acknowledgements}

The LHCf Collaboration acknowledges CERN staff and the ATLAS Collaboration for
their essential contributions to the successful operation of LHCf.
We thank S. Ostapchenko and T. Pierog for numerous discussions and for
confirming the results of the MC simulations.
We are grateful to C. Baus, T. Pierog and R. Ulrich for providing the
\textsc{crmc} program codes and useful comments.
This work has been partly supported by a Grant-in-Aid for Scientific research by
MEXT of Japan, a Grant-in-Aid for a JSPS Postdoctoral Fellowship for Research
Abroad, a Grant-in-Aid for Nagoya University GCOE ``QFPU'' from MEXT, and
Istituto Nazionale di Fisica Nucleare (INFN) in Italy.
Part of this work was performed using the computer resources provided by the
Institute for the Cosmic-Ray Research (University of Tokyo), CERN, and CNAF
(INFN).

\end{acknowledgements}

\section*{Appendix: Data tables}\label{sec:appendix}

The inclusive production rates of $\pizero$s measured by LHCf with all
corrections applied are summarized in
Tables~\ref{table:pttable_0}--~\ref{table:pztable_4}.
The ratios of $\pizero$ production rate of MC simulation to data are summarized
in Tables~\ref{table:ratio_7TeV_pt_0}--~\ref{table:ratio_5TeV_pz_4}.
The LHCf $\pt$ distributions for $\pPb$ collisions have the UPC component
subtracted.
The nuclear modification factor of $\pizero$s obtained from LHCf data are
summarized in Tables~\ref{table:nmftable_1}--~\ref{table:nmftable_4}.


\clearpage
\begin{table*}[htbp]
	\caption{Production rate for the $\pizero$ production in the rapidity range $8.8
	< y < 9.0$ in $\pp$ collisions and in the rapidity range $-8.8 > \ylab > -9.0$
	in $\pPb$ collisions. The rate and corresponding total uncertainty are in units
	of [GeV$^{-2}$]. \label{table:pttable_0}}
	\begin{ruledtabular}
\begin{tabular}{l c c c c c c}
	\multicolumn{1}{l}{$\pt$ range $[$GeV$]$}       &
	\multicolumn{2}{c}{$\pp$ at $\sqs=\tev{2.76}$}  &
	\multicolumn{2}{c}{$\pp$ at $\sqs=\tev{7}$}     &
	\multicolumn{2}{c}{$\pPb$ at $\snn=\tev{5.02}$} \\
	\cline{2-3} \cline{4-5} \cline{6-7}
     & Production rate & Total error
     & Production rate & Total error
     & Production rate & Total error \\
     \hline
$[$0.10,0.15$]$ & 3.71\etwo & $\pm$8.99\ethr & 2.56\eone & $\pm$3.75\etwo & 1.63\eone & $\pm$5.65\etwo \\
$[$0.15,0.20$]$ & 1.92\etwo & $\pm$3.36\ethr & 1.98\eone & $\pm$2.05\etwo & 1.86\eone & $\pm$4.19\etwo \\
$[$0.20,0.25$]$ & 3.47\ethr & $\pm$7.22\efou & 1.20\eone & $\pm$1.29\etwo & 1.52\eone & $\pm$2.76\etwo \\
$[$0.25,0.30$]$ & 5.67\efou & $\pm$2.49\efou & 6.96\etwo & $\pm$5.65\ethr & 1.06\eone & $\pm$1.52\etwo \\
$[$0.30,0.35$]$ &           &                & 3.94\etwo & $\pm$3.43\ethr & 5.59\etwo & $\pm$8.43\ethr \\
$[$0.35,0.40$]$ &           &                & 2.23\etwo & $\pm$2.28\ethr & 3.80\etwo & $\pm$6.14\ethr \\
\end{tabular}
	\end{ruledtabular}
\end{table*}

\begin{table*}[htbp]
	\caption{Production rate for the $\pizero$ production in the rapidity range $9.0
	< y < 9.2$ in $\pp$ collisions and in the rapidity range $-9.0 > \ylab > -9.2$
	in $\pPb$ collisions. The rate and corresponding total uncertainty are in units
	of [GeV$^{-2}$]. \label{table:pttable_1}}
	\begin{ruledtabular}
    \begin{tabular}{l c c c c c c}
	\multicolumn{1}{l}{$\pt$ range $[$GeV$]$}       &
	\multicolumn{2}{c}{$\pp$ at $\sqs=\tev{2.76}$}  &
	\multicolumn{2}{c}{$\pp$ at $\sqs=\tev{7}$}     &
	\multicolumn{2}{c}{$\pPb$ at $\snn=\tev{5.02}$} \\
	\cline{2-3} \cline{4-5} \cline{6-7}
     & Production rate & Total error
     & Production rate & Total error
     & Production rate & Total error \\
     \hline
$[$0.05,0.10$]$ & 4.26\etwo & $\pm$1.08\etwo & 2.98\eone & $\pm$4.72\etwo & 1.66\eone & $\pm$5.08\etwo \\
$[$0.10,0.15$]$ & 2.18\etwo & $\pm$3.74\ethr & 2.27\eone & $\pm$2.58\etwo & 1.78\eone & $\pm$3.66\etwo \\
$[$0.15,0.20$]$ & 4.20\ethr & $\pm$7.14\efou & 1.42\eone & $\pm$1.12\etwo & 1.69\eone & $\pm$2.97\etwo \\
$[$0.20,0.25$]$ & 2.89\efou & $\pm$1.59\efou & 8.48\etwo & $\pm$6.20\ethr & 1.14\eone & $\pm$2.10\etwo \\
$[$0.25,0.30$]$ &           &                & 5.50\etwo & $\pm$4.43\ethr & 5.31\etwo & $\pm$8.70\ethr \\
$[$0.30,0.35$]$ &           &                & 3.28\etwo & $\pm$3.02\ethr & 3.46\etwo & $\pm$5.49\ethr \\
\end{tabular}
	\end{ruledtabular}
\end{table*}

\begin{table*}[htbp]
	\caption{Production rate for the $\pizero$ production in the rapidity range $9.2
	< y < 9.4$ in $\pp$ collisions and in the rapidity range $-9.2 > \ylab > -9.4$
	in $\pPb$ collisions. The rate and corresponding total uncertainty are in units
	of [GeV$^{-2}$]. \label{table:pttable_2}}
	\begin{ruledtabular}
    \begin{tabular}{l c c c c c c}
	\multicolumn{1}{l}{$\pt$ range $[$GeV$]$}       &
	\multicolumn{2}{c}{$\pp$ at $\sqs=\tev{2.76}$}  &
	\multicolumn{2}{c}{$\pp$ at $\sqs=\tev{7}$}     &
	\multicolumn{2}{c}{$\pPb$ at $\snn=\tev{5.02}$} \\
	\cline{2-3} \cline{4-5} \cline{6-7}
     & Production rate & Total error
     & Production rate & Total error
     & Production rate & Total error \\
     \hline
$[$0.00,0.05$]$ & 4.46\etwo & $\pm$1.33\etwo & 2.74\eone & $\pm$5.08\etwo & 1.46\eone & $\pm$5.35\etwo \\
$[$0.05,0.10$]$ & 2.07\etwo & $\pm$4.12\ethr & 1.96\eone & $\pm$3.22\etwo & 1.71\eone & $\pm$3.98\etwo \\
$[$0.10,0.15$]$ & 7.15\ethr & $\pm$1.12\ethr & 1.54\eone & $\pm$1.76\etwo & 1.52\eone & $\pm$2.67\etwo \\
$[$0.15,0.20$]$ & 8.30\efou & $\pm$2.24\efou & 1.01\eone & $\pm$7.24\ethr & 1.39\eone & $\pm$2.42\etwo \\
$[$0.20,0.25$]$ & 4.11\efiv & $\pm$4.14\efiv & 5.53\etwo & $\pm$4.20\ethr & 6.48\etwo & $\pm$1.15\etwo \\
$[$0.25,0.30$]$ &           &                & 3.26\etwo & $\pm$2.77\ethr & 2.80\etwo & $\pm$4.62\ethr \\
$[$0.30,0.35$]$ &           &                & 1.59\etwo & $\pm$1.53\ethr & 1.57\etwo & $\pm$2.87\ethr \\
$[$0.35,0.40$]$ &           &                & 7.26\ethr & $\pm$9.04\efou & 8.06\ethr & $\pm$2.12\ethr \\
$[$0.40,0.45$]$ &           &                & 1.72\ethr & $\pm$3.83\efou & 4.43\ethr & $\pm$1.46\ethr \\
$[$0.45,0.50$]$ &           &                & 4.12\efou & $\pm$1.78\efou & 1.94\ethr & $\pm$9.94\efou \\
$[$0.50,0.55$]$ &           &                & 9.38\efiv & $\pm$9.38\efiv & 1.48\efou & $\pm$3.75\efou \\
$[$0.55,0.60$]$ &           &                & 9.53\efiv & $\pm$8.49\efiv & 1.28\efou & $\pm$2.08\efou \\
    \end{tabular}
	\end{ruledtabular}
\end{table*}

\begin{table*}[htbp]
	\caption{Production rate for the $\pizero$ production in the rapidity range $9.4
	< y < 9.6$ in $\pp$ collisions and in the rapidity range $-9.4 > \ylab > -9.6$
	in $\pPb$ collisions. The rate and corresponding total uncertainty are in units of
	[GeV$^{-2}$]. \label{table:pttable_3}}
	\begin{ruledtabular}
    \begin{tabular}{l c c c c c c}
	\multicolumn{1}{l}{$\pt$ range $[$GeV$]$}       &
	\multicolumn{2}{c}{$\pp$ at $\sqs=\tev{2.76}$}  &
	\multicolumn{2}{c}{$\pp$ at $\sqs=\tev{7}$}     &
	\multicolumn{2}{c}{$\pPb$ at $\snn=\tev{5.02}$} \\
	\cline{2-3} \cline{4-5} \cline{6-7}
     & Production rate & Total error
     & Production rate & Total error
     & Production rate & Total error \\
     \hline
$[$0.00,0.05$]$ & 1.71\etwo & $\pm$4.08\ethr & 1.95\eone & $\pm$3.27\etwo & 1.39\eone & $\pm$3.89\etwo \\
$[$0.05,0.10$]$ & 6.51\ethr & $\pm$1.29\ethr & 1.52\eone & $\pm$2.23\etwo & 1.36\eone & $\pm$2.90\etwo \\
$[$0.10,0.15$]$ & 7.27\efou & $\pm$1.97\efou & 1.08\eone & $\pm$8.30\ethr & 1.17\eone & $\pm$2.22\etwo \\
$[$0.15,0.20$]$ & 3.84\efiv & $\pm$3.87\efiv & 6.08\etwo & $\pm$4.41\ethr & 7.42\etwo & $\pm$1.35\etwo \\
$[$0.20,0.25$]$ &           &                & 3.37\etwo & $\pm$2.85\ethr & 3.08\etwo & $\pm$4.97\ethr \\
$[$0.25,0.30$]$ &           &                & 1.49\etwo & $\pm$1.49\ethr & 1.55\etwo & $\pm$2.91\ethr \\
$[$0.30,0.35$]$ &           &                & 5.36\ethr & $\pm$7.82\efou & 1.20\etwo & $\pm$2.53\ethr \\
$[$0.35,0.40$]$ &           &                & 1.60\ethr & $\pm$4.36\efou & 4.64\ethr & $\pm$1.83\ethr \\
$[$0.40,0.45$]$ &           &                & 2.72\efou & $\pm$1.45\efou & 9.01\efou & $\pm$7.91\efou \\
    \end{tabular}
	\end{ruledtabular}
\end{table*}

\begin{table*}[htbp]
	\caption{Production rate for the $\pizero$ production in the rapidity range $9.6
	< y < 9.8$ in $\pp$ collisions and in the rapidity range $-9.6 > \ylab > -9.8$
	in $\pPb$ collisions. The rate and corresponding total uncertainty are in units of
	[GeV$^{-2}$]. \label{table:pttable_4}}
	\begin{ruledtabular}
    \begin{tabular}{l c c c c c c}
	\multicolumn{1}{l}{$\pt$ range $[$GeV$]$}       &
	\multicolumn{2}{c}{$\pp$ at $\sqs=\tev{2.76}$}  &
	\multicolumn{2}{c}{$\pp$ at $\sqs=\tev{7}$}     &
	\multicolumn{2}{c}{$\pPb$ at $\snn=\tev{5.02}$} \\
	\cline{2-3} \cline{4-5} \cline{6-7}
     & Production rate & Total error
     & Production rate & Total error
     & Production rate & Total error \\
     \hline
$[$0.00,0.05$]$ & 3.56\ethr & $\pm$1.06\ethr & 1.63\eone & $\pm$1.98\etwo & 1.17\eone & $\pm$3.27\etwo \\
$[$0.05,0.10$]$ & 7.59\efou & $\pm$2.56\efou & 1.07\eone & $\pm$1.07\etwo & 1.09\eone & $\pm$2.36\etwo \\
$[$0.10,0.15$]$ & 4.71\efiv & $\pm$4.75\efiv & 6.40\etwo & $\pm$4.56\ethr & 6.56\etwo & $\pm$1.41\etwo \\
$[$0.15,0.20$]$ &           &                & 3.51\etwo & $\pm$2.77\ethr & 3.03\etwo & $\pm$5.17\ethr \\
$[$0.20,0.25$]$ &           &                & 1.54\etwo & $\pm$1.65\ethr & 1.42\etwo & $\pm$2.67\ethr \\
$[$0.25,0.30$]$ &           &                & 3.27\ethr & $\pm$6.32\efou & 6.67\ethr & $\pm$1.82\ethr \\
$[$0.30,0.35$]$ &           &                & 9.33\efou & $\pm$3.58\efou & 1.22\ethr & $\pm$8.18\efou \\
$[$0.35,0.40$]$ &           &                & 1.69\efou & $\pm$1.30\efou & $<1.73$\efou &             \\
    \end{tabular}
	\end{ruledtabular}
\end{table*}

\begin{table*}[htbp]
	\caption{Production rate for the $\pizero$ production in the rapidity range $9.8
	< y < 10.0$ in $\pp$ collisions and in the rapidity range $-9.8 > \ylab > -10.0$
	in $\pPb$ collisions. The rate and corresponding total uncertainty are in units of
	[GeV$^{-2}$]. \label{table:pttable_5}}
	\begin{ruledtabular}
    \begin{tabular}{l c c c c c c}
	\multicolumn{1}{l}{$\pt$ range $[$GeV$]$}       &
	\multicolumn{2}{c}{$\pp$ at $\sqs=\tev{2.76}$}  &
	\multicolumn{2}{c}{$\pp$ at $\sqs=\tev{7}$}     &
	\multicolumn{2}{c}{$\pPb$ at $\snn=\tev{5.02}$} \\
	\cline{2-3} \cline{4-5} \cline{6-7}
     & Production rate & Total error
     & Production rate & Total error
     & Production rate & Total error \\
     \hline
$[$0.00,0.05$]$ &           &                & 8.80\etwo & $\pm$1.05\etwo & 9.81\etwo & $\pm$3.22\etwo \\
$[$0.05,0.10$]$ &           &                & 5.90\etwo & $\pm$5.65\ethr & 6.41\etwo & $\pm$1.61\etwo \\
$[$0.10,0.15$]$ &           &                & 3.45\etwo & $\pm$2.73\ethr & 2.42\etwo & $\pm$5.51\ethr \\
$[$0.15,0.20$]$ &           &                & 1.45\etwo & $\pm$1.37\ethr & 1.29\etwo & $\pm$2.70\ethr \\
$[$0.20,0.25$]$ &           &                & 3.07\ethr & $\pm$5.82\efou & 7.63\ethr & $\pm$1.84\ethr \\
$[$0.25,0.30$]$ &           &                & 5.83\efou & $\pm$2.37\efou & 1.01\ethr & $\pm$7.40\efou \\
    \end{tabular}
	\end{ruledtabular}
\end{table*}

\begin{table*}[htbp]
	\caption{Production rate for the $\pizero$ production in the rapidity range
	$10.0 < y < 10.2$ in $\pp$ collisions and in the rapidity range $-10.0 > \ylab > -10.2$
	in $\pPb$ collisions. The rate and corresponding total uncertainty are in units of
	[GeV$^{-2}$]. \label{table:pttable_6}}
	\begin{ruledtabular}
    \begin{tabular}{l c c c c c c}
	\multicolumn{1}{l}{$\pt$ range $[$GeV$]$}       &
	\multicolumn{2}{c}{$\pp$ at $\sqs=\tev{2.76}$}  &
	\multicolumn{2}{c}{$\pp$ at $\sqs=\tev{7}$}     &
	\multicolumn{2}{c}{$\pPb$ at $\snn=\tev{5.02}$} \\
	\cline{2-3} \cline{4-5} \cline{6-7}
     & Production rate & Total error
     & Production rate & Total error
     & Production rate & Total error \\
     \hline
$[$0.00,0.05$]$ &           &                & 4.54\etwo & $\pm$5.63\ethr & 4.38\etwo & $\pm$1.80\etwo \\
$[$0.05,0.10$]$ &           &                & 3.06\etwo & $\pm$2.95\ethr & 2.51\etwo & $\pm$6.11\ethr \\
$[$0.10,0.15$]$ &           &                & 1.24\etwo & $\pm$1.12\ethr & 7.25\ethr & $\pm$2.38\ethr \\
$[$0.15,0.20$]$ &           &                & 2.04\ethr & $\pm$3.16\efou & 7.17\ethr & $\pm$1.82\ethr \\
$[$0.20,0.25$]$ &           &                & 5.33\efou & $\pm$2.41\efou & 3.00\efou & $\pm$5.46\efou \\
    \end{tabular}
	\end{ruledtabular}
\end{table*}

\begin{table*}[htbp]
	\caption{Production rate for the $\pizero$ production in the rapidity range
	$10.2 < y < 10.4$ in $\pp$ collisions and in the rapidity range $-10.2 > \ylab > -10.4$
	in $\pPb$ collisions. The rate and corresponding total uncertainty are in units of
	[GeV$^{-2}$]. \label{table:pttable_7}}
	\begin{ruledtabular}
    \begin{tabular}{l c c c c c c}
	\multicolumn{1}{l}{$\pt$ range $[$GeV$]$}       &
	\multicolumn{2}{c}{$\pp$ at $\sqs=\tev{2.76}$}  &
	\multicolumn{2}{c}{$\pp$ at $\sqs=\tev{7}$}     &
	\multicolumn{2}{c}{$\pPb$ at $\snn=\tev{5.02}$} \\
	\cline{2-3} \cline{4-5} \cline{6-7}
     & Production rate & Total error
     & Production rate & Total error
     & Production rate & Total error \\
     \hline
$[$0.00,0.05$]$ &           &                & 2.08\etwo & $\pm$2.31\ethr & 6.38\ethr & $\pm$6.42\ethr \\
$[$0.05,0.10$]$ &           &                & 1.10\etwo & $\pm$1.00\ethr & 9.20\ethr & $\pm$1.91\ethr \\
$[$0.10,0.15$]$ &           &                & 2.22\ethr & $\pm$2.16\efou & 5.33\ethr & $\pm$8.77\efou \\
$[$0.15,0.20$]$ &           &                & 8.39\efiv & $\pm$2.03\efiv & 4.87\efou & $\pm$1.64\efou \\
    \end{tabular}
	\end{ruledtabular}
\end{table*}

\begin{table*}[htbp]
	\caption{Production rate for the $\pizero$ production in the rapidity range
	$10.4 < y < 10.6$ in $\pp$ collisions and in the rapidity range $-10.4 > \ylab > -10.6$
	in $\pPb$ collisions. The rate and corresponding total uncertainty are in units of
	[GeV$^{-2}$]. \label{table:pttable_8}}
	\begin{ruledtabular}
    \begin{tabular}{l c c c c c c}
	\multicolumn{1}{l}{$\pt$ range $[$GeV$]$}       &
	\multicolumn{2}{c}{$\pp$ at $\sqs=\tev{2.76}$}  &
	\multicolumn{2}{c}{$\pp$ at $\sqs=\tev{7}$}     &
	\multicolumn{2}{c}{$\pPb$ at $\snn=\tev{5.02}$} \\
	\cline{2-3} \cline{4-5} \cline{6-7}
     & Production rate & Total error
     & Production rate & Total error
     & Production rate & Total error \\
     \hline
$[$0.00,0.05$]$ &           &                & 5.14\ethr & $\pm$7.21\efou & 8.01\ethr & $\pm$4.52\ethr \\
$[$0.05,0.10$]$ &           &                & 1.87\ethr & $\pm$2.11\efou & 1.78\ethr & $\pm$1.39\ethr \\
$[$0.10,0.15$]$ &           &                & 1.04\efou & $\pm$2.51\efiv & 1.04\efou & $\pm$3.45\efou \\
$[$0.15,0.20$]$ &           &                & 2.09\esix & $\pm$1.37\esix & 2.82\efou & $\pm$1.51\efou \\
    \end{tabular}
	\end{ruledtabular}
\end{table*}

\begin{table*}[htbp]
	\caption{Production rate for the $\pizero$ production in the rapidity range
	$10.6 < y < 10.8$ in $\pp$ collisions and in the rapidity range $-10.6 > \ylab > -10.8$
	in $\pPb$ collisions. The rate and corresponding total uncertainty are in units of
	[GeV$^{-2}$]. \label{table:pttable_9}}
	\begin{ruledtabular}
    \begin{tabular}{l c c c c c c}
	\multicolumn{1}{l}{$\pt$ range $[$GeV$]$}       &
	\multicolumn{2}{c}{$\pp$ at $\sqs=\tev{2.76}$}  &
	\multicolumn{2}{c}{$\pp$ at $\sqs=\tev{7}$}     &
	\multicolumn{2}{c}{$\pPb$ at $\snn=\tev{5.02}$} \\
	\cline{2-3} \cline{4-5} \cline{6-7}
     & Production rate & Total error
     & Production rate & Total error
     & Production rate & Total error \\
     \hline
$[$0.00,0.05$]$ &           &                & 6.81\efou & $\pm$1.68\efou & 1.17\ethr & $\pm$1.18\ethr \\
$[$0.05,0.10$]$ &           &                & 1.42\efou & $\pm$3.84\efiv & $<9.69$\efiv &             \\
$[$0.10,0.15$]$ &           &                & 2.81\esix & $\pm$1.57\esix & 2.85\efiv & $\pm$2.57\efiv \\
    \end{tabular}
	\end{ruledtabular}
\end{table*}


\clearpage
\begin{table*}[htbp]
	\caption{Production rate for the $\pizero$ production in the $\pt$ range $0.0 <
	\pt < \gev{0.2}$ in $\pp$ and $\pPb$ collisions. The rate and corresponding
	total uncertainty are in units of [GeV$^{-2}$].
	\label{table:pztable_0}}
	\begin{ruledtabular}
    \begin{tabular}{l c c c c c c}
	\multicolumn{1}{l}{$\pz$ range $[$GeV$]$}       &
	\multicolumn{2}{c}{$\pp$ at $\sqs=\tev{2.76}$}  &
	\multicolumn{2}{c}{$\pp$ at $\sqs=\tev{7}$}     &
	\multicolumn{2}{c}{$\pPb$ at $\snn=\tev{5.02}$} \\
	\cline{2-3} \cline{4-5} \cline{6-7}
     & Production rate & Total error
     & Production rate & Total error
     & Production rate & Total error \\
     \hline
$[$\hph625,\hph750$]$       & 4.55\etwo & $\pm$1.85\etwo &           &                &           &          \\
$[$\hph750,\hph875$]$       & 2.59\etwo & $\pm$5.79\ethr &           &                &           &          \\
$[$\hph875,1000$]$          & 1.19\etwo & $\pm$1.85\ethr & 1.82\eone & $\pm$2.21\etwo & 1.59\eone & $\pm$3.25\etwo \\
$[$1000,1125$]$             & 3.83\ethr & $\pm$5.87\efou & 1.52\eone & $\pm$1.47\etwo & 1.48\eone & $\pm$2.50\etwo \\
$[$1125,1250$]$             & 9.35\efou & $\pm$2.19\efou & 1.39\eone & $\pm$9.38\ethr & 1.48\eone & $\pm$2.41\etwo \\
$[$1250,1380$]$             & 1.16\efou & $\pm$7.08\efiv & 9.97\etwo & $\pm$6.89\ethr & 1.21\eone & $\pm$2.18\etwo \\
\end{tabular}
	\end{ruledtabular}
\end{table*}

\begin{table*}[htbp]
	\caption{Production rate for the $\pizero$ production in the $\pt$ range $0.2 <
	\pt < \gev{0.4}$ in $\pp$ and $\pPb$ collisions. The rate and corresponding
	total uncertainty are in units of [GeV$^{-2}$].
	\label{table:pztable_1}}
	\begin{ruledtabular}
    \begin{tabular}{l c c c c c c}
	\multicolumn{1}{l}{$\pz$ range $[$GeV$]$}       &
	\multicolumn{2}{c}{$\pp$ at $\sqs=\tev{2.76}$}  &
	\multicolumn{2}{c}{$\pp$ at $\sqs=\tev{7}$}     &
	\multicolumn{2}{c}{$\pPb$ at $\snn=\tev{5.02}$} \\
	\cline{2-3} \cline{4-5} \cline{6-7}
     & Production rate & Total error
     & Production rate & Total error
     & Production rate & Total error \\
     \hline
$[$\hph750,\hph875$]$ & 1.16\etwo & $\pm$3.73\ethr &           &                &           &                \\
$[$\hph875,1000$]$    & 4.09\ethr & $\pm$9.00\efou &           &                & 1.26\eone & $\pm$2.61\etwo \\
$[$1000,1125$]$       & 1.21\ethr & $\pm$2.89\efou & 8.47\etwo & $\pm$8.41\ethr & 1.25\eone & $\pm$1.78\etwo \\
$[$1125,1250$]$       & 5.55\efiv & $\pm$5.59\efiv & 7.00\etwo & $\pm$7.22\ethr & 9.80\etwo & $\pm$1.44\etwo \\
$[$1250,1380$]$       & 3.38\efiv & $\pm$3.41\efiv & 5.58\etwo & $\pm$4.31\ethr & 7.55\etwo & $\pm$1.15\etwo \\
$[$1380,1500$]$       &           &                & 4.55\etwo & $\pm$3.76\ethr & 4.95\etwo & $\pm$8.07\ethr \\
\end{tabular}
	\end{ruledtabular}
\end{table*}

\begin{table*}[htbp]
	\caption{Production rate for the $\pizero$ production in the $\pt$ range $0.4 <
	\pt < \gev{0.6}$ in $\pp$ and $\pPb$ collisions. The rate and corresponding
	total uncertainty are in units of [GeV$^{-2}$].
	\label{table:pztable_2}}
	\begin{ruledtabular}
    \begin{tabular}{l c c c c c c}
	\multicolumn{1}{l}{$\pz$ range $[$GeV$]$}       &
	\multicolumn{2}{c}{$\pp$ at $\sqs=\tev{2.76}$}  &
	\multicolumn{2}{c}{$\pp$ at $\sqs=\tev{7}$}     &
	\multicolumn{2}{c}{$\pPb$ at $\snn=\tev{5.02}$} \\
	\cline{2-3} \cline{4-5} \cline{6-7}
     & Production rate & Total error
     & Production rate & Total error
     & Production rate & Total error \\
     \hline
$[$1380,1500$]$ &            &          &           &                & 2.68\etwo & $\pm$5.38\ethr \\
$[$1500,1625$]$ &            &          & 1.15\etwo & $\pm$3.18\ethr & 1.45\etwo & $\pm$3.06\ethr \\
$[$1625,1750$]$ &            &          & 1.09\etwo & $\pm$2.06\ethr & 1.49\etwo & $\pm$2.48\ethr \\
$[$1750,1875$]$ &            &          & 9.86\ethr & $\pm$9.93\efou & 1.16\etwo & $\pm$1.82\ethr \\
$[$1875,2000$]$ &            &          & 6.78\ethr & $\pm$6.85\efou & 8.25\ethr & $\pm$1.27\ethr \\
$[$2000,2125$]$ &            &          & 4.66\ethr & $\pm$3.96\efou & 6.36\ethr & $\pm$1.05\ethr \\
$[$2125,2250$]$ &            &          & 4.08\ethr & $\pm$3.58\efou & 4.98\ethr & $\pm$9.15\efou \\
$[$2250,2375$]$ &            &          & 2.44\ethr & $\pm$2.38\efou & 4.15\ethr & $\pm$8.31\efou \\
$[$2375,2500$]$ &            &          & 1.73\ethr & $\pm$1.89\efou & 1.69\ethr & $\pm$6.44\efou \\
$[$2500,2625$]$ &            &          & 1.12\ethr & $\pm$1.86\efou & 2.43\ethr & $\pm$7.14\efou \\
$[$2625,2750$]$ &            &          & 5.57\efou & $\pm$1.47\efou & 1.24\ethr & $\pm$6.12\efou \\
$[$2750,2875$]$ &            &          & 2.93\efou & $\pm$1.07\efou & 5.21\efou & $\pm$4.99\efou \\
$[$2875,3000$]$ &            &          & 2.36\efou & $\pm$9.47\efiv & 1.37\ethr & $\pm$6.40\efou \\
    \end{tabular}
	\end{ruledtabular}
\end{table*}

\begin{table*}[htbp]
	\caption{Production rate for the $\pizero$ production in the $\pt$ range $0.6 <
	\pt < \gev{0.8}$ in $\pp$ and $\pPb$ collisions. The rate and corresponding
	total uncertainty are in units of [GeV$^{-2}$].
	\label{table:pztable_3}}
	\begin{ruledtabular}
    \begin{tabular}{l c c c c c c}
	\multicolumn{1}{l}{$\pz$ range $[$GeV$]$}       &
	\multicolumn{2}{c}{$\pp$ at $\sqs=\tev{2.76}$}  &
	\multicolumn{2}{c}{$\pp$ at $\sqs=\tev{7}$}     &
	\multicolumn{2}{c}{$\pPb$ at $\snn=\tev{5.02}$} \\
	\cline{2-3} \cline{4-5} \cline{6-7}
     & Production rate & Total error
     & Production rate & Total error
     & Production rate & Total error \\
     \hline
$[$1750,1875$]$ &            &          &           &                & 4.14\ethr & $\pm$1.08\ethr \\
$[$1875,2000$]$ &            &          &           &                & 3.80\ethr & $\pm$6.61\efou \\
$[$2000,2125$]$ &            &          & 2.47\ethr & $\pm$3.18\efou & 3.81\ethr & $\pm$5.91\efou \\
$[$2125,2250$]$ &            &          & 1.56\ethr & $\pm$2.03\efou & 2.24\ethr & $\pm$3.91\efou \\
$[$2250,2375$]$ &            &          & 1.26\ethr & $\pm$1.57\efou & 1.89\ethr & $\pm$3.19\efou \\
$[$2375,2500$]$ &            &          & 7.10\efou & $\pm$6.10\efiv & 1.47\ethr & $\pm$2.64\efou \\
$[$2500,2625$]$ &            &          & 5.30\efou & $\pm$4.94\efiv & 9.45\efou & $\pm$2.05\efou \\
$[$2625,2750$]$ &            &          & 3.82\efou & $\pm$4.62\efiv & 4.17\efou & $\pm$1.47\efou \\
$[$2750,2875$]$ &            &          & 2.55\efou & $\pm$4.86\efiv & 4.56\efou & $\pm$1.35\efou \\
$[$2875,3000$]$ &            &          & 1.20\efou & $\pm$3.30\efiv & 8.61\efiv & $\pm$9.66\efiv \\
$[$3000,3125$]$ &            &          & 5.73\efiv & $\pm$2.27\efiv & $<3.48$\efiv &             \\
$[$3125,3250$]$ &            &          & 3.21\efiv & $\pm$1.74\efiv & 2.74\efou & $\pm$1.24\efou \\
$[$3250,3375$]$ &            &          & 3.16\efiv & $\pm$1.79\efiv & 1.15\efou & $\pm$8.87\efiv \\
$[$3375,3500$]$ &            &          &           &                & 2.97\efiv & $\pm$5.38\efiv \\
    \end{tabular}
	\end{ruledtabular}
\end{table*}

\begin{table*}[htbp]
	\caption{Production rate for the $\pizero$ production in the $\pt$ range $0.8 <
	\pt < \gev{1.0}$ in $\pp$ and $\pPb$ collisions. The rate and corresponding
	total uncertainty are in units of [GeV$^{-2}$].
	\label{table:pztable_4}}
	\begin{ruledtabular}
    \begin{tabular}{l c c c c c c}
	\multicolumn{1}{l}{$\pz$ range $[$GeV$]$}       &
	\multicolumn{2}{c}{$\pp$ at $\sqs=\tev{2.76}$}  &
	\multicolumn{2}{c}{$\pp$ at $\sqs=\tev{7}$}     &
	\multicolumn{2}{c}{$\pPb$ at $\snn=\tev{5.02}$} \\
	\cline{2-3} \cline{4-5} \cline{6-7}
     & Production rate & Total error
     & Production rate & Total error
     & Production rate & Total error \\
     \hline
$[$2250,2375$]$ &            &          & 5.33\efou & $\pm$3.02\efou &           &                \\
$[$2375,2500$]$ &            &          & 3.19\efou & $\pm$1.41\efou & 5.46\efou & $\pm$2.57\efou \\
$[$2500,2625$]$ &            &          & 2.08\efou & $\pm$7.78\efiv & 4.85\efou & $\pm$1.37\efou \\
$[$2625,2750$]$ &            &          & 1.19\efou & $\pm$4.19\efiv & 2.69\efou & $\pm$1.04\efou \\
$[$2750,2875$]$ &            &          & 1.11\efou & $\pm$3.70\efiv & 2.70\efou & $\pm$9.97\efiv \\
$[$2875,3000$]$ &            &          & 3.35\efiv & $\pm$1.59\efiv & 1.84\efou & $\pm$7.34\efiv \\
$[$3000,3125$]$ &            &          & 2.98\efiv & $\pm$1.29\efiv & 6.77\efiv & $\pm$4.60\efiv \\
$[$3125,3250$]$ &            &          & 5.22\esix & $\pm$3.45\esix & $<1.10$\efiv &             \\
$[$3250,3375$]$ &            &          & 3.32\esix & $\pm$2.53\esix &           &                \\
$[$3375,3500$]$ &            &          &           &                & $<1.23$\efiv &             \\
    \end{tabular}
	\end{ruledtabular}
\end{table*}


\clearpage
\begin{table}[htbp]
    \caption{Ratio of $\pizero$ production rate of MC simulation to data in the
    rapidity range $8.8<y<9.0$ in $\pp$ collisions at $\sqs=\tev{7}$.}
    \label{table:ratio_7TeV_pt_0}
	\begin{ruledtabular}
    \begin{tabular}{c c c c c c}
    $\pt$ range & \textsc{dpmjet} & \textsc{qgsjet} & \textsc{sibyll} & \textsc{epos} & \textsc{pythia} \\
	$[$GeV$]$   & 3.06            & II-04           & 2.1             & \textsc{lhc}  & 8.185           \\
    \hline
$[$0.10, 0.15$]$ &  2.38 &  2.09 &  1.26 &  1.89 &  2.34 \\
$[$0.15, 0.20$]$ &  1.83 &  1.43 &  0.93 &  1.43 &  1.79 \\
$[$0.20, 0.25$]$ &  1.43 &  0.97 &  0.71 &  1.02 &  1.40 \\
$[$0.25, 0.30$]$ &  1.64 &  0.94 &  0.80 &  1.15 &  1.60 \\
$[$0.30, 0.35$]$ &  1.94 &  0.97 &  0.96 &  1.40 &  1.85 \\
$[$0.35, 0.40$]$ &  2.31 &  1.02 &  1.19 &  1.45 &  2.15 \\
$[$0.40, 0.45$]$ &  2.81 &  1.03 &  1.49 &  1.43 &  2.52 \\
$[$0.45, 0.50$]$ &  3.44 &  0.94 &  1.83 &  1.47 &  2.93 \\
$[$0.50, 0.55$]$ &  3.91 &  0.78 &  2.08 &  1.48 &  3.19 \\
$[$0.55, 0.60$]$ &  4.72 &  0.67 &  2.52 &  1.66 &  3.70 \\
$[$0.60, 0.65$]$ &  6.24 &  0.65 &  3.28 &  1.88 &  4.60 \\
$[$0.65, 0.70$]$ & 10.40 &  0.81 &  5.33 &  2.77 &  7.30 \\
$[$0.70, 0.75$]$ & 12.58 &  0.73 &  6.29 &  3.14 &  8.37 \\
$[$0.75, 0.80$]$ & 10.66 &  0.46 &  4.96 &  2.95 &  6.69 \\
$[$0.80, 0.85$]$ & 17.83 &  0.45 &  7.65 &  5.60 & 11.19 \\
$[$0.85, 0.90$]$ & 62.69 &  1.50 & 21.76 & 27.76 & 35.11 \\
$[$0.90, 0.95$]$ & 11.44 &  0.24 &  3.64 &  8.26 &  7.36 \\
$[$0.95, 1.00$]$ & 16.38 &  0.20 &  4.59 & 26.72 & 16.18 \\
    \end{tabular}
	\end{ruledtabular}
\end{table}

\begin{table}[htbp]
    \caption{Ratio of $\pizero$ production rate of MC simulation to data in the
    rapidity range $9.0<y<9.2$ in $\pp$ collisions at $\sqs=\tev{7}$.}
    \label{table:ratio_7TeV_pt_1}
	\begin{ruledtabular}
    \begin{tabular}{c c c c c c}
    $\pt$ range & \textsc{dpmjet} & \textsc{qgsjet} & \textsc{sibyll} & \textsc{epos} & \textsc{pythia} \\
	$[$GeV$]$   & 3.06            & II-04           & 2.1             & \textsc{lhc}  & 8.185           \\
    \hline
$[$0.05, 0.10$]$ &  1.96 &  1.95 &  1.00 &  1.56 &  1.92 \\
$[$0.10, 0.15$]$ &  1.45 &  1.27 &  0.69 &  1.14 &  1.40 \\
$[$0.15, 0.20$]$ &  1.27 &  0.96 &  0.58 &  0.93 &  1.21 \\
$[$0.20, 0.25$]$ &  1.37 &  0.87 &  0.62 &  0.90 &  1.31 \\
$[$0.25, 0.30$]$ &  1.39 &  0.73 &  0.62 &  1.00 &  1.28 \\
$[$0.30, 0.35$]$ &  1.55 &  0.70 &  0.69 &  1.00 &  1.36 \\
$[$0.35, 0.40$]$ &  2.20 &  0.80 &  0.99 &  1.07 &  1.86 \\
$[$0.40, 0.45$]$ &  3.31 &  0.80 &  1.41 &  1.21 &  2.59 \\
$[$0.45, 0.50$]$ &  4.47 &  0.68 &  1.78 &  1.42 &  3.23 \\
$[$0.50, 0.55$]$ &  6.99 &  0.67 &  2.60 &  1.90 &  4.56 \\
$[$0.55, 0.60$]$ &  7.84 &  0.51 &  2.68 &  1.77 &  4.50 \\
$[$0.60, 0.65$]$ & 15.60 &  0.66 &  4.77 &  3.27 &  8.00 \\
$[$0.65, 0.70$]$ & 14.14 &  0.46 &  3.70 &  3.58 &  6.31 \\
$[$0.70, 0.75$]$ & 36.46 &  0.79 &  8.05 & 14.00 & 13.20 \\
$[$0.75, 0.80$]$ & 18.03 &  0.33 &  2.97 & 11.88 &  8.13 \\    
    \end{tabular}
	\end{ruledtabular}
\end{table}

\begin{table}[htbp]
    \caption{Ratio of $\pizero$ production rate of MC simulation to data in the
    rapidity range $9.2<y<9.4$ in $\pp$ collisions at $\sqs=\tev{7}$.}
    \label{table:ratio_7TeV_pt_2}
	\begin{ruledtabular}
    \begin{tabular}{c c c c c c}
    $\pt$ range & \textsc{dpmjet} & \textsc{qgsjet} & \textsc{sibyll} & \textsc{epos} & \textsc{pythia} \\
	$[$GeV$]$   & 3.06            & II-04           & 2.1             & \textsc{lhc}  & 8.185           \\
    \hline
$[$0.00, 0.05$]$ &  1.53 &  1.71 &  0.72 &  1.20 &  1.46 \\
$[$0.05, 0.10$]$ &  1.41 &  1.40 &  0.64 &  1.12 &  1.34 \\
$[$0.10, 0.15$]$ &  1.23 &  1.04 &  0.52 &  0.93 &  1.15 \\
$[$0.15, 0.20$]$ &  1.18 &  0.84 &  0.50 &  0.76 &  1.12 \\
$[$0.20, 0.25$]$ &  1.39 &  0.78 &  0.56 &  0.95 &  1.24 \\
$[$0.25, 0.30$]$ &  1.53 &  0.72 &  0.60 &  1.00 &  1.28 \\
$[$0.30, 0.35$]$ &  2.06 &  0.71 &  0.76 &  0.88 &  1.59 \\
$[$0.35, 0.40$]$ &  2.91 &  0.56 &  0.94 &  0.88 &  2.07 \\
$[$0.40, 0.45$]$ &  7.74 &  0.80 &  2.17 &  1.95 &  4.87 \\
$[$0.45, 0.50$]$ & 18.99 &  1.13 &  4.27 &  3.54 & 10.20 \\
$[$0.50, 0.55$]$ & 40.53 &  1.21 &  7.25 &  7.54 & 17.05 \\
$[$0.55, 0.60$]$ & 13.68 &  0.32 &  2.01 &  3.24 &  4.50 \\    
    \end{tabular}
	\end{ruledtabular}
\end{table}

\begin{table}[htbp]
    \caption{Ratio of $\pizero$ production rate of MC simulation to data in the
    rapidity range $9.4<y<9.6$ in $\pp$ collisions at $\sqs=\tev{7}$.}
    \label{table:ratio_7TeV_pt_3}
	\begin{ruledtabular}
    \begin{tabular}{c c c c c c}
    $\pt$ range & \textsc{dpmjet} & \textsc{qgsjet} & \textsc{sibyll} & \textsc{epos} & \textsc{pythia} \\
	$[$GeV$]$   & 3.06            & II-04           & 2.1             & \textsc{lhc}  & 8.185           \\
    \hline
$[$0.00, 0.05$]$ &  1.24 &  1.35 &  0.52 &  0.98 &  1.13 \\
$[$0.05, 0.10$]$ &  1.25 &  1.19 &  0.49 &  0.94 &  1.14 \\
$[$0.10, 0.15$]$ &  1.15 &  0.91 &  0.44 &  0.74 &  1.05 \\
$[$0.15, 0.20$]$ &  1.29 &  0.80 &  0.48 &  0.83 &  1.14 \\
$[$0.20, 0.25$]$ &  1.47 &  0.74 &  0.52 &  0.99 &  1.19 \\
$[$0.25, 0.30$]$ &  2.15 &  0.71 &  0.66 &  0.80 &  1.57 \\
$[$0.30, 0.35$]$ &  3.75 &  0.57 &  0.89 &  0.94 &  2.45 \\
$[$0.35, 0.40$]$ &  7.32 &  0.54 &  1.29 &  1.38 &  4.14 \\
$[$0.40, 0.45$]$ & 20.42 &  0.64 &  2.43 &  3.26 &  8.97 \\    
    \end{tabular}
	\end{ruledtabular}
\end{table}

\begin{table}[htbp]
    \caption{Ratio of $\pizero$ production rate of MC simulation to data in the
    rapidity range $9.6<y<9.8$ in $\pp$ collisions at $\sqs=\tev{7}$.}
    \label{table:ratio_7TeV_pt_4}
	\begin{ruledtabular}
    \begin{tabular}{c c c c c c}
    $\pt$ range & \textsc{dpmjet} & \textsc{qgsjet} & \textsc{sibyll} & \textsc{epos} & \textsc{pythia} \\
	$[$GeV$]$   & 3.06            & II-04           & 2.1             & \textsc{lhc}  & 8.185           \\
    \hline
$[$0.00, 0.05$]$ &  1.00 &  1.02 &  0.35 &  0.70 &  0.88 \\
$[$0.05, 0.10$]$ &  1.15 &  1.03 &  0.42 &  0.72 &  1.04 \\
$[$0.10, 0.15$]$ &  1.28 &  0.87 &  0.44 &  0.75 &  1.10 \\
$[$0.15, 0.20$]$ &  1.46 &  0.78 &  0.45 &  0.97 &  1.14 \\
$[$0.20, 0.25$]$ &  2.09 &  0.67 &  0.53 &  0.67 &  1.46 \\
$[$0.25, 0.30$]$ &  5.93 &  0.72 &  1.02 &  1.25 &  3.78 \\
$[$0.30, 0.35$]$ & 10.39 &  0.50 &  1.14 &  1.60 &  5.78 \\
$[$0.35, 0.40$]$ & 16.08 &  0.31 &  1.01 &  3.17 &  7.07 \\    
    \end{tabular}
	\end{ruledtabular}
\end{table}

\begin{table}[htbp]
    \caption{Ratio of $\pizero$ production rate of MC simulation to data in the
    rapidity range $9.8<y<10.0$ in $\pp$ collisions at $\sqs=\tev{7}$.}
    \label{table:ratio_7TeV_pt_5}
	\begin{ruledtabular}
    \begin{tabular}{c c c c c c}
    $\pt$ range & \textsc{dpmjet} & \textsc{qgsjet} & \textsc{sibyll} & \textsc{epos} & \textsc{pythia} \\
	$[$GeV$]$   & 3.06            & II-04           & 2.1             & \textsc{lhc}  & 8.185           \\
    \hline
$[$0.00, 0.05$]$ &  1.20 &  1.13 &  0.40 &  0.62 &  1.03 \\
$[$0.05, 0.10$]$ &  1.38 &  1.03 &  0.43 &  0.76 &  1.14 \\
$[$0.10, 0.15$]$ &  1.54 &  0.89 &  0.43 &  1.03 &  1.17 \\
$[$0.15, 0.20$]$ &  2.26 &  0.75 &  0.49 &  0.67 &  1.54 \\
$[$0.20, 0.25$]$ &  6.10 &  0.65 &  0.77 &  1.11 &  3.99 \\
$[$0.25, 0.30$]$ & 12.79 &  0.43 &  0.91 &  2.01 &  7.39 \\    
    \end{tabular}
	\end{ruledtabular}
\end{table}

\begin{table}[htbp]
    \caption{Ratio of $\pizero$ production rate of MC simulation to data in the
    rapidity range $10.0<y<10.2$ in $\pp$ collisions at $\sqs=\tev{7}$.}
    \label{table:ratio_7TeV_pt_6}
	\begin{ruledtabular}
    \begin{tabular}{c c c c c c}
    $\pt$ range & \textsc{dpmjet} & \textsc{qgsjet} & \textsc{sibyll} & \textsc{epos} & \textsc{pythia} \\
	$[$GeV$]$   & 3.06            & II-04           & 2.1             & \textsc{lhc}  & 8.185           \\
    \hline
$[$0.00, 0.05$]$ &  1.57 &  1.14 &  0.42 &  1.06 &  1.20 \\
$[$0.05, 0.10$]$ &  1.73 &  1.10 &  0.44 &  1.11 &  1.24 \\
$[$0.10, 0.15$]$ &  2.74 &  0.95 &  0.51 &  0.77 &  1.83 \\
$[$0.15, 0.20$]$ &  9.16 &  0.97 &  0.90 &  1.55 &  6.16 \\
$[$0.20, 0.25$]$ & 10.38 &  0.31 &  0.55 &  1.74 &  6.50 \\    
    \end{tabular}
	\end{ruledtabular}
\end{table}

\begin{table}[htbp]
    \caption{Ratio of $\pizero$ production rate of MC simulation to data in the
    rapidity range $10.2<y<10.4$ in $\pp$ collisions at $\sqs=\tev{7}$.}
    \label{table:ratio_7TeV_pt_7}
	\begin{ruledtabular}
    \begin{tabular}{c c c c c c}
    $\pt$ range & \textsc{dpmjet} & \textsc{qgsjet} & \textsc{sibyll} & \textsc{epos} & \textsc{pythia} \\
	$[$GeV$]$   & 3.06            & II-04           & 2.1             & \textsc{lhc}  & 8.185           \\
    \hline
$[$0.00, 0.05$]$ &  2.22 &  1.29 &  0.45 &  0.88 &  1.42 \\
$[$0.05, 0.10$]$ &  3.11 &  1.05 &  0.49 &  0.75 &  1.99 \\
$[$0.10, 0.15$]$ &  8.51 &  0.95 &  0.71 &  1.35 &  5.94 \\
$[$0.15, 0.20$]$ & 52.42 &  1.64 &  2.39 &  9.00 & 37.20 \\    
    \end{tabular}
	\end{ruledtabular}
\end{table}

\begin{table}[htbp]
    \caption{Ratio of $\pizero$ production rate of MC simulation to data in the
    rapidity range $10.4<y<10.6$ in $\pp$ collisions at $\sqs=\tev{7}$.}
    \label{table:ratio_7TeV_pt_8}
	\begin{ruledtabular}
    \begin{tabular}{c c c c c c}
    $\pt$ range & \textsc{dpmjet} & \textsc{qgsjet} & \textsc{sibyll} & \textsc{epos} & \textsc{pythia} \\
	$[$GeV$]$   & 3.06            & II-04           & 2.1             & \textsc{lhc}  & 8.185           \\
    \hline
$[$0.00, 0.05$]$ &  5.60 &  1.17 &  0.56 &  0.97 &  3.56 \\
$[$0.05, 0.10$]$ &  9.64 &  1.05 &  0.64 &  1.48 &  7.16 \\
$[$0.10, 0.15$]$ & 37.19 &  1.39 &  1.24 &  7.51 & 27.39 \\
$[$0.15, 0.20$]$ & 19.23 &  0.00 &  0.00 & 11.78 &  0.00 \\    
    \end{tabular}
	\end{ruledtabular}
\end{table}

\begin{table}[htbp]
    \caption{Ratio of $\pizero$ production rate of MC simulation to data in the
    rapidity range $10.6<y<10.8$ in $\pp$ collisions at $\sqs=\tev{7}$.}
    \label{table:ratio_7TeV_pt_9}
	\begin{ruledtabular}
    \begin{tabular}{c c c c c c}
    $\pt$ range & \textsc{dpmjet} & \textsc{qgsjet} & \textsc{sibyll} & \textsc{epos} & \textsc{pythia} \\
	$[$GeV$]$   & 3.06            & II-04           & 2.1             & \textsc{lhc}  & 8.185           \\
    \hline
$[$0.00, 0.05$]$ & 14.17 &  0.75 &  0.40 &  2.06 & 12.35 \\
$[$0.05, 0.10$]$ & 17.43 &  0.32 &  0.50 &  4.31 & 11.74 \\
$[$0.10, 0.15$]$ &  7.14 &  0.00 &  0.00 &  2.16 &  0.54 \\    
    \end{tabular}
	\end{ruledtabular}
\end{table}

\clearpage
\begin{table}[htbp]
    \caption{Ratio of $\pizero$ production rate of MC simulation to data in the
    $\pt$ range $0.0<\pt<\gev{0.2}$ in $\pp$ collisions at $\sqs=\tev{7}$.}
    \label{table:ratio_7TeV_pz_0}
	\begin{ruledtabular}
    \begin{tabular}{c c c c c c}
    $\pz$ range & \textsc{dpmjet} & \textsc{qgsjet} & \textsc{sibyll} & \textsc{epos} & \textsc{pythia} \\
	$[$GeV$]$   & 3.06            & II-04           & 2.1             & \textsc{lhc}  & 8.185           \\
    \hline
$[$\hph875, 1000$]$ &  1.20 &  1.10 &  0.52 &  0.93 &  1.12 \\
$[$1000, 1125$]$ &  1.12 &  1.00 &  0.45 &  0.82 &  1.04 \\
$[$1125, 1250$]$ &  0.97 &  0.83 &  0.37 &  0.65 &  0.88 \\
$[$1250, 1375$]$ &  1.09 &  0.88 &  0.40 &  0.61 &  0.98 \\
$[$1375, 1500$]$ &  1.20 &  0.89 &  0.41 &  0.58 &  1.03 \\
$[$1500, 1625$]$ &  1.28 &  0.88 &  0.41 &  0.76 &  1.07 \\
$[$1625, 1750$]$ &  1.35 &  0.87 &  0.41 &  1.20 &  1.07 \\
$[$1750, 1875$]$ &  1.48 &  0.92 &  0.41 &  1.04 &  1.09 \\
$[$1875, 2000$]$ &  1.69 &  1.00 &  0.42 &  0.82 &  1.17 \\
$[$2000, 2125$]$ &  2.00 &  0.99 &  0.44 &  0.67 &  1.34 \\
$[$2125, 2250$]$ &  2.56 &  0.98 &  0.51 &  0.67 &  1.65 \\
$[$2250, 2375$]$ &  2.88 &  0.70 &  0.43 &  0.63 &  1.82 \\
$[$2375, 2500$]$ &  5.17 &  1.03 &  0.63 &  1.00 &  3.27 \\
$[$2500, 2625$]$ &  6.87 &  0.98 &  0.65 &  1.25 &  4.50 \\
$[$2625, 2750$]$ &  8.61 &  0.92 &  0.64 &  1.35 &  6.24 \\
$[$2750, 2875$]$ & 11.68 &  0.81 &  0.65 &  1.48 &  8.78 \\
$[$2875, 3000$]$ & 16.87 &  0.87 &  0.69 &  2.10 & 13.53 \\
$[$3000, 3125$]$ & 16.61 &  0.45 &  0.60 &  2.15 & 14.06 \\
$[$3125, 3250$]$ & 21.18 &  0.58 &  0.66 &  4.61 & 19.13 \\
$[$3250, 3375$]$ & 44.10 &  0.27 &  0.66 & 17.76 & 10.21 \\
$[$3375, 3500$]$ & 20.03 &  0.00 &  0.05 & 11.78 &  0.99 \\    
    \end{tabular}
	\end{ruledtabular}
\end{table}

\begin{table}[htbp]
    \caption{Ratio of $\pizero$ production rate of MC simulation to data in the
	$\pt$ range $0.2<\pt<\gev{0.4}$ in $\pp$ collisions at $\sqs=\tev{7}$.}
    \label{table:ratio_7TeV_pz_1}
	\begin{ruledtabular}
    \begin{tabular}{c c c c c c}
    $\pz$ range & \textsc{dpmjet} & \textsc{qgsjet} & \textsc{sibyll} & \textsc{epos} & \textsc{pythia} \\
	$[$GeV$]$   & 3.06            & II-04           & 2.1             & \textsc{lhc}  & 8.185           \\
    \hline
$[$1000, 1125$]$ &  1.39 &  0.81 &  0.72 &  0.98 &  1.37 \\
$[$1125, 1250$]$ &  1.36 &  0.75 &  0.67 &  0.93 &  1.31 \\
$[$1250, 1375$]$ &  1.39 &  0.73 &  0.66 &  0.95 &  1.30 \\
$[$1375, 1500$]$ &  1.41 &  0.70 &  0.63 &  0.99 &  1.29 \\
$[$1500, 1625$]$ &  1.50 &  0.72 &  0.65 &  1.05 &  1.32 \\
$[$1625, 1750$]$ &  1.50 &  0.68 &  0.61 &  0.95 &  1.25 \\
$[$1750, 1875$]$ &  1.64 &  0.71 &  0.62 &  0.85 &  1.31 \\
$[$1875, 2000$]$ &  1.87 &  0.71 &  0.65 &  0.76 &  1.41 \\
$[$2000, 2125$]$ &  2.12 &  0.63 &  0.65 &  0.69 &  1.54 \\
$[$2125, 2250$]$ &  2.92 &  0.62 &  0.79 &  0.80 &  2.02 \\
$[$2250, 2375$]$ &  3.89 &  0.59 &  0.86 &  0.95 &  2.57 \\
$[$2375, 2500$]$ &  4.93 &  0.58 &  0.93 &  1.07 &  3.11 \\
$[$2500, 2625$]$ & 10.24 &  0.92 &  1.59 &  2.04 &  6.40 \\
$[$2625, 2750$]$ & 12.52 &  0.81 &  1.57 &  1.92 &  7.62 \\
$[$2750, 2875$]$ & 18.24 &  0.73 &  1.75 &  2.33 & 10.62 \\
$[$2875, 3000$]$ & 22.25 &  0.68 &  1.61 &  2.99 & 13.15 \\
$[$3000, 3125$]$ & 30.98 &  0.55 &  1.45 &  4.81 & 16.99 \\
$[$3125, 3250$]$ & 36.45 &  0.46 &  1.81 &  9.28 & 21.07 \\
$[$3250, 3375$]$ & 39.91 &  0.25 &  1.05 & 13.29 &  6.50 \\
$[$3375, 3500$]$ & 39.63 &  0.05 &  0.00 & 15.28 &  2.05 \\    
    \end{tabular}
	\end{ruledtabular}
\end{table}

\begin{table}[htbp]
    \caption{Ratio of $\pizero$ production rate of MC simulation to data in the
    $\pt$ range $0.4<\pt<\gev{0.6}$ in $\pp$ collisions at $\sqs=\tev{7}$.}
    \label{table:ratio_7TeV_pz_2}
	\begin{ruledtabular}
    \begin{tabular}{c c c c c c}
    $\pz$ range & \textsc{dpmjet} & \textsc{qgsjet} & \textsc{sibyll} & \textsc{epos} & \textsc{pythia} \\
	$[$GeV$]$   & 3.06            & II-04           & 2.1             & \textsc{lhc}  & 8.185           \\
    \hline
$[$1500, 1625$]$ &  2.66 &  0.89 &  1.65 &  1.35 &  2.46 \\
$[$1625, 1750$]$ &  2.41 &  0.72 &  1.41 &  1.11 &  2.14 \\
$[$1750, 1875$]$ &  2.27 &  0.60 &  1.25 &  0.96 &  1.94 \\
$[$1875, 2000$]$ &  2.82 &  0.63 &  1.44 &  1.10 &  2.30 \\
$[$2000, 2125$]$ &  3.51 &  0.64 &  1.64 &  1.21 &  2.75 \\
$[$2125, 2250$]$ &  3.42 &  0.48 &  1.45 &  1.10 &  2.51 \\
$[$2250, 2375$]$ &  4.90 &  0.56 &  1.85 &  1.45 &  3.36 \\
$[$2375, 2500$]$ &  5.95 &  0.53 &  1.92 &  1.52 &  3.74 \\
$[$2500, 2625$]$ &  7.72 &  0.55 &  2.16 &  1.61 &  4.48 \\
$[$2625, 2750$]$ & 13.14 &  0.70 &  3.03 &  2.22 &  6.82 \\
$[$2750, 2875$]$ & 20.33 &  0.73 &  3.66 &  3.38 &  9.43 \\
$[$2875, 3000$]$ & 19.78 &  0.52 &  2.69 &  3.09 &  8.08 \\    
    \end{tabular}
	\end{ruledtabular}
\end{table}

\begin{table}[htbp]
    \caption{Ratio of $\pizero$ production rate of MC simulation to data in the
    $\pt$ range $0.6<\pt<\gev{0.8}$ in $\pp$ collisions at $\sqs=\tev{7}$.}
    \label{table:ratio_7TeV_pz_3}
	\begin{ruledtabular}
    \begin{tabular}{c c c c c c}
    $\pz$ range & \textsc{dpmjet} & \textsc{qgsjet} & \textsc{sibyll} & \textsc{epos} & \textsc{pythia} \\
	$[$GeV$]$   & 3.06            & II-04           & 2.1             & \textsc{lhc}  & 8.185           \\
    \hline
$[$2000, 2125$]$ &  3.18 &  0.47 &  2.32 &  1.25 &  2.80 \\
$[$2125, 2250$]$ &  4.30 &  0.53 &  2.85 &  1.52 &  3.60 \\
$[$2250, 2375$]$ &  4.50 &  0.46 &  2.81 &  1.44 &  3.51 \\
$[$2375, 2500$]$ &  6.52 &  0.55 &  3.71 &  1.83 &  4.98 \\
$[$2500, 2625$]$ &  7.37 &  0.50 &  3.69 &  1.93 &  5.06 \\
$[$2625, 2750$]$ &  8.01 &  0.45 &  3.51 &  1.85 &  5.05 \\
$[$2750, 2875$]$ &  9.74 &  0.37 &  3.53 &  2.03 &  5.40 \\
$[$2875, 3000$]$ & 15.66 &  0.46 &  4.41 &  3.52 &  7.32 \\
$[$3000, 3125$]$ & 22.78 &  0.50 &  4.98 &  5.70 &  8.80 \\
$[$3125, 3250$]$ & 18.98 &  0.39 &  3.92 &  7.66 &  7.07 \\
$[$3250, 3375$]$ & 10.02 &  0.13 &  0.89 &  5.85 &  2.73 \\    
    \end{tabular}
	\end{ruledtabular}
\end{table}

\begin{table}[htbp]
    \caption{Ratio of $\pizero$ production rate of MC simulation to data in the
    $\pt$ range $0.8<\pt<\gev{1.0}$ in $\pp$ collisions at $\sqs=\tev{7}$.}
    \label{table:ratio_7TeV_pz_4}
	\begin{ruledtabular}
    \begin{tabular}{c c c c c c}
    $\pz$ range & \textsc{dpmjet} & \textsc{qgsjet} & \textsc{sibyll} & \textsc{epos} & \textsc{pythia} \\
	$[$GeV$]$   & 3.06            & II-04           & 2.1             & \textsc{lhc}  & 8.185           \\
    \hline
$[$2250, 2375$]$ &  4.75 &  0.45 &  4.34 &  1.72 &  4.54 \\
$[$2375, 2500$]$ &  6.50 &  0.55 &  5.55 &  2.45 &  5.91 \\
$[$2500, 2625$]$ &  8.07 &  0.57 &  6.18 &  2.67 &  6.95 \\
$[$2625, 2750$]$ & 11.27 &  0.59 &  7.67 &  3.30 &  9.16 \\
$[$2750, 2875$]$ &  9.22 &  0.35 &  5.65 &  2.56 &  7.02 \\
$[$2875, 3000$]$ & 22.52 &  0.65 & 10.80 &  7.06 & 15.46 \\
$[$3000, 3125$]$ & 16.96 &  0.32 &  6.48 &  6.12 & 10.31 \\
$[$3125, 3250$]$ & 46.87 &  0.90 & 15.67 & 23.77 & 26.10 \\
$[$3250, 3375$]$ & 33.52 &  0.43 &  7.16 & 30.40 & 17.58 \\    
    \end{tabular}
	\end{ruledtabular}
\end{table}

\clearpage
\begin{table}[htbp]
    \caption{Ratio of $\pizero$ production rate of MC simulation to data in the
    rapidity range $8.8<y<9.0$ in $\pp$ collisions at $\sqs=\tev{2.76}$.}
    \label{table:ratio_2TeV_pt_0}
	\begin{ruledtabular}
    \begin{tabular}{c c c c c c}
    $\pt$ range & \textsc{dpmjet} & \textsc{qgsjet} & \textsc{sibyll} & \textsc{epos} & \textsc{pythia} \\
	$[$GeV$]$   & 3.06            & II-04           & 2.1             & \textsc{lhc}  & 8.185           \\
    \hline
$[$0.10, 0.15$]$ &  1.67 &  1.13 &  0.61 &  1.24 &  1.45 \\
$[$0.15, 0.20$]$ &  2.04 &  0.93 &  0.62 &  0.87 &  1.59 \\
$[$0.20, 0.25$]$ &  6.68 &  1.23 &  1.36 &  1.57 &  4.90 \\
$[$0.25, 0.30$]$ & 20.32 &  1.32 &  2.43 &  3.07 & 14.56 \\    
    \end{tabular}
	\end{ruledtabular}
\end{table}

\begin{table}[htbp]
    \caption{Ratio of $\pizero$ production rate of MC simulation to data in the
    rapidity range $9.0<y<9.2$ in $\pp$ collisions at $\sqs=\tev{2.76}$.}
    \label{table:ratio_2TeV_pt_1}
	\begin{ruledtabular}
    \begin{tabular}{c c c c c c}
    $\pt$ range & \textsc{dpmjet} & \textsc{qgsjet} & \textsc{sibyll} & \textsc{epos} & \textsc{pythia} \\
	$[$GeV$]$   & 3.06            & II-04           & 2.1             & \textsc{lhc}  & 8.185           \\
    \hline
$[$0.05, 0.10$]$ &  1.47 &  1.08 &  0.49 &  1.07 &  1.20 \\
$[$0.10, 0.15$]$ &  1.85 &  0.91 &  0.51 &  0.77 &  1.40 \\
$[$0.15, 0.20$]$ &  5.73 &  1.01 &  0.93 &  1.18 &  4.16 \\
$[$0.20, 0.25$]$ & 34.50 &  2.00 &  3.01 &  4.31 & 26.06 \\    
    \end{tabular}
	\end{ruledtabular}
\end{table}

\begin{table}[htbp]
    \caption{Ratio of $\pizero$ production rate of MC simulation to data in the
    rapidity range $9.2<y<9.4$ in $\pp$ collisions at $\sqs=\tev{2.76}$.}
    \label{table:ratio_2TeV_pt_2}
	\begin{ruledtabular}
    \begin{tabular}{c c c c c c}
    $\pt$ range & \textsc{dpmjet} & \textsc{qgsjet} & \textsc{sibyll} & \textsc{epos} & \textsc{pythia} \\
	$[$GeV$]$   & 3.06            & II-04           & 2.1             & \textsc{lhc}  & 8.185           \\
    \hline
$[$0.00, 0.05$]$ &  1.22 &  0.89 &  0.34 &  0.72 &  0.91 \\
$[$0.05, 0.10$]$ &  2.00 &  0.98 &  0.46 &  0.69 &  1.41 \\
$[$0.10, 0.15$]$ &  3.52 &  0.64 &  0.46 &  0.65 &  2.54 \\
$[$0.15, 0.20$]$ & 10.45 &  0.63 &  0.73 &  1.43 &  8.89 \\
$[$0.20, 0.25$]$ & 15.42 &  0.34 &  0.33 &  4.37 &  7.87 \\    
    \end{tabular}
	\end{ruledtabular}
\end{table}

\begin{table}[htbp]
    \caption{Ratio of $\pizero$ production rate of MC simulation to data in the
    rapidity range $9.4<y<9.6$ in $\pp$ collisions at $\sqs=\tev{2.76}$.}
    \label{table:ratio_2TeV_pt_3}
	\begin{ruledtabular}
    \begin{tabular}{c c c c c c}
    $\pt$ range & \textsc{dpmjet} & \textsc{qgsjet} & \textsc{sibyll} & \textsc{epos} & \textsc{pythia} \\
	$[$GeV$]$   & 3.06            & II-04           & 2.1             & \textsc{lhc}  & 8.185           \\
    \hline
$[$0.00, 0.05$]$ &  2.12 &  0.72 &  0.31 &  0.47 &  1.36 \\
$[$0.05, 0.10$]$ &  3.65 &  0.70 &  0.39 &  0.59 &  2.86 \\
$[$0.10, 0.15$]$ & 11.29 &  0.68 &  0.69 &  1.48 &  9.99 \\
$[$0.15, 0.20$]$ &  9.89 &  0.19 &  0.18 &  3.04 &  4.37 \\    
    \end{tabular}
	\end{ruledtabular}
\end{table}

\begin{table}[htbp]
    \caption{Ratio of $\pizero$ production rate of MC simulation to data in the
    rapidity range $9.6<y<9.8$ in $\pp$ collisions at $\sqs=\tev{2.76}$.}
    \label{table:ratio_2TeV_pt_4}
	\begin{ruledtabular}
    \begin{tabular}{c c c c c c}
    $\pt$ range & \textsc{dpmjet} & \textsc{qgsjet} & \textsc{sibyll} & \textsc{epos} & \textsc{pythia} \\
	$[$GeV$]$   & 3.06            & II-04           & 2.1             & \textsc{lhc}  & 8.185           \\
    \hline
$[$0.00, 0.05$]$ &  4.63 &  0.51 &  0.38 &  0.56 &  4.35 \\
$[$0.05, 0.10$]$ &  8.44 &  0.48 &  0.36 &  1.06 &  7.80 \\
$[$0.10, 0.15$]$ &  5.46 &  0.06 &  0.13 &  1.66 &  1.81 \\    
    \end{tabular}
	\end{ruledtabular}
\end{table}

\clearpage
\begin{table}[htbp]
    \caption{Ratio of $\pizero$ production rate of MC simulation to data in the
    $\pt$ range $0.0<\pt<\gev{0.2}$ in $\pp$ collisions at $\sqs=\tev{2.76}$.}
    \label{table:ratio_2TeV_pz_0}
	\begin{ruledtabular}
    \begin{tabular}{c c c c c c}
    $\pz$ range & \textsc{dpmjet} & \textsc{qgsjet} & \textsc{sibyll} & \textsc{epos} & \textsc{pythia} \\
	$[$GeV$]$   & 3.06            & II-04           & 2.1             & \textsc{lhc}  & 8.185           \\
    \hline
$[$\hph625, \hph750$]$ &  1.34 &  0.94 &  0.46 &  1.05 &  1.11 \\
$[$\hph750, \hph875$]$ &  1.66 &  0.89 &  0.45 &  0.63 &  1.22 \\
$[$\hph875, 1000$]$ &  2.63 &  0.65 &  0.42 &  0.56 &  1.79 \\
$[$1000, 1125$]$ &  5.37 &  0.70 &  0.53 &  0.83 &  4.42 \\
$[$1125, 1250$]$ & 12.08 &  0.60 &  0.63 &  1.30 & 11.26 \\
$[$1250, 1380$]$ & 18.30 &  0.41 &  0.43 &  4.88 & 10.87 \\    
    \end{tabular}
	\end{ruledtabular}
\end{table}

\clearpage
\begin{table}[htbp]
    \caption{Ratio of $\pizero$ production rate of MC simulation to data in the
    $\pt$ range $0.2<\pt<\gev{0.4}$ in $\pp$ collisions at $\sqs=\tev{2.76}$.}
    \label{table:ratio_2TeV_pz_1}
	\begin{ruledtabular}
    \begin{tabular}{c c c c c c}
    $\pz$ range & \textsc{dpmjet} & \textsc{qgsjet} & \textsc{sibyll} & \textsc{epos} & \textsc{pythia} \\
	$[$GeV$]$   & 3.06            & II-04           & 2.1             & \textsc{lhc}  & 8.185           \\
    \hline
$[$\hph750, \hph875$]$ &  2.56 &  0.84 &  0.93 &  0.93 &  2.03 \\
$[$\hph875, 1000$]$ &  5.10 &  0.79 &  1.24 &  1.28 &  3.66 \\
$[$1000, 1125$]$ & 11.69 &  0.87 &  1.73 &  1.82 &  7.99 \\
$[$1125, 1250$]$ & 138.29 &  4.19 & 10.41 & 16.28 & 90.57 \\
$[$1250, 1380$]$ & 44.14 &  0.45 &  1.61 & 13.18 & 15.99 \\    
    \end{tabular}
	\end{ruledtabular}
\end{table}

\clearpage
\begin{table}[htbp]
    \caption{Ratio of $\pizero$ production rate of MC simulation to data in the
    rapidity range $-8.8>\ylab>-9.0$ in $\pPb$ collisions at $\snn=\tev{5.02}$.}
    \label{table:ratio_5TeV_pt_0}
	\begin{ruledtabular}
    \begin{tabular}{c c c c}
    $\pt$ range & \textsc{dpmjet} & \textsc{qgsjet} & \textsc{epos} \\
	$[$GeV$]$   & 3.06            & II-04           & \textsc{lhc}  \\
    \hline
$[$0.10, 0.15$]$ &  1.39 &  1.41 &  1.15 \\
$[$0.15, 0.20$]$ &  0.99 &  0.92 &  0.85 \\
$[$0.20, 0.25$]$ &  0.90 &  0.77 &  0.77 \\
$[$0.25, 0.30$]$ &  0.83 &  0.62 &  0.66 \\
$[$0.30, 0.35$]$ &  0.94 &  0.60 &  0.70 \\
$[$0.35, 0.40$]$ &  0.91 &  0.53 &  0.63 \\
$[$0.40, 0.45$]$ &  1.03 &  0.55 &  0.69 \\
$[$0.45, 0.50$]$ &  1.35 &  0.65 &  0.85 \\
$[$0.50, 0.55$]$ &  1.28 &  0.56 &  0.79 \\
$[$0.55, 0.60$]$ &  1.62 &  0.58 &  0.93 \\
$[$0.60, 0.65$]$ &  1.89 &  0.60 &  1.07 \\
$[$0.65, 0.70$]$ &  2.11 &  0.58 &  1.15 \\
$[$0.70, 0.75$]$ &  2.03 &  0.51 &  1.11 \\
$[$0.75, 0.80$]$ &  3.44 &  0.79 &  1.77 \\
$[$0.80, 0.85$]$ &  5.00 &  1.05 &  2.68 \\
$[$0.85, 0.90$]$ &  2.83 &  0.56 &  1.75 \\
$[$0.90, 0.95$]$ & 10.56 &  1.88 &  6.44 \\
$[$0.95, 1.00$]$ & 29.43 &  4.41 & 23.12 \\    
    \end{tabular}
	\end{ruledtabular}
\end{table}

\begin{table}[htbp]
    \caption{Ratio of $\pizero$ production rate of MC simulation to data in the
    rapidity range $-9.0>\ylab>-9.2$ in $\pPb$ collisions at $\snn=\tev{5.02}$.}
    \label{table:ratio_5TeV_pt_1}
	\begin{ruledtabular}
    \begin{tabular}{c c c c}
    $\pt$ range & \textsc{dpmjet} & \textsc{qgsjet} & \textsc{epos} \\
	$[$GeV$]$   & 3.06            & II-04           & \textsc{lhc}  \\
    \hline
$[$0.05, 0.10$]$ &  1.41 &  1.57 &  1.15 \\
$[$0.10, 0.15$]$ &  1.05 &  1.03 &  0.87 \\
$[$0.15, 0.20$]$ &  0.86 &  0.78 &  0.73 \\
$[$0.20, 0.25$]$ &  0.85 &  0.71 &  0.71 \\
$[$0.25, 0.30$]$ &  0.98 &  0.68 &  0.73 \\
$[$0.30, 0.35$]$ &  0.96 &  0.57 &  0.65 \\
$[$0.35, 0.40$]$ &  1.14 &  0.64 &  0.74 \\
$[$0.40, 0.45$]$ &  1.30 &  0.65 &  0.79 \\
$[$0.45, 0.50$]$ &  1.43 &  0.62 &  0.81 \\
$[$0.50, 0.55$]$ &  1.51 &  0.49 &  0.77 \\
$[$0.55, 0.60$]$ &  2.80 &  0.75 &  1.32 \\
$[$0.60, 0.65$]$ &  3.33 &  0.73 &  1.52 \\
$[$0.65, 0.70$]$ &  2.59 &  0.50 &  1.13 \\
$[$0.70, 0.75$]$ &  6.67 &  1.14 &  2.88 \\
$[$0.75, 0.80$]$ &  4.88 &  0.87 &  2.48 \\    
    \end{tabular}
	\end{ruledtabular}
\end{table}

\begin{table}[htbp]
    \caption{Ratio of $\pizero$ production rate of MC simulation to data in the
    rapidity range $-9.2>\ylab>-9.4$ in $\pPb$ collisions at $\snn=\tev{5.02}$.}
    \label{table:ratio_5TeV_pt_2}
	\begin{ruledtabular}
    \begin{tabular}{c c c c}
    $\pt$ range & \textsc{dpmjet} & \textsc{qgsjet} & \textsc{epos} \\
	$[$GeV$]$   & 3.06            & II-04           & \textsc{lhc}  \\
    \hline
$[$0.00, 0.05$]$ &  1.39 &  1.68 &  1.12 \\
$[$0.05, 0.10$]$ &  1.08 &  1.16 &  0.88 \\
$[$0.10, 0.15$]$ &  0.92 &  0.87 &  0.75 \\
$[$0.15, 0.20$]$ &  0.79 &  0.70 &  0.67 \\
$[$0.20, 0.25$]$ &  0.88 &  0.69 &  0.68 \\
$[$0.25, 0.30$]$ &  1.09 &  0.69 &  0.75 \\
$[$0.30, 0.35$]$ &  1.16 &  0.67 &  0.75 \\
$[$0.35, 0.40$]$ &  1.32 &  0.66 &  0.75 \\
$[$0.40, 0.45$]$ &  1.41 &  0.56 &  0.71 \\
$[$0.45, 0.50$]$ &  1.80 &  0.49 &  0.76 \\
$[$0.50, 0.55$]$ &  5.02 &  1.09 &  1.86 \\
$[$0.55, 0.60$]$ &  4.57 &  0.81 &  1.63 \\    
    \end{tabular}
	\end{ruledtabular}
\end{table}

\begin{table}[htbp]
    \caption{Ratio of $\pizero$ production rate of MC simulation to data in the
    rapidity range $-9.4>\ylab>-9.6$ in $\pPb$ collisions at $\snn=\tev{5.02}$.}
    \label{table:ratio_5TeV_pt_3}
	\begin{ruledtabular}
    \begin{tabular}{c c c c}
    $\pt$ range & \textsc{dpmjet} & \textsc{qgsjet} & \textsc{epos} \\
	$[$GeV$]$   & 3.06            & II-04           & \textsc{lhc}  \\
    \hline
$[$0.00, 0.05$]$ &  1.12 &  1.27 &  0.90 \\
$[$0.05, 0.10$]$ &  0.98 &  0.99 &  0.78 \\
$[$0.10, 0.15$]$ &  0.86 &  0.79 &  0.71 \\
$[$0.15, 0.20$]$ &  0.85 &  0.71 &  0.67 \\
$[$0.20, 0.25$]$ &  1.02 &  0.67 &  0.70 \\
$[$0.25, 0.30$]$ &  1.14 &  0.68 &  0.71 \\
$[$0.30, 0.35$]$ &  0.95 &  0.46 &  0.50 \\
$[$0.35, 0.40$]$ &  1.27 &  0.42 &  0.54 \\
$[$0.40, 0.45$]$ &  2.64 &  0.59 &  0.89 \\    
    \end{tabular}
	\end{ruledtabular}
\end{table}

\begin{table}[htbp]
    \caption{Ratio of $\pizero$ production rate of MC simulation to data in the
    rapidity range $-9.6>\ylab>-9.8$ in $\pPb$ collisions at $\snn=\tev{5.02}$.}
    \label{table:ratio_5TeV_pt_4}
	\begin{ruledtabular}
    \begin{tabular}{c c c c}
    $\pt$ range & \textsc{dpmjet} & \textsc{qgsjet} & \textsc{epos} \\
	$[$GeV$]$   & 3.06            & II-04           & \textsc{lhc}  \\
    \hline
$[$0.00, 0.05$]$ &  0.96 &  1.02 &  0.78 \\
$[$0.05, 0.10$]$ &  0.88 &  0.84 &  0.71 \\
$[$0.10, 0.15$]$ &  0.92 &  0.80 &  0.73 \\
$[$0.15, 0.20$]$ &  1.04 &  0.73 &  0.72 \\
$[$0.20, 0.25$]$ &  1.19 &  0.73 &  0.72 \\
$[$0.25, 0.30$]$ &  1.36 &  0.62 &  0.64 \\
$[$0.30, 0.35$]$ &  2.73 &  0.77 &  0.96 \\
$[$0.35, 0.40$]$ &  9.27 &  1.71 &  2.57 \\    
    \end{tabular}
	\end{ruledtabular}
\end{table}

\begin{table}[htbp]
    \caption{Ratio of $\pizero$ production rate of MC simulation to data in the
    rapidity range $-9.8>\ylab>-10.0$ in $\pPb$ collisions at $\snn=\tev{5.02}$.}
    \label{table:ratio_5TeV_pt_5}
	\begin{ruledtabular}
    \begin{tabular}{c c c c}
    $\pt$ range & \textsc{dpmjet} & \textsc{qgsjet} & \textsc{epos} \\
	$[$GeV$]$   & 3.06            & II-04           & \textsc{lhc}  \\
    \hline
$[$0.00, 0.05$]$ &  0.86 &  0.85 &  0.68 \\
$[$0.05, 0.10$]$ &  0.92 &  0.83 &  0.71 \\
$[$0.10, 0.15$]$ &  1.18 &  0.89 &  0.83 \\
$[$0.15, 0.20$]$ &  1.27 &  0.80 &  0.75 \\
$[$0.20, 0.25$]$ &  1.23 &  0.54 &  0.53 \\
$[$0.25, 0.30$]$ &  2.98 &  0.73 &  0.90 \\    
    \end{tabular}
	\end{ruledtabular}
\end{table}

\begin{table}[htbp]
    \caption{Ratio of $\pizero$ production rate of MC simulation to data in the
    rapidity range $-10.0>\ylab>-10.2$ in $\pPb$ collisions at $\snn=\tev{5.02}$.}
    \label{table:ratio_5TeV_pt_6}
	\begin{ruledtabular}
    \begin{tabular}{c c c c}
    $\pt$ range & \textsc{dpmjet} & \textsc{qgsjet} & \textsc{epos} \\
	$[$GeV$]$   & 3.06            & II-04           & \textsc{lhc}  \\
    \hline
$[$0.00, 0.05$]$ &  1.00 &  0.90 &  0.74 \\
$[$0.05, 0.10$]$ &  1.14 &  0.91 &  0.81 \\
$[$0.10, 0.15$]$ &  1.77 &  1.15 &  1.00 \\
$[$0.15, 0.20$]$ &  1.28 &  0.55 &  0.52 \\
$[$0.20, 0.25$]$ &  4.66 &  1.06 &  1.31 \\    
    \end{tabular}
	\end{ruledtabular}
\end{table}

\begin{table}[htbp]
    \caption{Ratio of $\pizero$ production rate of MC simulation to data in the
    rapidity range $-10.2>\ylab>-10.4$ in $\pPb$ collisions at $\snn=\tev{5.02}$.}
    \label{table:ratio_5TeV_pt_7}
	\begin{ruledtabular}
    \begin{tabular}{c c c c}
    $\pt$ range & \textsc{dpmjet} & \textsc{qgsjet} & \textsc{epos} \\
	$[$GeV$]$   & 3.06            & II-04           & \textsc{lhc}  \\
    \hline
$[$0.00, 0.05$]$ &  1.73 &  1.43 &  1.25 \\
$[$0.05, 0.10$]$ &  1.52 &  1.06 &  0.85 \\
$[$0.10, 0.15$]$ &  1.56 &  0.68 &  0.60 \\
$[$0.15, 0.20$]$ &  4.02 &  0.89 &  1.04 \\    
    \end{tabular}
	\end{ruledtabular}
\end{table}

\begin{table}[htbp]
    \caption{Ratio of $\pizero$ production rate of MC simulation to data in the
    rapidity range $-10.4>\ylab>-10.6$ in $\pPb$ collisions at $\snn=\tev{5.02}$.}
    \label{table:ratio_5TeV_pt_8}
	\begin{ruledtabular}
    \begin{tabular}{c c c c}
    $\pt$ range & \textsc{dpmjet} & \textsc{qgsjet} & \textsc{epos} \\
	$[$GeV$]$   & 3.06            & II-04           & \textsc{lhc}  \\
    \hline
$[$0.00, 0.05$]$ &  1.40 &  0.92 &  0.75 \\
$[$0.05, 0.10$]$ &  2.68 &  1.18 &  0.93 \\
$[$0.10, 0.15$]$ &  5.58 &  1.32 &  1.45 \\
$[$0.15, 0.20$]$ &  1.30 &  0.15 &  0.39 \\    
    \end{tabular}
	\end{ruledtabular}
\end{table}

\begin{table}[htbp]
    \caption{Ratio of $\pizero$ production rate of MC simulation to data in the
    rapidity range $-10.6>\ylab>-10.8$ in $\pPb$ collisions at $\snn=\tev{5.02}$.}
    \label{table:ratio_5TeV_pt_9}
	\begin{ruledtabular}
    \begin{tabular}{c c c c}
    $\pt$ range & \textsc{dpmjet} & \textsc{qgsjet} & \textsc{epos} \\
	$[$GeV$]$   & 3.06            & II-04           & \textsc{lhc}  \\
    \hline
$[$0.00, 0.05$]$ &  2.17 &  0.97 &  0.98 \\
$[$0.05, 0.10$]$ &  5.77 &  1.52 &  1.61 \\
$[$0.10, 0.15$]$ &  4.60 &  0.65 &  1.47 \\    
    \end{tabular}
	\end{ruledtabular}
\end{table}

\clearpage
\begin{table}[htbp]
    \caption{Ratio of $\pizero$ production rate of MC simulation to data in the
    $\pt$ range $0.0<\pt<\gev{0.2}$ in $\pPb$ collisions at $\snn=\tev{5.02}$.}
    \label{table:ratio_5TeV_pz_0}
	\begin{ruledtabular}
    \begin{tabular}{c c c c}
    $\pz$ range & \textsc{dpmjet} & \textsc{qgsjet} & \textsc{epos} \\
	$[$GeV$]$   & 3.06            & II-04           & \textsc{lhc}  \\
    \hline
$[$\hph875, 1000$]$ &  0.97 &  0.97 &  0.80 \\
$[$1000, 1125$]$ &  0.88 &  0.86 &  0.73 \\
$[$1125, 1250$]$ &  0.78 &  0.74 &  0.65 \\
$[$1250, 1375$]$ &  0.80 &  0.74 &  0.65 \\
$[$1375, 1500$]$ &  0.85 &  0.77 &  0.69 \\
$[$1500, 1625$]$ &  0.93 &  0.81 &  0.70 \\
$[$1625, 1750$]$ &  1.11 &  0.92 &  0.79 \\
$[$1750, 1875$]$ &  1.20 &  0.92 &  0.84 \\
$[$1875, 2000$]$ &  1.55 &  1.13 &  1.14 \\
$[$2000, 2125$]$ &  1.27 &  0.93 &  0.86 \\
$[$2125, 2250$]$ &  1.77 &  1.26 &  1.04 \\
$[$2250, 2375$]$ &  1.61 &  1.09 &  0.86 \\
$[$2375, 2500$]$ &  1.38 &  0.83 &  0.66 \\
$[$2500, 2625$]$ &  1.48 &  0.78 &  0.66 \\
$[$2625, 2750$]$ &  1.77 &  0.79 &  0.73 \\
$[$2750, 2875$]$ &  1.73 &  0.69 &  0.66 \\
$[$2875, 3000$]$ &  1.77 &  0.63 &  0.60 \\
$[$3000, 3125$]$ &  2.23 &  0.71 &  0.70 \\
$[$3125, 3250$]$ &  5.20 &  1.47 &  1.54 \\
$[$3250, 3375$]$ &  5.92 &  1.40 &  1.66 \\
$[$3375, 3500$]$ &  6.44 &  1.57 &  1.65 \\
$[$3500, 3625$]$ & 10.06 &  1.75 &  2.85 \\
$[$3625, 3750$]$ &  8.63 &  1.53 &  3.11 \\
$[$3750, 3875$]$ & 25.57 &  1.70 & 10.48 \\
$[$3875, 4000$]$ & 10.74 &  0.70 &  7.90 \\    
    \end{tabular}
	\end{ruledtabular}
\end{table}

\begin{table}[htbp]
    \caption{Ratio of $\pizero$ production rate of MC simulation to data in the
    $\pt$ range $0.2<\pt<\gev{0.4}$ in $\pPb$ collisions at $\snn=\tev{5.02}$.}
    \label{table:ratio_5TeV_pz_1}
	\begin{ruledtabular}
    \begin{tabular}{c c c c}
    $\pz$ range & \textsc{dpmjet} & \textsc{qgsjet} & \textsc{epos} \\
	$[$GeV$]$   & 3.06            & II-04           & \textsc{lhc}  \\
    \hline
$[$\hph875, 1000$]$ &  0.88 &  0.68 &  0.73 \\
$[$1000, 1125$]$ &  0.76 &  0.58 &  0.62 \\
$[$1125, 1250$]$ &  0.77 &  0.57 &  0.61 \\
$[$1250, 1375$]$ &  0.79 &  0.56 &  0.60 \\
$[$1375, 1500$]$ &  0.91 &  0.61 &  0.65 \\
$[$1500, 1625$]$ &  1.02 &  0.65 &  0.71 \\
$[$1625, 1750$]$ &  0.94 &  0.58 &  0.64 \\
$[$1750, 1875$]$ &  1.02 &  0.61 &  0.68 \\
$[$1875, 2000$]$ &  1.23 &  0.73 &  0.81 \\
$[$2000, 2125$]$ &  1.30 &  0.76 &  0.78 \\
$[$2125, 2250$]$ &  1.28 &  0.72 &  0.72 \\
$[$2250, 2375$]$ &  1.31 &  0.68 &  0.67 \\
$[$2375, 2500$]$ &  1.24 &  0.57 &  0.59 \\
$[$2500, 2625$]$ &  1.16 &  0.46 &  0.52 \\
$[$2625, 2750$]$ &  1.89 &  0.61 &  0.72 \\
$[$2750, 2875$]$ &  2.25 &  0.63 &  0.78 \\
$[$2875, 3000$]$ &  2.07 &  0.53 &  0.66 \\
$[$3000, 3125$]$ &  3.27 &  0.68 &  0.88 \\
$[$3125, 3250$]$ &  5.21 &  0.92 &  1.25 \\
$[$3250, 3375$]$ &  4.39 &  0.68 &  1.03 \\
$[$3375, 3500$]$ &  7.73 &  0.99 &  1.79 \\
$[$3500, 3625$]$ &  4.46 &  0.36 &  1.02 \\
$[$3625, 3750$]$ &  2.53 &  0.23 &  0.93 \\
$[$3750, 3875$]$ &  1.81 &  0.11 &  0.58 \\
$[$3875, 4000$]$ &  1.07 &  0.05 &  0.74 \\    
    \end{tabular}
	\end{ruledtabular}
\end{table}

\begin{table}[htbp]
    \caption{Ratio of $\pizero$ production rate of MC simulation to data in the
    $\pt$ range $0.4<\pt<\gev{0.6}$ in $\pPb$ collisions at $\snn=\tev{5.02}$.}
    \label{table:ratio_5TeV_pz_2}
	\begin{ruledtabular}
    %
	\end{ruledtabular}
\end{table}

\begin{table}[htbp]
    \caption{Ratio of $\pizero$ production rate of MC simulation to data in the
    $\pt$ range $0.6<\pt<\gev{0.8}$ in $\pPb$ collisions at $\snn=\tev{5.02}$.}
    \label{table:ratio_5TeV_pz_3}
	\begin{ruledtabular}
    %
	\end{ruledtabular}
\end{table}


\clearpage
\begin{table*}[htbp]
\caption{Nuclear modification factor for the $\pizero$s in the rapidity ranges
$-8.8>\ylab>-9.0$, $-9.0>\ylab>-9.2$, and $-9.2>\ylab>-9.4$ in $\pPb$ collisions
at $\snn=\tev{5.02}$. The uncertainties include the both statistical and
systematic uncertainties.}
	\label{table:nmftable_1}
	\begin{ruledtabular}
    %
	\end{ruledtabular}
\end{table*}

\begin{table*}[htbp]
\caption{Nuclear modification factor for the $\pizero$s in the rapidity ranges
$-9.4>\ylab>-9.6$, $-9.6>\ylab>-9.8$, and $-9.8>\ylab>-10.0$ in $\pPb$
collisions at $\snn=\tev{5.02}$. The uncertainties include the both statistical
and systematic uncertainties.}
	\label{table:nmftable_2}
	\begin{ruledtabular}
    %
	\end{ruledtabular}
\end{table*}

\begin{table*}[htbp]
\caption{Nuclear modification factor for the $\pizero$s in the rapidity ranges
$-10.0>\ylab>-10.2$, $-10.2>\ylab>-10.4$, and $-10.4>\ylab>-10.6$ in $\pPb$
collisions at $\snn=\tev{5.02}$. The uncertainties include the both statistical
and systematic uncertainties.}
	\label{table:nmftable_3}
	\begin{ruledtabular}
    %
	\end{ruledtabular}
\end{table*}

\begin{table*}[htbp]
\caption{Nuclear modification factor for the $\pizero$s in the rapidity ranges
$-10.6>\ylab>-10.8$ in $\pPb$ collisions at $\snn=\tev{5.02}$. The uncertainties
include the both statistical and systematic uncertainties.}
	\label{table:nmftable_4}
	\begin{ruledtabular}
	%
	\end{ruledtabular}
\end{table*}


\end{document}